\documentclass[runningheads]{llncs}
\usepackage{graphicx}
\usepackage{tikz}
\usepackage{comment}
\usepackage{amsmath,amssymb} % define this before the line numbering.
\usepackage{color}
\usepackage{booktabs}
\usetikzlibrary{spy}
\usepackage{algorithm}
\usepackage[noend]{algpseudocode}
\usepackage{custom/mycommands}
\usepackage{custom/mydefs}

\usepackage[capitalize]{cleveref}
\crefname{section}{Sec.}{Secs.}
\Crefname{section}{Section}{Sections}
\Crefname{table}{Table}{Tables}
\crefname{table}{Tab.}{Tabs.}

\usepackage[accsupp]{axessibility}  % Improves PDF readability for those with disabilities.

\begin{document}
% \renewcommand\thelinenumber{\color[rgb]{0.2,0.5,0.8}\normalfont\sffamily\scriptsize\arabic{linenumber}\color[rgb]{0,0,0}}
% \renewcommand\makeLineNumber {\hss\thelinenumber\ \hspace{6mm} \rlap{\hskip\textwidth\ \hspace{6.5mm}\thelinenumber}}
% \linenumbers
\pagestyle{headings}
\mainmatter
\def\ECCVSubNumber{537}  % Insert your submission number here

\title{Implicit Neural Representations for Image Compression} % Replace with your title

% INITIAL SUBMISSION 
\begin{comment}
\titlerunning{ECCV-22 submission ID \ECCVSubNumber} 
\authorrunning{ECCV-22 submission ID \ECCVSubNumber} 
\author{Anonymous ECCV submission}
\institute{Paper ID \ECCVSubNumber}
\end{comment}
%******************

% CAMERA READY SUBMISSION
% \begin{comment}
\titlerunning{Implicit Neural Representations for Image Compression}
% If the paper title is too long for the running head, you can set
% an abbreviated paper title here
%
\author{Yannick Strümpler$^*$  \inst{1} \and
Janis Postels$^*$  \inst{1} \and
Ren Yang \inst{1} \and
Luc Van Gool\inst{1} \and
Federico Tombari \inst{2, 3}}
\authorrunning{Y. Strümpler et al.}
% First names are abbreviated in the running head.
% If there are more than two authors, 'et al.' is used.
%
\institute{ETH Zurich \and Technical University of Munich \and Google}
% \end{comment}
%******************
\maketitle

\def\thefootnote{*}\footnotetext{Equal contribution}\def\thefootnote{\arabic{footnote}}

%%%%%%%%% ABSTRACT

\begin{abstract}
Implicit Neural Representations (INRs) gained attention as a novel and effective representation for various data types. 
%
% Thus far, prior work mostly focused on optimizing their reconstruction performance. 
Recently, prior work applied INRs to image compressing. Such compression algorithms are promising candidates as a general purpose approach for any coordinate-based data modality. 
%
%So far the focus has primarily been on optimizing the absolute performance of such networks.
%
%This work investigates INRs from a novel perspective, i.e., as a tool for image compression.
However, in order to live up to this promise current INR-based compression algorithms need to improve their rate-distortion performance by a large margin.
%
%We investigate a new direction, that optimizes such a representation in a source compression setting under the fundamental rate-distortion tradeoff. 
%
This work progresses on this problem. First, we propose meta-learned initializations for INR-based compression which improves rate-distortion performance. As a side effect it also leads to faster convergence speed.
Secondly, we introduce a simple yet highly effective change to the network architecture compared to prior work on INR-based compression. Namely, we combine SIREN networks with positional encodings which improves rate distortion performance. 
% To this end, we propose the first comprehensive image compression pipeline based on INRs including quantization, quantization-aware retraining and entropy coding.
%
%To this end, we use a SIREN network with an input encoding, overfit this to a single data sample and quantize the weights after training. To further improve rate-distortion performance, we employ regularization during overfitting and supplement the quantization step with adaptive rounding and quantization aware retraining. 
%
% Encoding with INRs, i.e., overfitting to a data sample, is typically orders of magnitude slower.
%
% To mitigate this drawback, we leverage meta-learned initializations based on MAML to reach the encoding in fewer gradient updates which also generally improves rate-distortion performance of INRs. 
%
%especially if the distribution of the data is known apriori.
%
Our contributions to source compression with INRs vastly outperform prior work. We show that our INR-based compression algorithm, meta-learning combined with SIREN and positional encodings, outperforms JPEG2000 and Rate-Distortion Autoencoders on Kodak with 2x reduced dimensionality for the first time and closes the gap on full resolution images. 
To underline the generality of INR-based source compression, we further perform experiments on 3D shape compression where our method greatly outperforms Draco - a traditional compression algorithm.
%is competitive with common compression algorithms designed specifically for images and closes the gap to state-of-the-art learned approaches based on Rate-Distortion Autoencoders.
%
% Moreover, we provide an extensive ablation study on the importance of individual components of our method which we hope facilitates future research on this novel approach to image compression.
%
%We evaluate our method in an image compression setting and show that our compression method is competitive with common compression algorithms designed specifically for the images.
%
%Moreover, this method provides the advantage that it can be adapted to data types for which there are no storage efficient algorithms so far.
\end{abstract}

%%%%%%%%% BODY TEXT
\section{Introduction}
Living in a world where digitalization is ubiquitous and important decisions are based on big data analytics, the problem of how to store information effectively is more important than ever. Source compression is the generalized term for representing data in a compact form, that either preserves all the information (lossless compression) or sacrifices some information for even smaller file sizes (lossy compression). It is a key component to tackle the flood of image and video data that is uploaded, transmitted and downloaded from the internet every day. 
While lossless compression is arguably more desirable, it has a fundamental theoretical limit, namely Shannon's entropy~\cite{shannon}.
Therefore, lossy compression aims at trading off a file's quality with its size - called rate-distortion trade-off. 
\begin{figure}[]
    \centering
    \includegraphics[width=\linewidth]{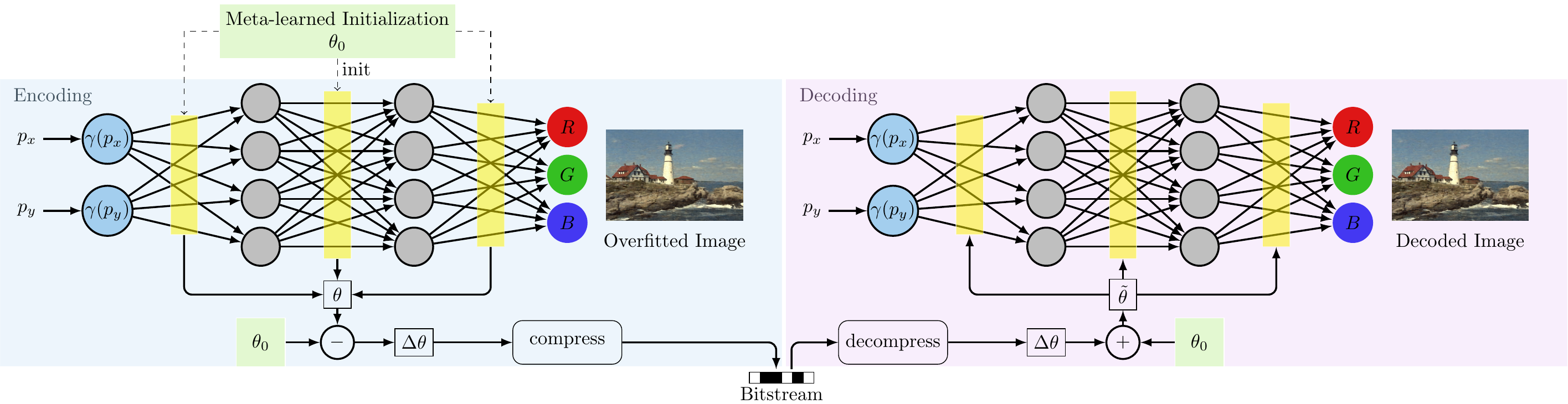}
    \caption{Method Overview: We summarize our approach to use \acp{inr} for compression by using the model weights $\theta$ as the representation for an image. We also visualize how a meta-learned initialization $\theta_0$ is used in the encoding and decoding process in order to compress only the weight update $\Delta\theta$ into the bitstream.}
    \label{fig:method_overview}
\end{figure}
%
%Using lossless compression may be the more desirable option in order not to loose information, however, the reduction in file size that is achievable has a theoretical limit. In many cases it is sufficient to have a representation of the data that is "good enough" for the intended purpose and achieving a high compression ratio has priority. The fundamental problem in lossy compression is finding the right tradeoff between file size and quality, typically referred to as the rate-distortion tradeoff.
%For the most common domains such as audio, image and video data, there are hand-designed algorithms that achieve a good rate-distortion tradeoff. 

Apart from traditional hand-designed algorithms tuned for particular data modalities, e.g. audio, images or video, machine learning research has recently developed promising learned approaches to source compression by leveraging the power of neural networks.
Such methods typically build on the well-known autoencoder \cite{hinton1994autoencoders} by implementing a constrained version of it. 
These so-called \acp{rdae} \cite{pmlr-v80-alemi18a,ball2016endtoend,mentzer2018conditional,habibian2019video} jointly optimize the quality of the decoded data sample and its encoded file size.
%
%A branch of machine learning research has focused on leveraging the power of neural networks to achieve more storage efficient compression techniques. A typical architecture is the so-called autoencoder, that consists of an encoder and decoder trained jointly. The encoder's output is the compressed code that represents the data. This compressed code is used by the decoder to reconstruct the original data as closely as possible.

This work sidesteps the prevalent approach of \acp{rdae} and investigates a novel paradigm for source compression - particularly focusing on image compression.
Recently, \acp{inr} gained popularity as a flexible, multi-purpose data representation that is able to produce high-fidelity samples on images \cite{sitzmann2020implicit}, 3D shapes \cite{park2019deepsdf,sitzmann2020implicit} and scenes \cite{mildenhall2020nerf}. 
In general, \acp{inr} represent data that lives on an underlying regular grid by learning a mapping between the grid's coordinates and the corresponding data values (e.g. RGB values) and have even been hypothesized to yield well compressed representations \cite{sitzmann2020implicit}. 
%
%Consequently, a naturally arising question is: How good are these \acp{inr} in terms of rate-distortion performance?
Due to their generality and concurrent early attempts to leverage them for compression~\cite{coin,chen2021nerv,dupont2022coin++}, \acp{inr} denote a promising candidate as a general purpose compression algorithm.
%
%However, to this date \acp{inr} have been surprisingly absent from research on source compression. 
%
%To the best of our knowledge, only COIN \cite{coin} and the concurrent NeRV \cite{chen2021nerv} considers \acp{inr} to compress images, resp. videos.

Currently there are two main challenges for \ac{inr}-based compression algorithms: 
(1) Straightforward approaches struggle to compete even with the simplest traditional algorithms \cite{coin}. 
(2) Since \acp{inr} encode data by overfitting to particular instances, the encoding time is perceived impractical. 
%
% To this end, we propose a comprehensive image compression pipeline on the basis on \acp{inr}.
To this end, we make two contributions.
Firstly, we propose meta-learned for \ac{inr}-based compression. We exploit recent advances in meta-learning for \acp{inr} \cite{sitzmann2019metasdf,tancik2020meta} based on Model-Agnostic Meta-Learning (MAML) \cite{pmlr-v70-finn17a} to find weight initializations that can compress data with fewer gradient updates as well as yield better rate-distortion performance. 
Secondly, we combine SIREN with positional encodings for \ac{inr}-based compression which greatly improves rate-distrortion performance.
While we focus on images, we emphasize that our proposed method can easily be adapted to any coordinate-based data modality.
Overall, we introduce a compression pipeline that vastly outperforms the recently proposed COIN~\cite{coin} and is competitive with traditional compression algorithms for images. Moreover, we demonstrate that meta-learned \acp{inr} already outperform JPEG2000 and a few \acp{rdae} on downsampled images.
Lastly, we emphasize the generality of \ac{inr}-based image compression by directly applying our approach to 3D data compression where we outperform the traditional algorithm Draco. 
\section{Related Work}

\textbf{Learned Image Compression.} Learned image compression was introduced in \cite{ball2016endtoend} by proposing an end-to-end autoencoder and entropy model that jointly optimizes rate and distortion. In the following, \cite{balle2018variational} extends this approach by adding a scale hyperprior, and then \cite{minnen2018-neurips,mentzer2018conditional,lee2019context} propose employing autoregressive entropy models to further improve the compression performance. Later, Hu~\etal~\cite{hu2020coarse} propose a coarse-to-fine hierarchical hyperprior, and Cheng~\etal~\cite{cheng2020image} achieve further improvements by adding attention modules and using a \ac{gmm} to estimate the distribution of latent representations. The current state-of-the art is achieved by \cite{xie2021enhanced}: They propose an invertible convolutional network, and apply residual feature enhancement as pre-processing and post-processing. Moreover, there are also plenty of methods aiming at variable rate compression, mainly including RNN-based autoencoders~\cite{Toderici2016Variable,toderici2017full,johnston2018improved} and conditional autoencoders~\cite{choi2019variable}. Besides, \cite{agustsson2019generative,mentzer2020high} propose image compression with Generative Adversarial Networks (GAN) to optimize perceptual quality. \\
\textbf{Implicit Neural Representations.} One of the early works on \acp{inr} is DeepSDF \cite{Park_2019_CVPR} which is a neural network representation for 3D shapes. In particular, they use a \ac{sdf} to represent the shape by a field where every point in space holds the distance to the shape's surface. Concurrently to DeepSDF, multiple works propose similar approaches to represent 3D shapes with \acp{inr}, \eg, the occupancy network~\cite{mescheder2019occupancy} and the implicit field decoder~\cite{chen2018implicit_decoder}. Besides, \acp{inr} have also been used for scene representation~\cite{mildenhall2020nerf}, image representation~\cite{chen2021learning,Skorokhodov_2021_CVPR} and compact representation~\cite{davies2021effectiveness}. \\
\textbf{Model Compression.} In the past decades, there has been a plethora of works on model compression~\cite{menghani2021efficient}.
%
% For instance, \cite{DBLP:journals/corr/HanMD15} proposes a model compression method that first prunes the weights based on the $L_1$ norm, and then assigns the weights to cluster centers which are stored in a codebook. Finally, the encoded weights and the codebook are losslessly compressed using Huffman coding.
%
For instance,\cite{DBLP:journals/corr/HanMD15} proposes sequentially applying pruning, quantization and entropy coding combined with retraining in between the steps.
%
% Later, \cite{agustsson2017soft} suggests an end-to-end learning framework that uses a soft-to-hard quantization relaxation in the rate-distortion optimization of model compression, and compresses the quantized weights using entropy coding.
%
Later, \cite{agustsson2017soft} suggests an end-to-end learning approach using a rate-distortion objective.
To optimize performance under quantization, several works
~\cite{HabiJN20,haq,Dong2019HAWQHA,uhlich2019mixed} use mixed-precision quantization, while others~\cite{kim2020frostnet,fan2020training,chai2021quantizationguided,LouizosRBGW19,Nagel2019DataFreeQT,nagel2020down} propose post-quantization optimization techniques. \\
\textbf{Model Weights for Instance-Adaptive Compression.}
Recently, \cite{rozendaal2021overfitting} suggests finetuning the decoder weights of an \ac{rdae} on a per-instance basis and appending the weight update to the latent vector, thereby improving \acp{rdae}. It is related to our work in that model weights are included in the representation, however the \ac{rdae} architecture fundamentally differs from ours. Most recently, Dupont~\etal~\cite{coin} propose the first \ac{inr}-based image compression approach COIN, which overfits an \ac{inr}'s model weights to represent single images and compresses the \ac{inr} using quantization. Importantly, COIN does not use meta-learning for initializing \acp{inr}, positional encodings for SIREN, post-quantization retraining and entropy coding. Furthermore, \cite{bird20213d} recently proposed a compression algorithm for entire scenes based on compressing the weights of NeRF~\cite{mildenhall2020nerf}. Moreover, concurrently NeRV~\cite{chen2021nerv} proposed to compress videos using \acp{inr}. While they use another data modality and neither use post-quantization retraining nor meta-learned initializations, 
% we value the growing interest in the young field of \acp{inr}-based source compression.
their work shows the potential of \ac{inr}-based compression of coordinate-based data.
In another concurrent work, \cite{dupont2022coin++} also proposes to apply meta-learning to \ac{inr}-based compression in an effort to extend COIN. However, unlike this work they do not outperform JPEG on the full resolution images on KODAK. Their performance is similar to our method absent of meta-learning and positional encodings (see \myfigref{fig:encodings}).

\section{Method}
\subsection{Background}
\paragraph{\acp{inr}} store coordinate-based data such as images, videos and 3D shapes by representing data as a continuous function from coordinates to values. For example, an image is a function of a horizontal and vertical coordinate $(p_x, p_y)$ and maps to a color vector within a color space such as RGB:
\begin{equation}\label{eq:inr_map}
    I: (p_x, p_y) \rightarrow (R,G,B)
\end{equation}
This mapping can be approximated by a neural network $f_{\theta}$, typically a \ac{mlp} with parameters $\theta$, such that $I(p_x, p_y) \approx f_{\theta}(p_x, p_y)$. Since these functions are continuous, \acp{inr} are resolution agnostic, \ie they can be evaluated on arbitrary coordinates within the normalized range $[-1, 1]$. To express a pixel based image tensor $\vx$ , we evaluate the image function on a uniformly spaced coordinate grid $\vp$ such that $\vx = I(\vp) \in \R^{W \times H \times 3}$ with
\begin{equation}
\begin{split}
\vp_{ij} &= \left(\dfrac{2i}{W-1} - 1, \dfrac{2j}{H-1} - 1\right) \in [-1,1]^2 
\\\forall \, i &\in \{0, \dots, W-1\}, j \in \{0, \dots, H-1\}.
\end{split}    
\end{equation}
Note that each coordinate vector is mapped independently:  
\begin{equation}
    f_\theta(\vp) = \begin{bmatrix}
    f_\theta(\vp_{11})  & \dots & f_\theta(\vp_{1H})\\
    \vdots  &\ddots &\vdots\\
     f_\theta(\vp_{W1})  & \dots &f_\theta(\vp_{WH})\\
    \end{bmatrix}.
\end{equation}

\paragraph{Rate-distortion Autoencoders.} The predominant approach in learned source compression are \acp{rdae}: An encoder network produces a compressed representation, typically called a latent vector $\vz \in \R^d$, which a jointly trained decoder network uses to reconstruct the original input. Early approaches enforce compactness of $\vz$ by limiting its dimension $d$~\cite{hintonautoencoder}. Newer methods constrain the representation by adding an entropy estimate, the so-called rate loss, of $\vz$ to the loss. This rate term, reflecting the storage requirement of $\vz$, is minimized jointly with a distortion term, that quantifies the compression error.

\subsection{Image Compression using \acp{inr}}
In contrast to \acp{rdae}, \acp{inr} store all information implicitly in the network weights $\theta$. The input to the \ac{inr} itself, \ie the coordinate, does not contain any information. The encoding process is equivalent to training the \ac{inr}. The decoding process is equivalent to loading a set of weights into the network and evaluating on a coordinate grid. We can summarize this as:
\begin{equation}
     \arg \min_{\theta} \cL(\vx,f_\theta(\vp)) = \theta^{\star} \xrightarrow[\text{transmit } \theta^{\star}]{} \vhx = f_{\theta^\star}(\vp)  .
\end{equation}
Thus, we only need to store $\theta^\star$ to reconstruct a distorted version of the original image $\vx$. With our approach, we describe a method to find $\theta^\star$ to achieve compact storage and good reconstruction at the same time.

\paragraph{Architecture.}
We use SIREN, namely a \ac{mlp} using sine activations with a frequency $\omega = 30$ as proposed originally in \cite{sitzmann2020implicit}, which has recently shown good performance on image data. We adopt the initialization scheme suggested by the authors. Since we aim to evaluate our method at multiple bitrates, we vary the model size to obtain a rate-distortion curve. We also provide an ablation on how to vary the model size to achieve optimal rate-distortion performance (see supplementary material) and on the architecture of the \ac{inr} (see \mysecref{ssec:inputenc_activation}).

\paragraph{Input Encoding.}
An input encoding transforms the input coordinate to a higher dimension, which has been shown to improve perceptual quality~\cite{mildenhall2020nerf,tancik2020fourfeat}.
Notably, to the best of our knowledge we are the first to combine SIREN with an input encoding - previously input encodings have only been used for \acp{inr} based on the \ac{relu} activation functions.
We apply an adapted version of the positional encoding presented in \cite{mildenhall2020nerf}, where we introduce the scale parameter $\sigma$ to adjust the frequency spacing (similarly to \cite{tancik2020fourfeat}) and concatenate the frequency terms with the original coordinate $p$ (as in the SIREN codebase\footnote[1]{\url{https://github.com/vsitzmann/siren}}):
\begin{equation}
\begin{split}
    \gamma (p) = & (p, \sin(\sigma^0 \pi p), \cos(\sigma^0 \pi p), \dots, \\ &\sin(\sigma^{L-1} \pi p), \cos(\sigma^{L-1} \pi p)).
\end{split}
    \label{nerfencodingscaled}
\end{equation}
where $L$ is the number of frequencies used. We investigate the impact of the input encoding in \mysecref{ssec:inputenc_activation}.

\subsection{Compression Pipeline for \acp{inr}}
\begin{figure*}[!t]
    \centering
    \includegraphics[width=0.9\linewidth]{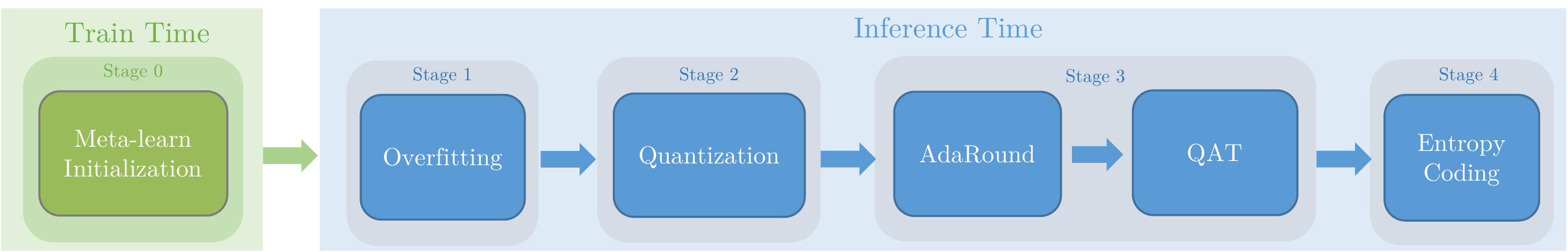}
    \caption{Overview of \ac{inr}-based compression pipeline. Blue: The basic compression pipeline comprising overfitting, quantization, AdaRound, \ac{qat} and entropy coding. Green: Additional meta-learning of initializations at training time.}
    \label{fig:algometa}
\end{figure*}

This section introduces our \ac{inr}-based compression pipeline. First, we describe our basic approach based on randomly initialized \acp{inr} (\mysecref{ssec:basic_approach}). Then, we propose meta-learned initializations to improve the rate-distortion performance and encoding time of \ac{inr}-based compression (\mysecref{ssec:meta_learned_approach}). The entire pipeline is depicted in \myfigref{fig:algometa} and a higher level overview is shown in \myfigref{fig:method_overview}.

\textbf{Basic Approach using Random Initialization.}\label{ssec:basic_approach} \textbf{Stage 1: Overfitting.}
First, we overfit the \ac{inr} $f_\theta$ to a data sample at test time. This is equivalent to calling the encoder of other learned methods. We call this step overfitting to emphasize that the \ac{inr} is trained to 
only represent a single image. Given an image $\vx$ and a coordinate grid $\vp$, we minimize the objective:
\begin{equation}
    \label{optobj}
    \arg \min_{\theta} \cL_\text{MSE}(\vx,f_\theta(\vp)).
\end{equation}
We use the \ac{mse} as the loss function to measure similarity of the ground-truth target and the \ac{inr}s output:
\begin{equation}
    \label{mseloss}
    \cL_\text{MSE}(\vx,\vhx) = \sum_i^W \sum_j^H \dfrac{\norm{\vx_{ij} - \vhx_{ij}}_2^2}{WH}.
\end{equation}
Note that $\vx_{ij} \in \R^3$ is the color vector of a single pixel. \\
\textbf{Regularization.}
In image compression, we aim at minimizing distortion (\eg, \ac{mse}) as well as bitrate simultaneously.
%Achieving a low \ac{mse} during overfitting is important for the overall performance. We do however have to keep in mind, that our final goal is not only a small \ac{mse} but also a low bitrate, which depends on the entropy of the model weights. 
Since the model entropy is not differentiable, we can not directly use it in gradient-based optimization. One option that has been used in literature is to use a differentiable entropy estimator during training~\cite{agustsson2017soft}. We however choose to use a regularization term that approximately induces lower entropy. 
In particular, we apply $L_1$ regularization to the model weights. Overall, this yields the following optimization objective:
%In particular we choose the $L_1$ loss function which we apply to the model weights as a regularization term in the full loss function:
\begin{equation}
    \label{fullloss}
    \cL(\vx, f_\theta(\vp)) = \cL_\text{MSE}(\vx, f_\theta(\vp)) + \lambda \norm{\theta}_1
\end{equation}
where $\lambda$ determines the importance of the $L_1$ regularization which induces sparsity. 
Our regularization term is related to the sparsity loss employed in \cite{rozendaal2021overfitting}: we have the same goal of limiting the entropy of the weights, however we apply this to an \ac{inr}, whereas they apply it to a traditional explicit decoder. \\
\textbf{Stage 2: Quantization.}
Typically, the model weights resulting from overfitting are single precision floating point numbers requiring 32 bits per weight. To reduce the memory requirement, we quantize the weights using the AI Model Efficiency Toolkit (AIMET)\footnote[1]{\url{https://quic.github.io/}}. We employ quantization specific to each weight tensor such that the uniformly-spaced quantization grid is adjusted to the value range of the tensor. The bitwidth determines the number of discrete levels, \ie quantization bins. 
We find empirically that bitwidths in the range of 7-8 lead to optimal rate-distortion performance for our models as shown in the supplement. \\
%We use bitwidths in the range of 7-8, which is where we typically find a sweetspot. \\ 
\textbf{Stage 3: Post-Quantization Optimization.}
Quantization reduces the models performance by rounding the weights to their nearest quantization bin. We leverage two methods to mitigate this effect. First, we employ \emph{AdaRound}~\cite{nagel2020down}, which is a second-order optimization method to decide whether to round a weight up or down. The core idea is that the traditional nearest rounding is not always the best choice, as shown in \cite{nagel2020down}. Subsequently, we fine-tune the quantized weights using \ac{qat}. This step aims to reverse part of the quantization error. Quantization is non-differentiable and we thus rely on the \ac{ste}~\cite{bengio2013estimating} for the gradient computation, essentially bypassing the quantization operation during backpropagation. \\
\textbf{Stage 4: Entropy Coding.} Finally, we perform entropy coding to further losslessly compress weights.
%The final step to obtain the bitstream of the compressed weights is entropy coding. 
In particular, we use a binarized arithmetic coding algorithm to losslessly compress the quantized weights.

\textbf{Meta-learned Initializations for Compressing INRs.}\label{ssec:meta_learned_approach}
%  We summarize the whole procedure of computing a compressed \ac{inr} combined with meta-learning the initialization in \myfigref{fig:algometa}.
Directly applying \acp{inr} to compression has two severe limitations: firstly, it requires overfitting a model from scratch to a data sample during the encoding step. Secondly, it does not allow embedding inductive biases into the compression algorithm (\eg, knowledge of a particular image distribution). To this end, we apply meta-learning, i.e. Model Agnostic Meta-Learning (MAML)~\cite{finn2017model}, for learning a weight initialization that is close to the weight values and entails information of the distribution of images.
Previous work on meta-learning for \ac{inr}s has aimed at improving mainly convergence speed~\cite{tancik2020meta}. The learned initialization $\theta_0$ is claimed to be closer in weight space to the final \ac{inr}. We want to exploit this fact for compression under the hypothesis that the update $\Delta \theta = \theta - \theta_0$ requires less storage than the full weight tensor $\theta$. We thus fix $\theta_0$ and include it in the decoder such that it is sufficient to transmit $\Delta \theta$, or, to be precise, the quantized update $\Delta \Tilde{\theta}$. The decoder can then reconstruct the image by computing:
\begin{equation}
    \Tilde{\theta} = \theta_0 + \Delta \Tilde{\theta}, \quad \vhx = f_{\Tilde{\theta}}(\vp).
\end{equation}
We expect the value range occupied by the weight updates $\Delta\theta$ to be significantly smaller than that of the full weights $\theta$. The range between the lowest and highest quantization bin can thus be smaller when quantizing the weight updates. At a fixed bitwidth, the stepsize in-between quantization bins will be smaller in the case of weight updates and, thus, the average rounding error is also smaller.

Note that the initialization is only learned once per distribution $\cD$ prior to overfitting a single image. Thus, we introduce it as Stage 0. Stage 0 happens at training time, is performed on many images and is not part of inference. Stages 1-4 happen at inference time and aim at compressing a single image. Consequently, using meta-learned initializations does not increase inference time.

% \paragraph{Learning the Initializations.}
% We learn the initialization for a distribution of images with \myalgoref{alg:metasiren}. We use the variant that learns not only the initialization, but also the learning rate parameters $\alpha$. In particular, we learn the learning rates of the inner loop on a \emph{per parameter per step} basis, that means each parameters has its own learning rate for each of the $k$ steps.

\textbf{Integration into a Compression Pipeline.}
When we want to encode only the update $\Delta\theta$, we need to adjust our compression pipeline accordingly. During overfitting we change the objective to:
\begin{equation}
    \label{fullloss}
    \cL(\vx, f_\theta(\vp)) = \cL_\text{MSE}(\vx, f_\theta(\vp)) + \lambda \norm{\Delta\theta}_1
\end{equation}
thus, the regularization term now induces the model weights to stay close to the initialization. Also, we directly apply quantization to the update $\Delta\theta$. In order to perform AdaRound and \ac{qat}, we apply a decomposition to all linear layers in the \ac{mlp} to separate initial values from the update:
\begin{equation}
\begin{split}
       \MW \vx + \vb &= (\MW_0 + \Delta\MW) \vx + (\vb_0 + \Delta\vb)\\ &= \underbrace{(\MW_0 \vx + \vb_0)}_{\text{fix}} +  \underbrace{(\Delta\MW \vx + \Delta\vb)}_{\text{quantize \& retrain}}.
\end{split}
\end{equation}
This is necessary, because optimizing the rounding and \ac{qat} require the original input-output function of each linear layer. Splitting it up into two parallel linear layers, we can fix the linear layer containing $\MW_0$ and $\vb_0$ and apply quantization, AdaRound and \ac{qat} to the update parameters $\Delta\MW $ and $\Delta\vb$.

\subsubsection{\acp{inr} for 3D Shape Compression}
The proposed \ac{inr}-based compression pipeline is applicable to any coordinate based data modality with minimal modification. We demonstrate this for 3D shapes. A 3D shape can be represented as a signed distance function:
\begin{equation}
    SDF: (p_x, p_y, p_z) \rightarrow d
\end{equation}
\ie we assign a signed distance $d$ between each point $(p_x, p_y, p_z)$ in 3D space and the shape surface. Here, the sign of the distance indicates whether we are inside (negative) or outside of the shape (positive). We can now simply train our \ac{inr} to approximate the SDF:
\begin{equation}
    f_{\theta}(\vp) \approx SDF(\vp).
\end{equation}
When training \acp{inr} to estimate SDFs accurate predictions close to the surface are most important. Therefore, we adopt the sampling strategy proposed in \cite{takikawa2021nglod}.

\section{Experiments}

\textbf{Datasets.} The \textbf{Kodak}~\cite{kodak} dataset is a collection of 24 images containing various objects, people or landscapes. This dataset has a resolution of $768 \times 512$ pixels (vertical $\times$ horizontal).
The \textbf{DIV2K} dataset introduced in \cite{Agustsson_2017_CVPR_Workshops} contains 1000 high resolution images with a width of $\approx 2000$ pixels. The dataset is split into 800 training, 100 validation and 100 test images. For our purpose of meta-learning the initialization, we resize the DIV2K images to the same resolution as Kodak ($768 \times 512$).
\textbf{CelebA}~\cite{liu2015faceattributes} is a dataset containing over 200'000 images of celebrities
%Because we are evaluating compression performance, we use the uncompressed PNG version of this dataset.
%
with a resolution of $178 \times 218$. We evaluate our method on 100 images that are randomly sampled from the test set. For our 3D shape compression experiment we use 5 high resolution meshes from the  \textbf{Stanford 3D Scanning Repository}~\cite{stanford_graphics}, which we normalize such that they fit into a unit cube prior to training. More details are in the supplement. \\
%
% We do not use the full test set for computational reasons.
\textbf{Metrics.}
We evaluate two metrics to analyze performance in terms of rate and distortion. We measure the rate as the total number of bits required to store the representation divided by the number of pixels $W \cdot H$ of the image:
\begin{equation}
    \text{bitrate} = \dfrac{\text{total number of bits}}{W H} \quad [\text{bpp}].
\end{equation}
We measure distortion in terms of \ac{mse} and convert it to the \ac{psnr} using the formula:
\begin{equation}
    \text{PSNR} = 10 \log_{10} \left(\dfrac{1}{MSE} \right) \quad [\text{dB}].
\end{equation}
\\
\textbf{Baselines.} We compare our method against traditional codecs, \ac{inr} based compression and learned approaches based on \acp{rdae}.
\begin{itemize}
    \setlength\itemsep{0em}
    \item Traditional image compression codecs: JPEG, JPEG2000, BPG
    \item \ac{inr}-based image compression: Dupont~\etal~\cite{coin} (COIN)
    \item \ac{rdae}-based image compression: Ball\'{e}~\etal~\cite{ball2016endtoend}, Xie~\etal~\cite{xie2021enhanced}
    \item 3D mesh compression: Draco \footnote{\url{https://github.com/google/draco}}
\end{itemize}

\textbf{Optimization and Hyperparameters.}
We use a default set of hyperparameters throughout the experiment section unless mentioned otherwise. In particular, we use \acp{inr} with 3 hidden layers and sine activations combined with the positional encoding using $\sigma=1.4$. On the higher resolution Kodak dataset, we set the number of frequencies to $L=16$, whereas on CelebA we set $L=12$. We vary the number of hidden units per layer $M$, \ie the width of the \ac{mlp}, to evaluate performance at different rate-distortion operating points.
% For CelebA images we choose $M \in \{ 24, 32, 48, 64\}$ and for Kodak images we choose $M \in \{32, 48, 64, 128\}$. The regularization parameter is $\lambda = 10^{-5}$ as default.
We refer to our method with random initialization as the \emph{basic} approach whereas the method including meta-learned initialization is called \emph{meta-learned}. We found the optimal bitwidth to be $b=7$ for the \emph{meta-learned} approach and $b=8$ for the \emph{basic} approach. For additional details on the training and hyperparameters we refer to the supplementary material.

%  We determined the hyperparameters by performing a grid search and choosing a value that works best at different bitrates.
% We implement our method in PyTorch~\cite{pytorch} and train the \acp{inr} using Adam~\cite{kingma2015adam} during overfitting and \ac{qat}. We use mixed precision training to reduce the memory requirement and improve training speed. In general, the results shown are obtained  using AdaRound and \ac{qat} combined.  Since we change the axis scaling between experiments, we include the JPEG and JPEG2000 baselines in all plots to provide a reference.\\

\subsection{Comparison with State-of-the-Art}\label{sec:sota_comparison}
%\paragraph{Comparison of Rate-Distortion Curves.} 
% For this section we train with default hyperparameters resulting in our best performing models for the \emph{basic} approach and the \emph{meta-learned} approach. 
% For the basic approach we use $8$-bit quantization and for the meta-learned approach 7-bit quantization. These are the best bitwidth settings for the two approaches.
% We compare the rate-distortion performance of our proposed methods to conventional codecs (JPEG, JPEG2000 and BPG) and various learned autoencoder methods on CelebA (\myfigref{celebcomp}) and Kodak (\myfigref{kodakcomp}). For the Kodak dataset we also report the performance of the \ac{inr} based image compression approach COIN~\cite{coin}. 

\textbf{Full resolution.} \myfigref{celeba_kodakcomp} depicts our results on CelebA/Kodak respectively. The proposed \emph{basic} approach can already outperform COIN clearly over the whole range of bitrates. It is also better than JPEG for most bitrates, except the highest setting on CelebA. With our proposed \emph{meta-learned} approach we improve over the \emph{basic} approach at all bitrates. Between the two datasets, the difference is noticeably greater on the CelebA dataset. At the lowest bitrate examined the meta-learned approach reaches the performance of JPEG2000, however our approach cannot keep up with JPEG2000 at higher bitrates. On the CelebA dataset, the meta-learned approach also almost reaches the performance of an autoencoder with a factorized prior~\cite{ball2016endtoend} at lower bitrates. Towards higher bitrates, the advantage of the autoencoder becomes clearer. BPG as well as the state-of-the-art \ac{rdae}~\cite{xie2021enhanced} clearly outperform our method on both datasets. 
\begin{figure}[!t]
\centering
\includegraphics[width=0.38\linewidth]{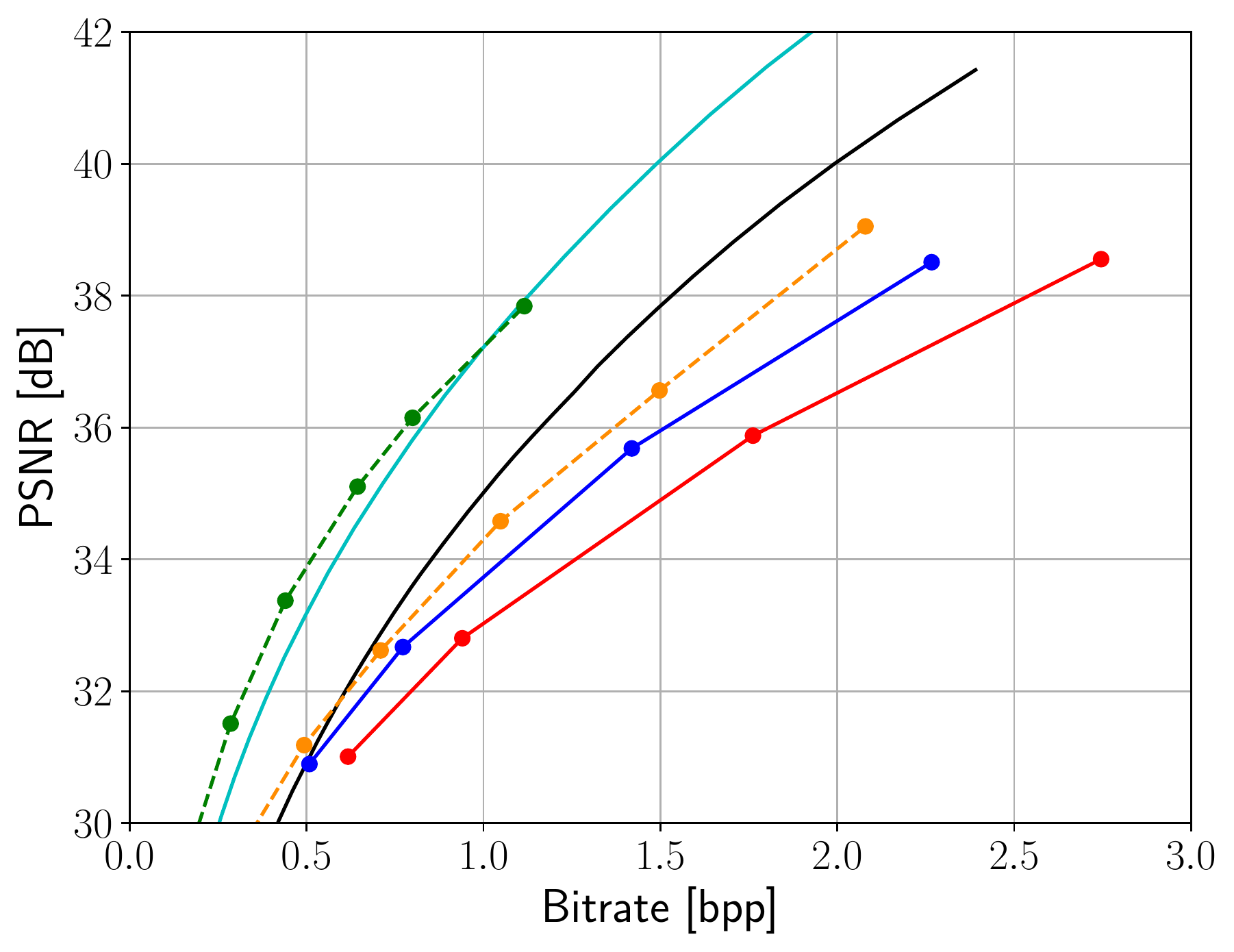}
\includegraphics[width=0.60\linewidth]{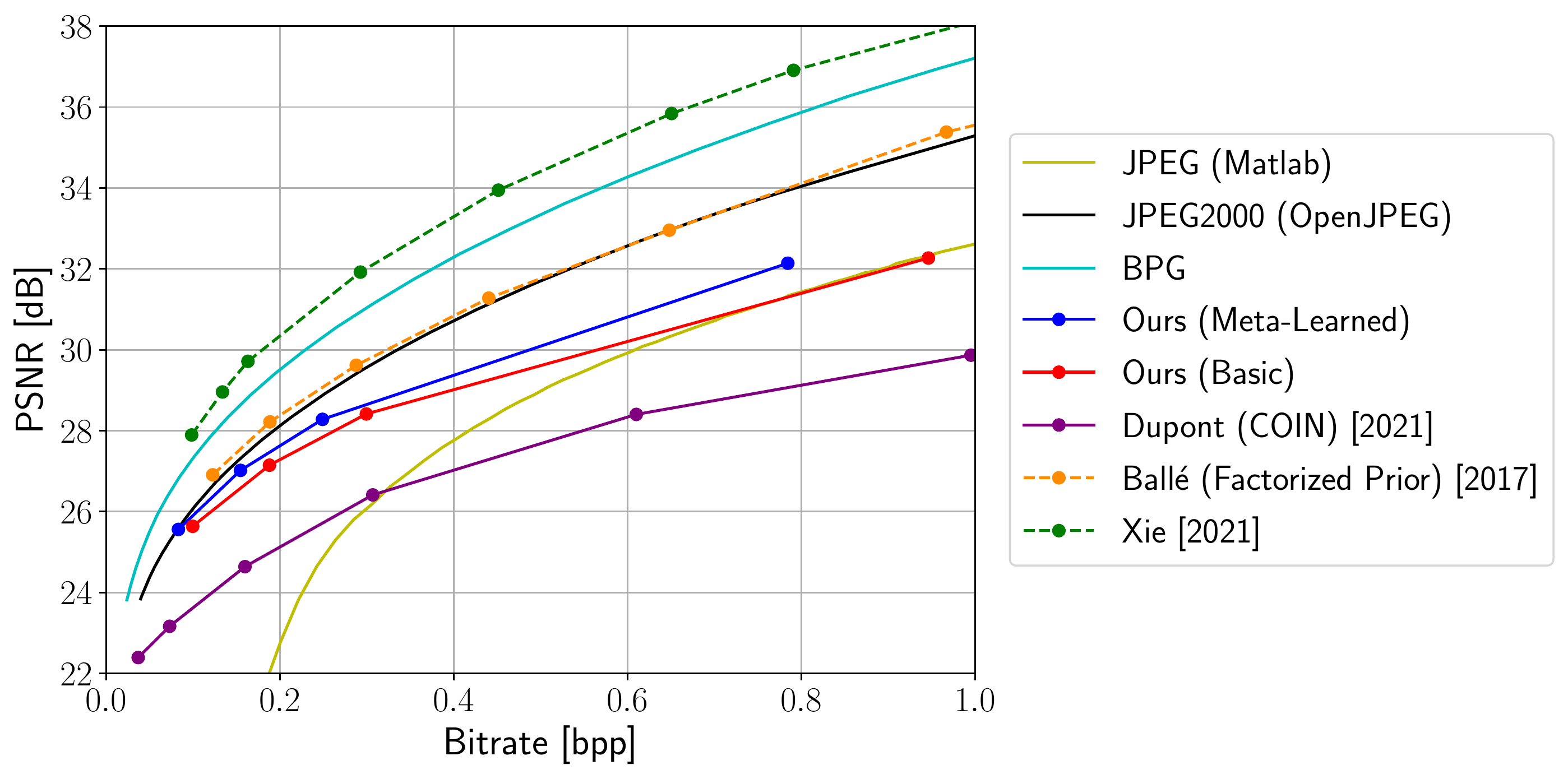}
\caption{Performance overview over image compression approaches including conventional (solid line), learned autoencoder (dashed line) and learned INR methods (solid line with dots) evaluated on the \textbf{CelebA} (left) \textbf{Kodak} (right) dataset.}
\label{celeba_kodakcomp}
\end{figure}

\textbf{Reduced image resolution.} We further compare our \emph{basic} and \emph{meta-learned} approach with other methods on Kodak with reduced resolution (2x/4x). These image are comprised of $384 \times 256$, resp. $192 \times 128$, pixels. We observe that the \emph{meta-learned} approach again performs strictly better than the \emph{basic} approach. Moreover,  our \emph{meta-learned} approach demonstrates competitive performance for this image resolution outperforming all other methods, except BPG and Xie \etal, over the entire range of bitrates.
\begin{figure}[!t]
\centering
\includegraphics[width=0.38\linewidth]{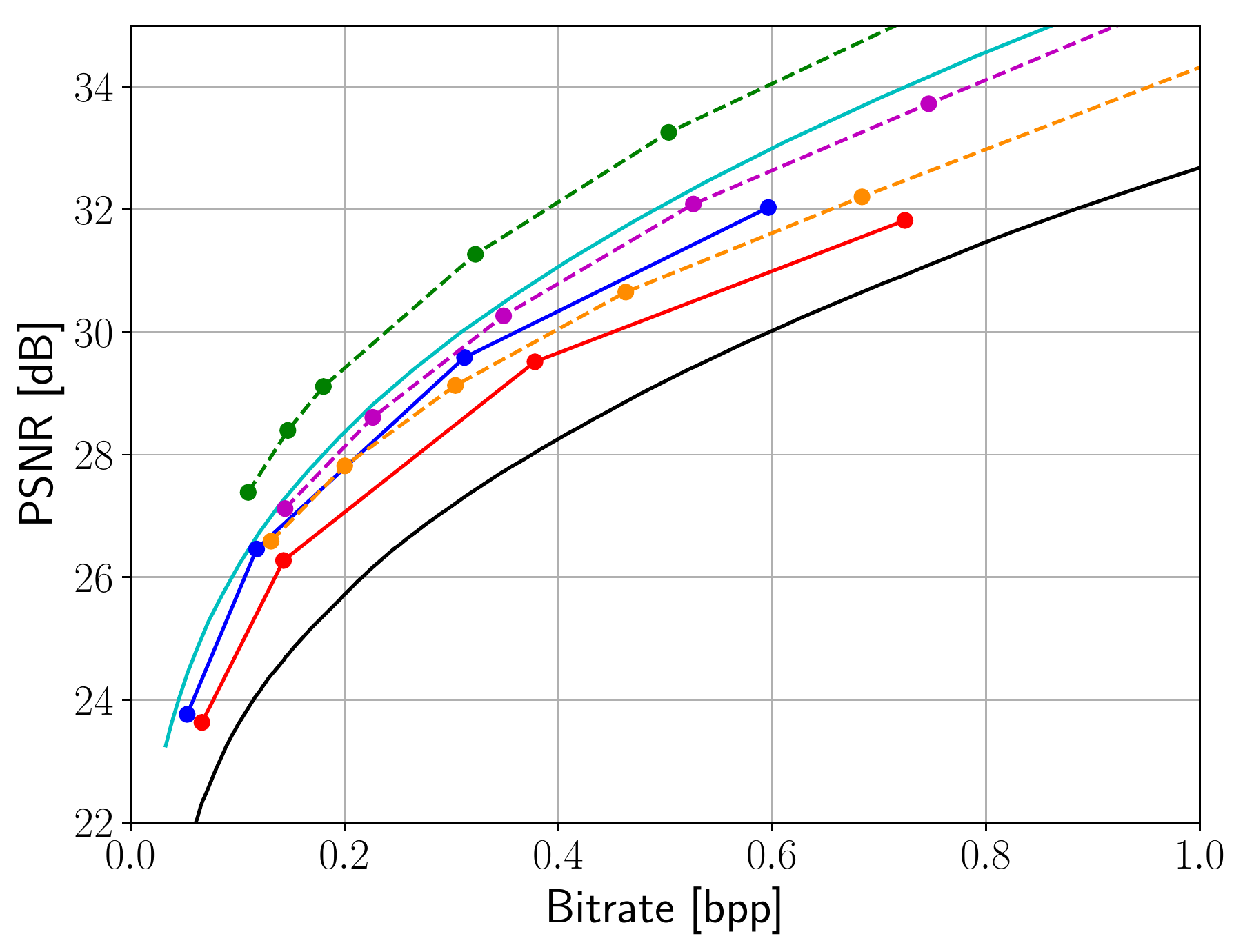}
\includegraphics[width=0.6\linewidth]{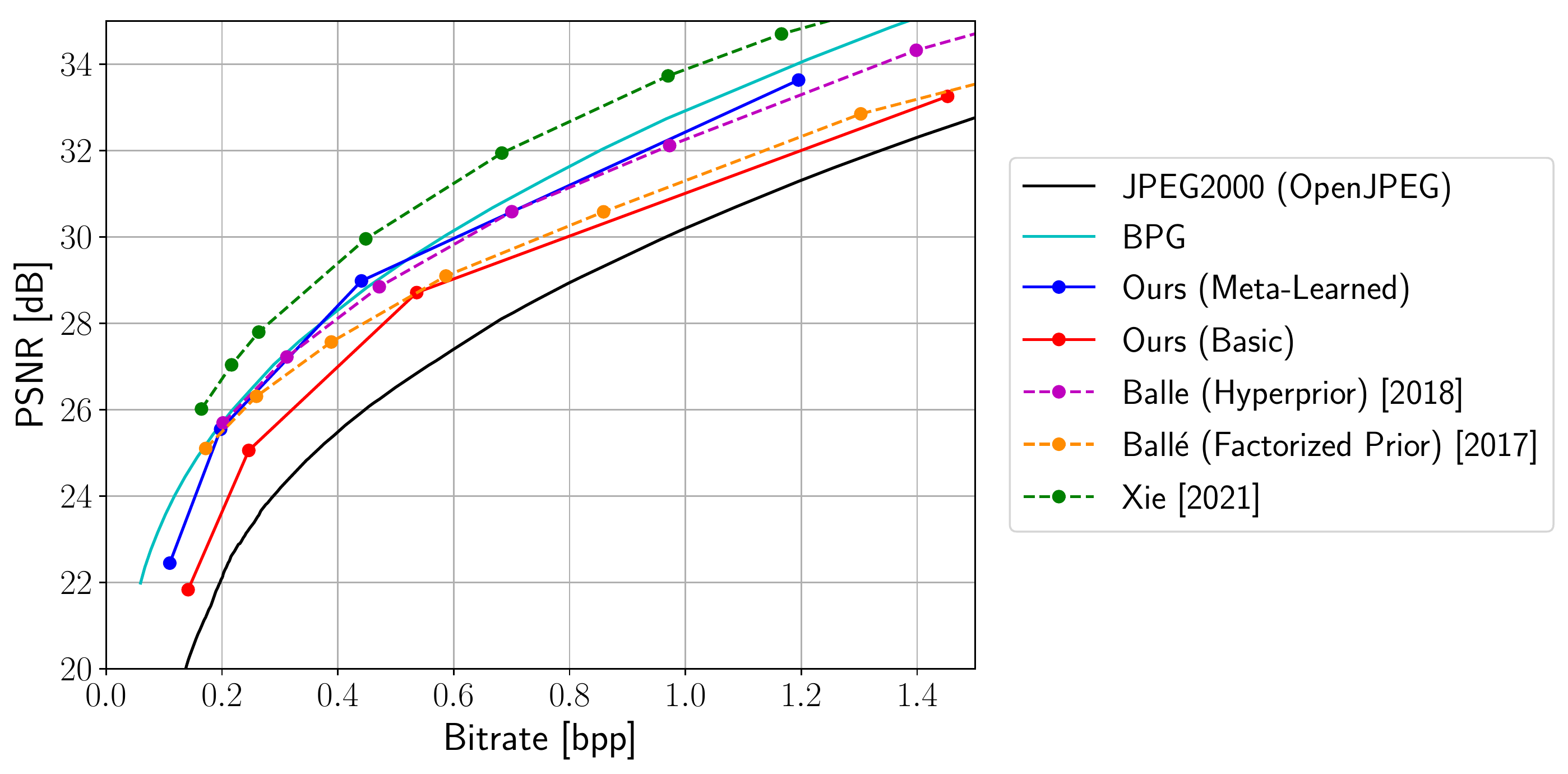}
\caption{Image compression approaches including conventional (solid line), \acp{rdae} (dashed line) and learned \ac{inr}-based methods (solid line with dots) evaluated on the \textbf{Kodak} dataset with image resolution reduced by a factor of two (left) and four (right). Meta-learned \acp{inr} show competitive performance in this regime.}
\label{celeba_kodakcomp}
\end{figure}

\begin{figure}[!t]
\centering

\includegraphics[width=0.7\linewidth]{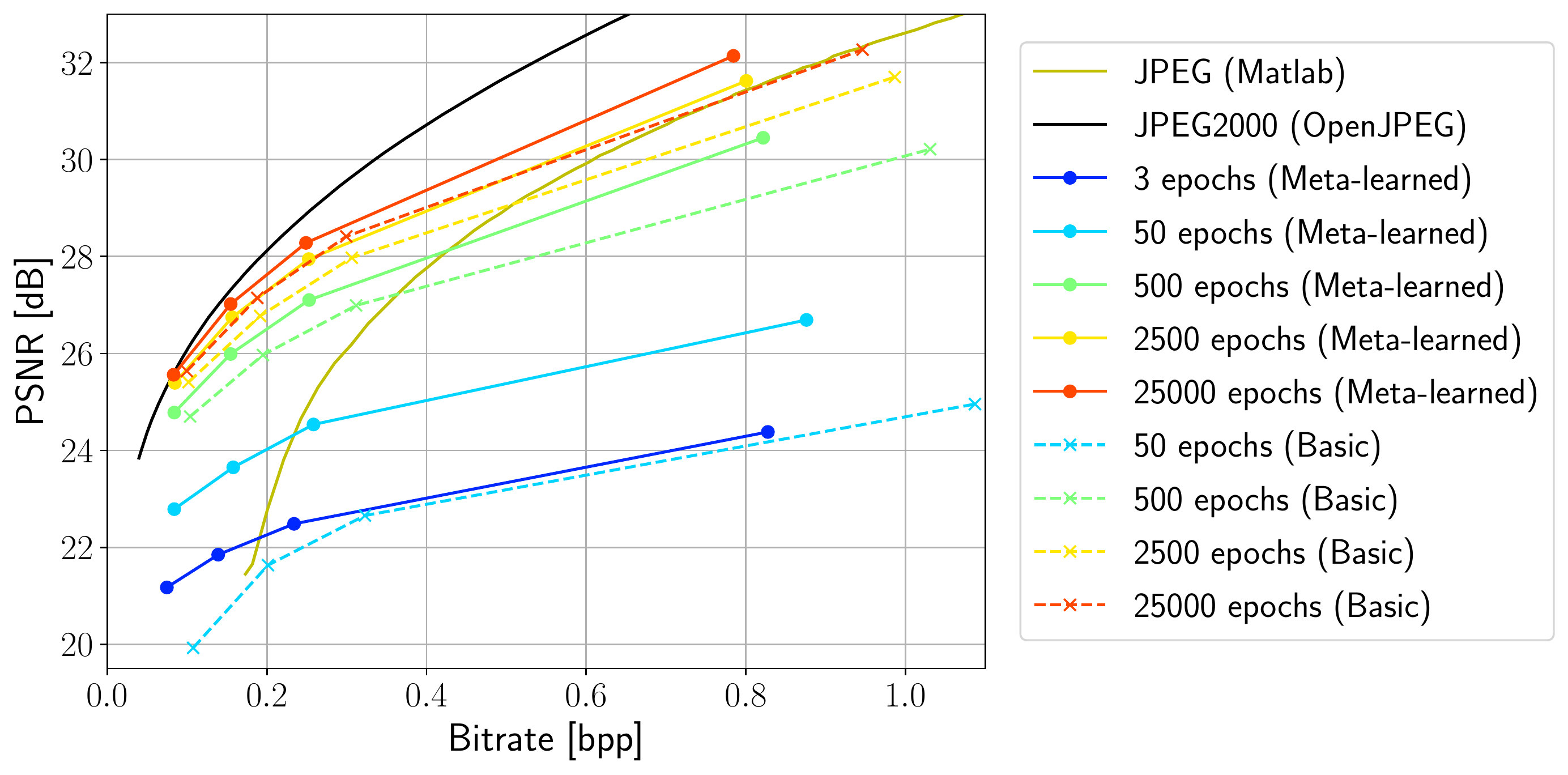}

\caption{Comparing the convergence speed of the meta-learned and basic approach evaluated on the Kodak dataset. The meta-learned approach converges faster, which is especially apparent in the beginning of the overfitting. After only 2500 epochs it reaches the same performance as the basic approach after 25000 epochs.}
\label{fig:convergence}
\end{figure}

\subsection{Visual Comparison to JPEG and JPEG2000}
\begin{figure*}[t!]
	\setlength{\linewidth}{\textwidth}
	\setlength{\hsize}{\textwidth}
% 	\vspace{-9mm}
	\centering
	\renewcommand{\tabcolsep}{0.05pt}
	\renewcommand{\arraystretch}{0.1}
	\renewcommand{\imgwidth}{0.23}
	\renewcommand{\spyloc}{(-1.2, -1.3)}
	\renewcommand{\spypos}{(-0.8, 0.1)}
	\newcommand{\spysize}{1.0cm}
	\begin{tabular}{ccccc}	
	    \footnotesize original & 
		\footnotesize JPEG & 
		\footnotesize JPEG2000 &
		\footnotesize Ours (Meta-learned)  \\	
		\begin{tikzpicture}[spy using outlines={circle,magenta,magnification=2,size=\spysize, connect spies}]
			\node {\includegraphics[ width=\imgwidth\linewidth]{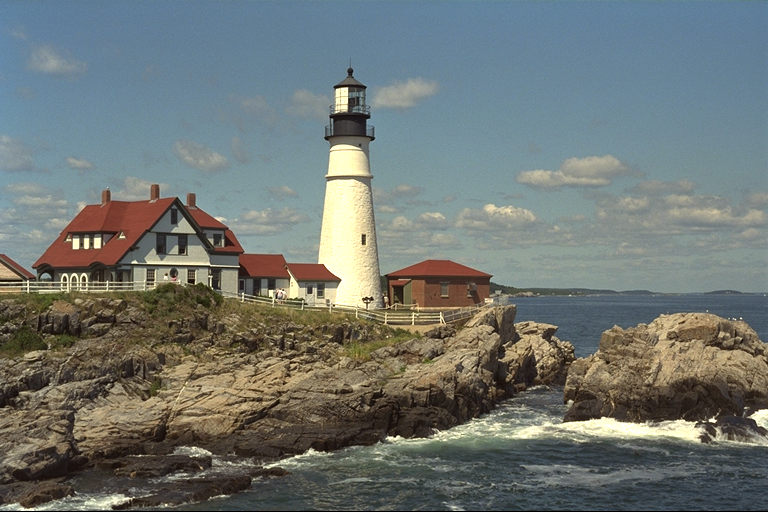}};
			\spy on \spypos in node [right] at \spyloc;%(1.6, 0.7);
		\end{tikzpicture} &			
		\begin{tikzpicture}[spy using outlines={circle,magenta,magnification=2,size=\spysize, connect spies}]
		 
			\node[label={[font=\footnotesize, xshift=0.7cm, yshift=-2.8cm, align=center] $0.1720$ bpp\\ $20.67$ dB} ] {\includegraphics[width=\imgwidth\linewidth]{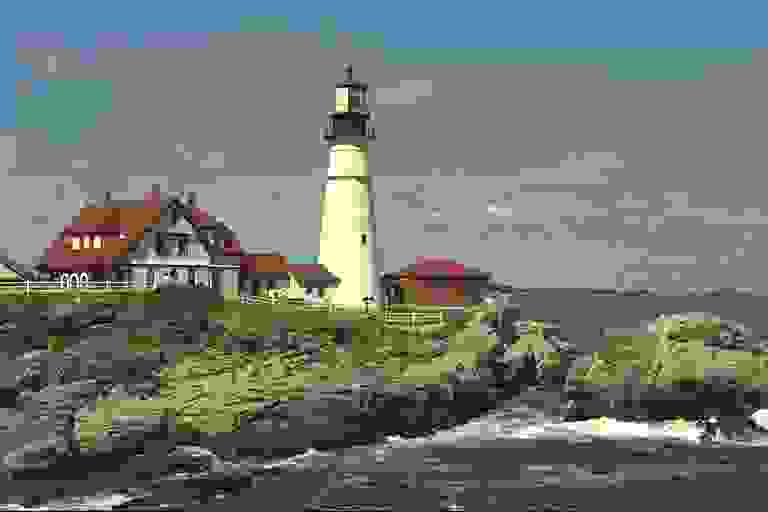}};
			\spy  on \spypos in node [right] at \spyloc;
		\end{tikzpicture} &			
		\begin{tikzpicture}[spy using outlines={circle,magenta,magnification=2,size=\spysize, connect spies}]
			\node[label={[font=\footnotesize, xshift=0.7cm, yshift=-2.8cm, align=center] $0.0835$ bpp\\ $24.43$ dB} ] {\includegraphics[width=\imgwidth\linewidth]{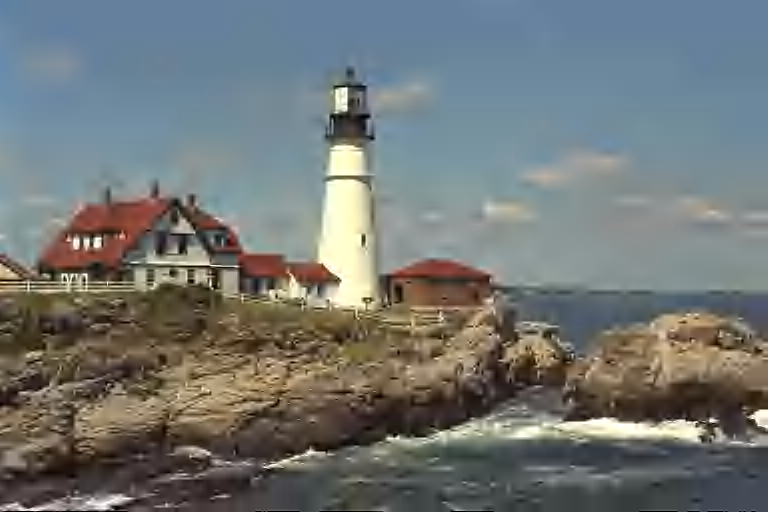}};
			\spy  on \spypos in node [right] at \spyloc;
		\end{tikzpicture} &
		% chair
		\begin{tikzpicture}[spy using outlines={circle,magenta,magnification=2,size=\spysize, connect spies}]
			\node[label={[font=\footnotesize, xshift=0.7cm, yshift=-2.8cm, align=center] $0.0829$ bpp\\ $24.43$ dB} ] {\includegraphics[width=\imgwidth\linewidth]{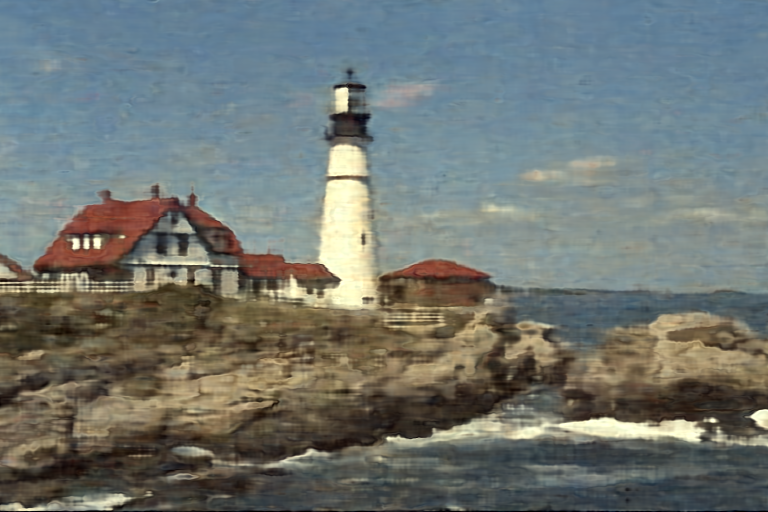}};
			\spy on \spypos in node [right] at \spyloc;
		\end{tikzpicture}
	\end{tabular}
	
	\renewcommand{\tabcolsep}{0.0pt}
	\renewcommand{\arraystretch}{0.1}
	\renewcommand{\imgwidth}{0.23}
	\renewcommand{\spyloc}{(-0.8, -1.8)}
	\renewcommand{\spypos}{(-0.35, -0.2)}
	\renewcommand{\spysize}{1.0cm}
	\begin{tabular}{ccccc}	
		\begin{tikzpicture}[spy using outlines={circle,magenta,magnification=2,size=\spysize, connect spies}]
			\node {\includegraphics[ width=\imgwidth\linewidth]{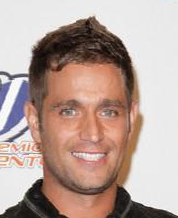}};
			\spy on \spypos in node [right] at \spyloc;%(1.6, 0.7);
		\end{tikzpicture} &			
		\begin{tikzpicture}[spy using outlines={circle,magenta,magnification=2,size=\spysize, connect spies}]
			\node {\includegraphics[width=\imgwidth\linewidth]{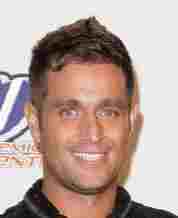}};
			\spy  on \spypos in node [right] at \spyloc;
		\end{tikzpicture} &			
		\begin{tikzpicture}[spy using outlines={circle,magenta,magnification=2,size=\spysize, connect spies}]
			\node {\includegraphics[width=\imgwidth\linewidth]{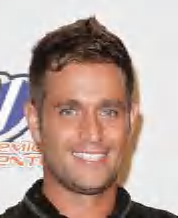}};
			\spy  on \spypos in node [right] at \spyloc;
		\end{tikzpicture} &
		% chair
		\begin{tikzpicture}[spy using outlines={circle,magenta,magnification=2,size=\spysize, connect spies}]
			\node {\includegraphics[width=\imgwidth\linewidth]{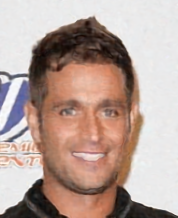}};
			\spy on \spypos in node [right] at \spyloc;
		\end{tikzpicture} 			\\
	%	\vspace{-5mm}
		\footnotesize & 
		\footnotesize $0.526$ bpp & 
		\footnotesize $0.514$ bpp &
		\footnotesize $0.508$ bpp  \\
		\footnotesize & 
		\footnotesize $28.58$ dB & 
		\footnotesize  $31.68$ dB &
		\footnotesize  $32.56$ dB \\		
	\end{tabular}
	\caption{Visual comparison of images compressed with JPEG (quality factor 1/13), JPEG2000 (compression factor 287/47) and our meta-learned approach on Kodak/CelebA (top/bottom). We use a model with a hidden dimension of $M=32$/$24$. JPEG introduces heavy block artifacts and loss of color information resulting in the worst image in comparison. JPEG2000 shows blurring and blocking around edges. Our method maintains better local contrast but shows periodic artifacts visible in the sky as well as smearing at some edges.}
\label{fig:imagecomparisonkodak}
\end{figure*}

We compare compressed images of our meta-learned approach with the codecs JPEG and JPEG2000 in \myfigref{fig:imagecomparisonkodak}. We visually confirm that our model significantly improves over JPEG: Our model produces an overall more pleasing image with better detail and less artifacts although we operate at a lower bitrate on both images. For the Kodak image in \myfigref{fig:imagecomparisonkodak} we achieve a slightly lower bitrate at the same distortion compared to JPEG2000. Visually, the JPEG2000 image shows more artifacts around edges and in regions with high frequency details. The sky is however rendered better on the JPEG2000 image because our model introduces periodic artifacts.  For the CelebA image in \myfigref{fig:imagecomparisonkodak} our method achieves a lower bitrate and higher \ac{psnr} than the JPEG2000 image. JPEG2000 again shows artifacts around edges (for example around the letters in the background) and smoothes out transitions from lighter to darker areas on the face. Our method produces a more natural tonal transition.

% \begin{figure}[!t]
% \centering
% \includegraphics[width=0.95\linewidth]{img/celeb100.pdf}
% \caption{Performance overview over image compression approaches including conventional (solid line), learned autoencoder (dashed line) and learned INR methods (solid line with dots) evaluated on the \textbf{CelebA} dataset.}
% \label{celebcomp}
% \end{figure}

\subsection{Convergence Speed}

In \myfigref{fig:convergence} we show how the basic and meta-learned approach compare over different numbers of epochs. Especially in the beginning of the overfitting, the meta-learned approach shows significantly faster convergence. Already after the first 3 epochs, we obtain better performance than what the basic approach achieves after 50 epochs. Convergence slows down as we approach the final performance of the respective model, while the meta-learned approach maintains the advantage: It achieves the same performance after 2500 epochs as the basic approach after 25000 epochs. This amounts to a reduction in training time  of 90\%.

\subsection{Choosing Input Encoding and Activation}\label{ssec:inputenc_activation}
An important architecture choice is the combination of input encoding and the activation function used.  We compare against the Gaussian encoding proposed in \cite{tancik2020fourfeat}. For this encoding we use the same number of frequencies as hidden dimensions ($L=M$) as in \cite{tancik2020fourfeat} and  a standard deviation of $\sigma = 4$.
 We train models with different hidden dimensions ($M \in \{32, 48, 64, 96, 128\}$) and different input encodings on the Kodak dataset starting from random initializations using the regularization parameter $\lambda = 10^{-6}$.

\begin{figure}[!t]
\centering
\begin{subfigure}[b]{.49\linewidth}
\includegraphics[width=\linewidth]{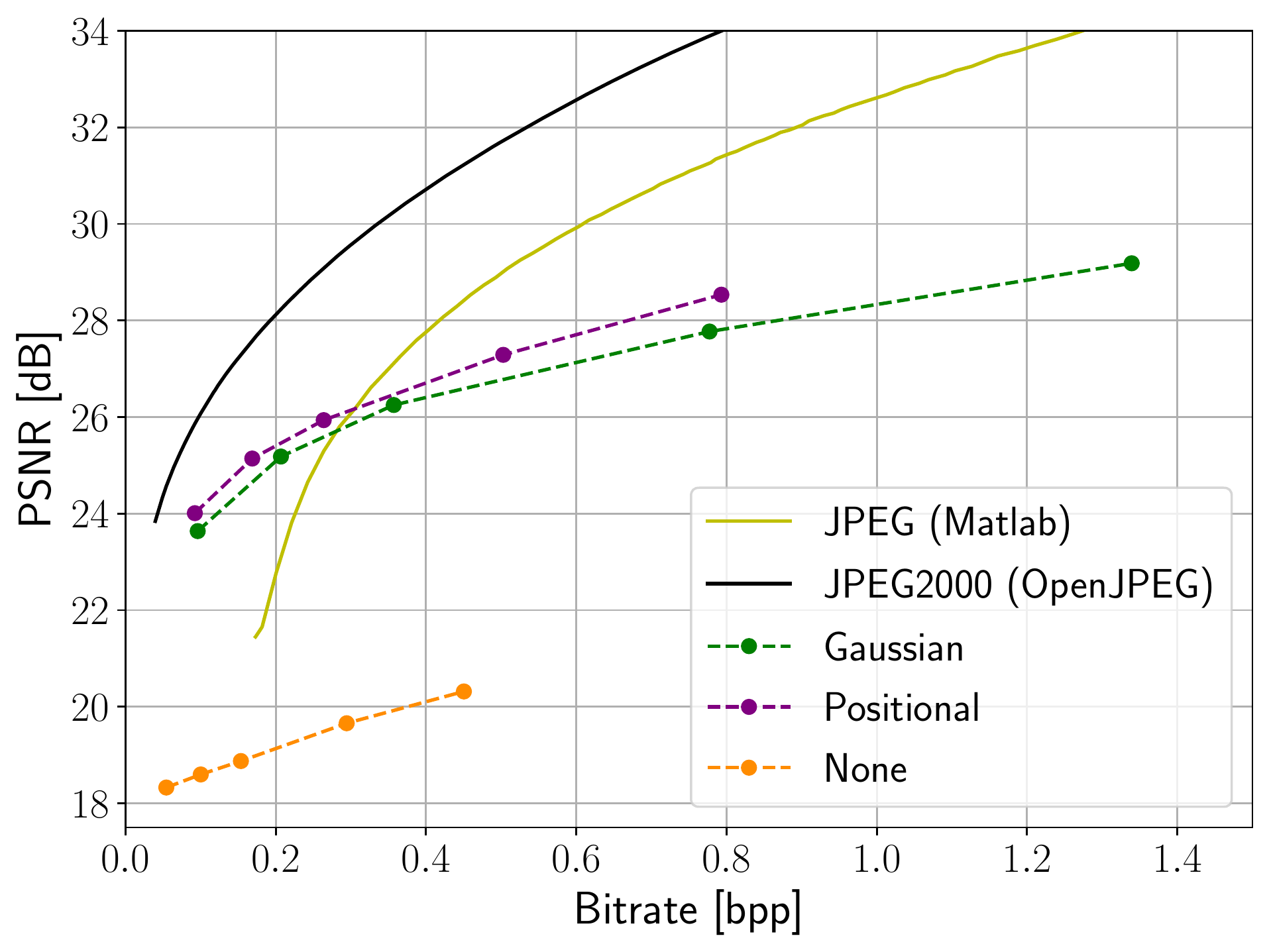}
\caption{\ac{relu}}
\label{fig:reluenc}
\end{subfigure}
\hfil
\begin{subfigure}[b]{.49\linewidth}
\includegraphics[width=\linewidth]{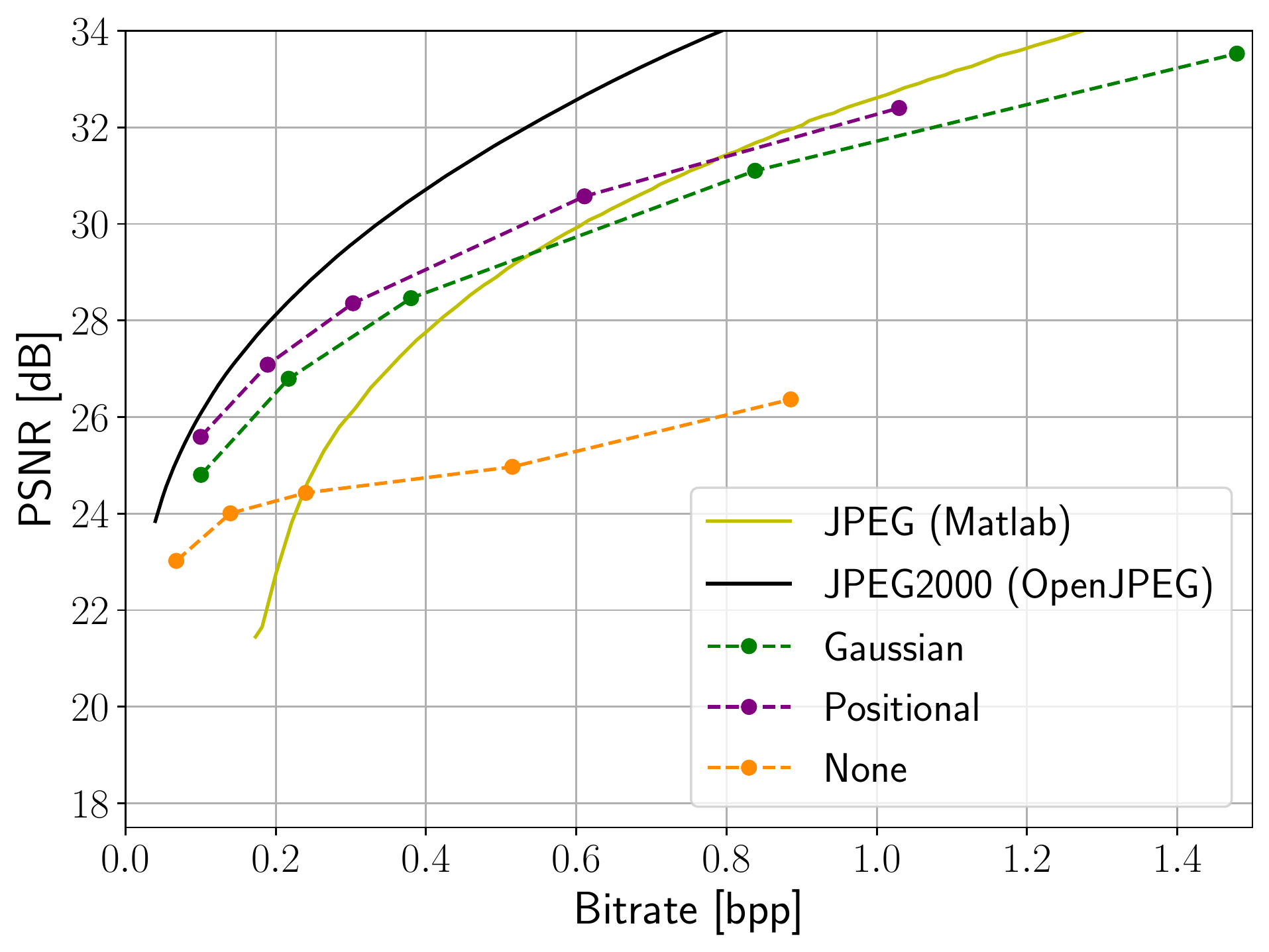}
\caption{Sine}
\label{fig:sineenc}
\end{subfigure}
\caption{Rate-distortion performance of different combinations of input encoding and activation function on the Kodak dataset.}
\label{fig:encodings}
\vspace{-1em}
\end{figure}

Looking at \myfigref{fig:reluenc}, compared to \myfigref{fig:sineenc} we can see that the sine activation outperforms  the \ac{relu} activation in every configuration, especially at higher bitrates. The best overall input encoding is \emph{positional} encoding beating \emph{Gaussian} for both activations. The \ac{mlp} without input encoding and sine activations, the SIREN architecture, performs significantly better than its \ac{relu} counterpart but still cannot reach the performance of the models with input encoding. 

Importantly, we investigate whether positional encoding improves SIREN in general or rather renders it more robust to quantization. Therefore, we measure the quantization error of our basic approach for different bitwidths. The result is depicted in \myfigref{quant_error}. ReLU and sine activations both show a reduced quantization error when trained with positional encoding. However, the effect is most obvious in the case of SIREN. Comparing the PSNR-delta of SIREN with and without positional encoding in \myfigref{quant_error} with \myfigref{fig:encodings} (b) reveals that applying positional encoding makes SIREN predominantly more robust to quantization.

\subsection{3D Shape Compression}\label{sec:3d}
\begin{figure}[!t]
\begin{minipage}{.385\textwidth}
    \centering
    \includegraphics[width=1.0\linewidth]{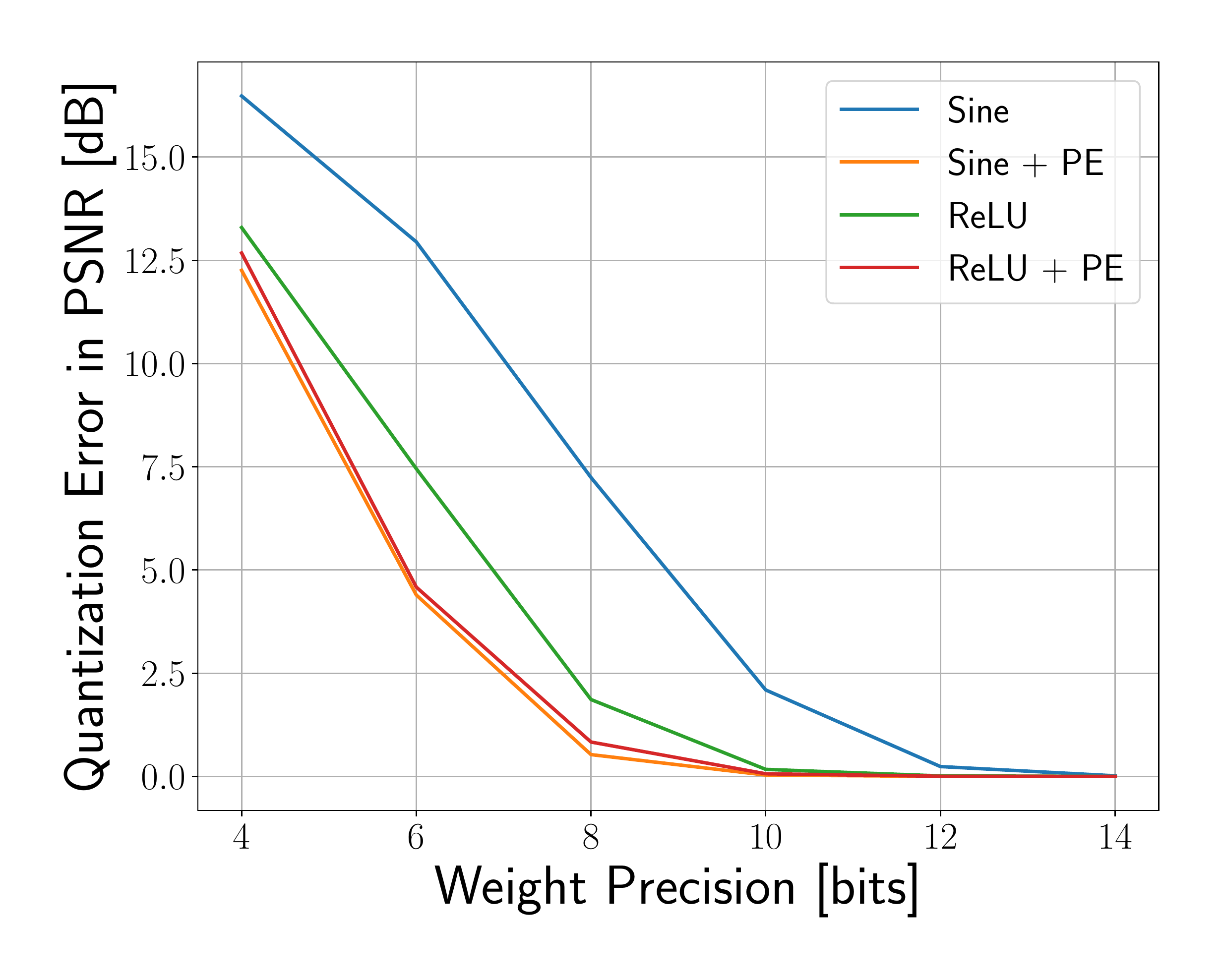}
    \caption{Quantization error of our basic model using ReLU/sine activations with/without positional encoding (PE).}
    \label{quant_error}
\end{minipage}%
\hspace{2mm}
\begin{minipage}{.59\textwidth}
    \centering
    \includegraphics[width=1.0\linewidth]{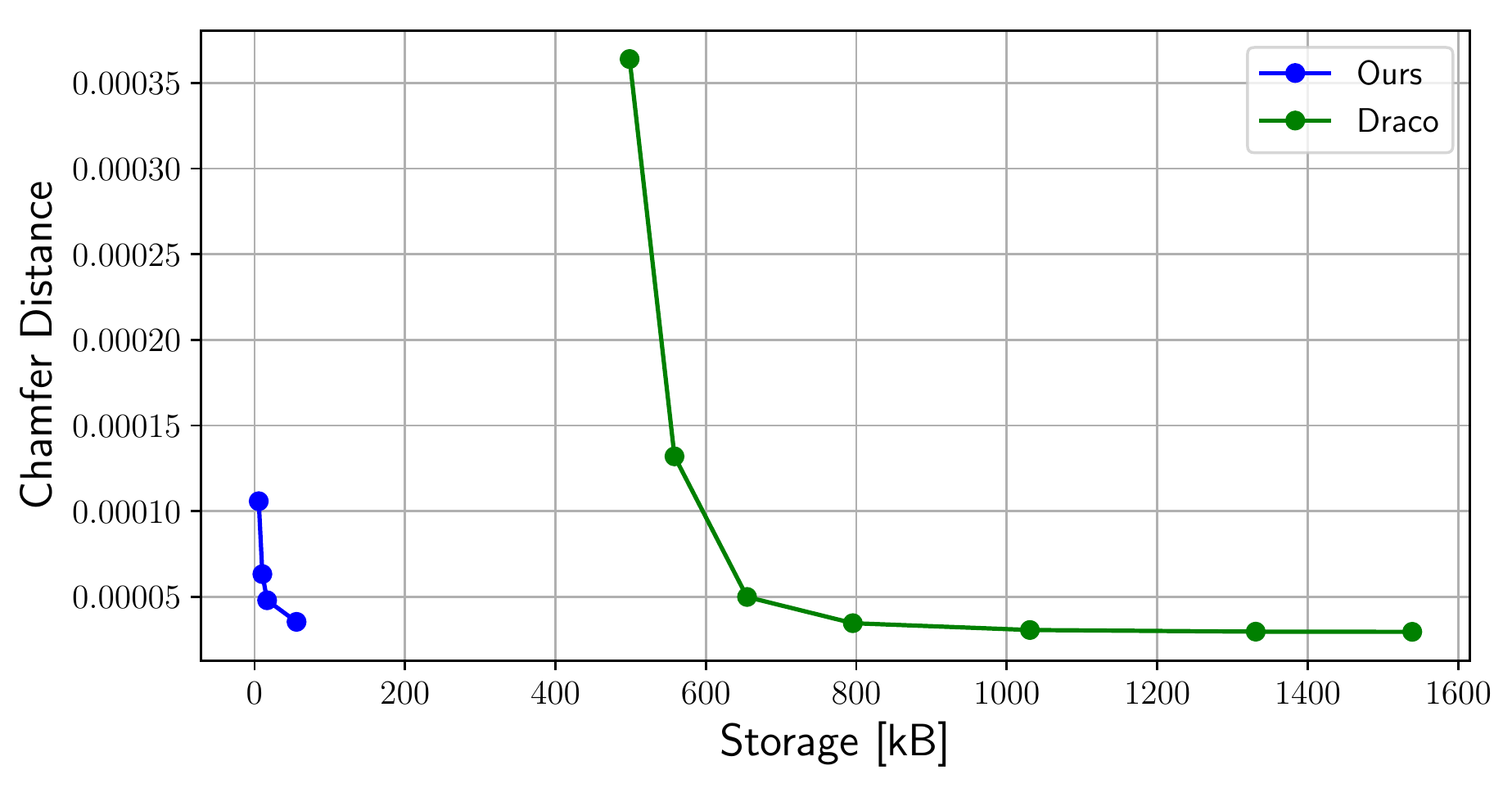}
    \caption{Rate-Distortion performance for 3D shape compression of our method (basic) and the traditional algorithm Draco. We clearly outperform Draco.}
    \label{3dcompression}
\end{minipage}
\end{figure}

To demonstrate that our algorithm is applicable to coordinate-based data beyond images, we provide an additional experiment showing its performance on the task of 3D shape compression. Since the main goal of this experiment is to show the transferability of \acp{inr}-based compression, we only train our basic approach without meta-learning on 3D shapes. We plot the chamfer distance averaged over all shapes against the storage required in \myfigref{3dcompression} and compare to the algorithm Draco which is based on mesh quantization. We focus on the comparison with mesh-based compression algorithms because they also preserve a continuous surface unlike the alternative approach of point cloud compression. We require much fewer bits to encode a shape of similar quality than Draco. Further details regarding this experiment are in the supplement.

\section{Conclusion}
Overall, \acp{inr} demonstrated great potential as a compressed representation for images. 
Our main contributions, the use of meta-learned initializations and SIREN combined with positional encodings, largely improve rate-distortion performance compared to previous methods~\cite{coin} performing image compression based on \acp{inr}. Moreover, our approach is the first \acp{inr}-based method that is competitive with traditional codecs over a large portion of bitrates.

Meta-learned initializations are superior to random initializations. Specifically, they reduce the bitrate at the same reconstruction quality. This supports the hypothesis that weight updates are more compressible. In particular, the performance gain is larger on the CelebA dataset, where the initializations are trained on an image distribution that is more similar to the test set. Moreover, the distribution of faces has less variation than the distribution of natural scenes which eases learning a single strong initialization. Consequently, we make our compression algorithm adaptive to a certain distribution by including \textit{a priori} knowledge into the initialization.

%Next to the performance benefit,
Moreover, meta-learned initializations are a potential solution for long encoding times of \ac{inr}-based compression: Our meta-learned approach can reduce training time by up to 90\% at a fixed performance.

% However, if we exploit meta-learned initializations for faster convergence we have to accept reduction in distortion performance.
% Nevertheless, this is an important step in the right direction and a promising direction for future research. 
%Future work may mitigate this issue by advances in meta-learning algorithms.\\

We also highlight the importance of applying input encodings in \ac{inr}-based compression (see \myfigref{fig:encodings}). This demonstrates significance of choosing the correct inductive biases for compression and is another promising future research avenue. Furthermore, the observation that input encodings render \acp{inr} more robust to quantization (see \myfigref{quant_error}) has potential applications beyond compression.

Interestingly, the here proposed \ac{inr}-based compression technique is competitive on lower resolution images (see \mysecref{sec:sota_comparison}).
However, the performance falls short of \acp{rdae} and BPG on higher resolution images. We hypothesize that processing pixels independently has inefficient scaling properties. Therefore, it is crucial for future research to develop novel architectures for \acp{inr} beyond the \ac{mlp} that mitigate the current deficits at at high resolution images.

Lastly, our basic approach outperforms the traditional algorithm Draco on 3D mesh compression (see \mysecref{sec:3d}). Thus conducting further research into 3D shape compression based on \acp{inr} denotes a promising direction.

% We have demonstrated the capabilities of \acp{inr} for source compression on images, but there is further potential of applying our findings to other data shapes in the future. One of the advantages of \acp{inr} is their flexibility: If we want to compress three dimensional data, we just increase the dimension of the input by another coordinate. Hand-designed algorithms specifically designed for images like JPEG2000 cannot easily adapted to 3D data because transforming the input to frequency space, which is the basis of most hand-designed algorithms, is much more challenging for higher dimensional data. \\

% A challenge that may be addressed in the future is how to scale \acp{inr} to achieve competitive rate-distortion performance at higher bitrates. The general trend we have seen, is that the advantage of conventional and learned autoencoder based methods increases as we compress at higher bitrates. We have observed that for \ac{mlp} based architectures we have to increase the hidden dimension of the layers to achieve less distortion. The number of parameters however grows quadratically with the hidden dimension, which increases the bitrate more significantly than it reduces distortion.

\section{Acknowledgements}
This work was partially supported by Google.

\clearpage
% ---- Bibliography ----
%
% BibTeX users should specify bibliography style 'splncs04'.
% References will then be sorted and formatted in the correct style.
%
\bibliographystyle{splncs04}
\bibliography{bib}

\clearpage
\section{Supplementary Material}
In \mysecref{sec:metalearning} we provide a description of the meta-learning algorithm and in \mysecref{sec:overfitmeta} we explain in detail how we overfit starting from meta-learned initializations.  We motivate our architecture choice in \mysecref{sec:numlayers} and show the influence of $L_1$ regularization in \mysecref{sec:l1reg}.  Additional ablation studies on the influence of quantization bitwidth and the generalization potential of meta-learned initializations are shown in \mysecref{sec:bitwidth} and \mysecref{sec:generalization} respectively. A further runtime comparison between our method, JPEG and SOTA \acp{rdae} is presented in \mysecref{sec:runtime_jpeg_perf}. In \mysecref{sec:hyper} we present a complete overview of the hyperparameters used in the different stages of the compression pipeline. Finally, we provide additional qualitative examples comparing our method to JPEG and JPEG2000 in \mysecref{sec:imagex} and Draco in \mysecref{sec:3dshapevisual}.
\subsection{Meta-Learning Algorithm}\label{sec:metalearning}
\paragraph{Meta-learned Initializations for Implicit Neural Representations}

\myalgoref{alg:metasiren} is an adapted version of the algorithm presented in \cite{sitzmann2019metasdf}. We modify the notation to match ours and change the objective to image regression instead of signed distance function regression. Generally, the algorithm consists of two loops, the outer loop (lines 7-15) and the inner loop (lines 11-13). The outer loop index $i$ is denoted as a superscript, whereas the inner loop index $j$ is denoted as a subscript. The outer loop is executed for a predefined number of iterations $n$. First of all, we sample an image $\vx^i$ from the data distribution. We then define a coordinate vector $\vp$ with a coordinate grid of the same resolution as the image $\vx^i$. We initialize the inner loop parameters $\phi^i_0$  to the current outer loop parameters $\theta$. For $i=1$ this is just random initialization, afterwards these are the meta-learned parameters. We start the inner loop on line 9. For $k$ iterations we compute the \ac{mse} loss between the image $\vx^i$ and the output of the \ac{inr} parameterized by the inner loop parameters  $\phi^i_j$. On line 13 we perform a gradient update of the inner loop parameters. $\alpha$ contains the learning rates for the inner loop gradient update.  A simple choice is using the same static learning rate for all parameters in  $\phi^i_j$.
The power of meta-learning is that we can also meta-learn the learning rate of the inner loop, thus $\alpha$ is an optimization variable just like $\theta$. We can take this even a step further by meta-learning a learning rate for every individual inner loop parameter and every step $j$. This variant is referred to by the authors \cite{sitzmann2019metasdf} as a \emph{per parameter per step} learning rate type \footnote{\url{https://github.com/vsitzmann/metasdf}}. We are effectively learning $k$ times as many learning rates as model parameters. We use the Hadamard product on line 13 to denote that the product between the learning rates in $\alpha$ and the gradient is performed componentwise. The subscript $j$ of $\alpha$ denotes that we have different learning rates in each step $j$.

After $k$ iterations of the inner loop, we recompute the loss for the inner loop parameters of the last step. On line 15 we perform a gradient update of the meta-learned model parameters and the learning rates $\alpha$. This is a gradient of gradients: backpropagation is applied to the computation graph of the inner loop that itself contains the inner loop gradient updates. We continue on line 8 by sampling the next image and repeat the process. After all outer loops have finished, we return the meta-learned weights $\theta$ and learning rates $\alpha$.

\begin{algorithm}[!t]
    \caption{MetaSiren (modified version of MetaSDF\cite{sitzmann2019metasdf} Algorthim 1)}
    \label{alg:metasiren}
    \begin{algorithmic}[1]
        \State \textbf{Required Inputs:}
            \State $\quad \cD$: dataset for meta-learning
            \State $\quad \alpha_{init}$: initial learning rates for inner loop
        \Procedure{trainInitialization}{$\cD, \alpha_{init}$}
        \State Randomly initialize $\theta$
        \State $\alpha \gets \alpha_{init}$
        \For{$i \in [1,n]$}
            \State \text{Sample training image} $\vx^i \sim \cD$
            \State Get coordinates $\vp = \text{coord}( \vx^i) \in [-1,1]^{W \times H}$
            \State Initialize  $\phi^i_0 \gets \theta$, $\cL \gets 0$
            \For{$ j \in [0, k-1]$}
                \State $\cL \gets \text{MSE}(f_{\phi^i_j}(\vp), \vx^i)$
                \State $\phi^i_{j+1} \gets \phi^i_{j} - \alpha_j \odot \nabla_{\phi^i_{j}} \cL$
            \EndFor
            \State $\cL \gets \text{MSE}(f_{\phi^i_k}(\vp), \vx^i)$
            \State $\theta, \alpha \gets (\theta, \alpha) - \beta \nabla_{(\theta, \alpha)} \cL$
        \EndFor
        \State \textbf{return} $\theta, \alpha$
        \EndProcedure
    \end{algorithmic}
\end{algorithm}

\begin{algorithm}[!t]
    \caption{Overfit \ac{inr} starting from a meta-learned initialization}
    \label{alg:overfitmeta}
    \begin{algorithmic}[1]
        \State \textbf{Required Inputs:}
            \State $\quad \vx$: the image to overfit
            \State $\quad \vp$: coordinate grid at desired resolution
            \State $\quad \theta_0$: meta-learned initialization
            \State $\quad \alpha$: meta-learned learning rates
        \Procedure{overfitMeta}{$\vx,\vp, \theta_0, \alpha$}
        \For{$ j \in [0, k-1]$}
            \State $\theta_{j+1} \gets \theta_{j} - \alpha_j \odot \nabla_{\theta_{j}} \text{MSE}(f_{\theta_j}(\vp), \vx)$
        \EndFor
        \State $\theta \gets \theta_k$
        \State $\theta^\star \gets \arg \min_{\theta} \cL(\vx,f_\theta(\vp))$
        \State \textbf{return} $\theta^\star$
        \EndProcedure
    \end{algorithmic}
\end{algorithm}

\subsection{Overfitting from Meta-Learned Initializations}\label{sec:overfitmeta}
\begin{figure}[!t]
\centering
\includegraphics[width=0.7\linewidth]{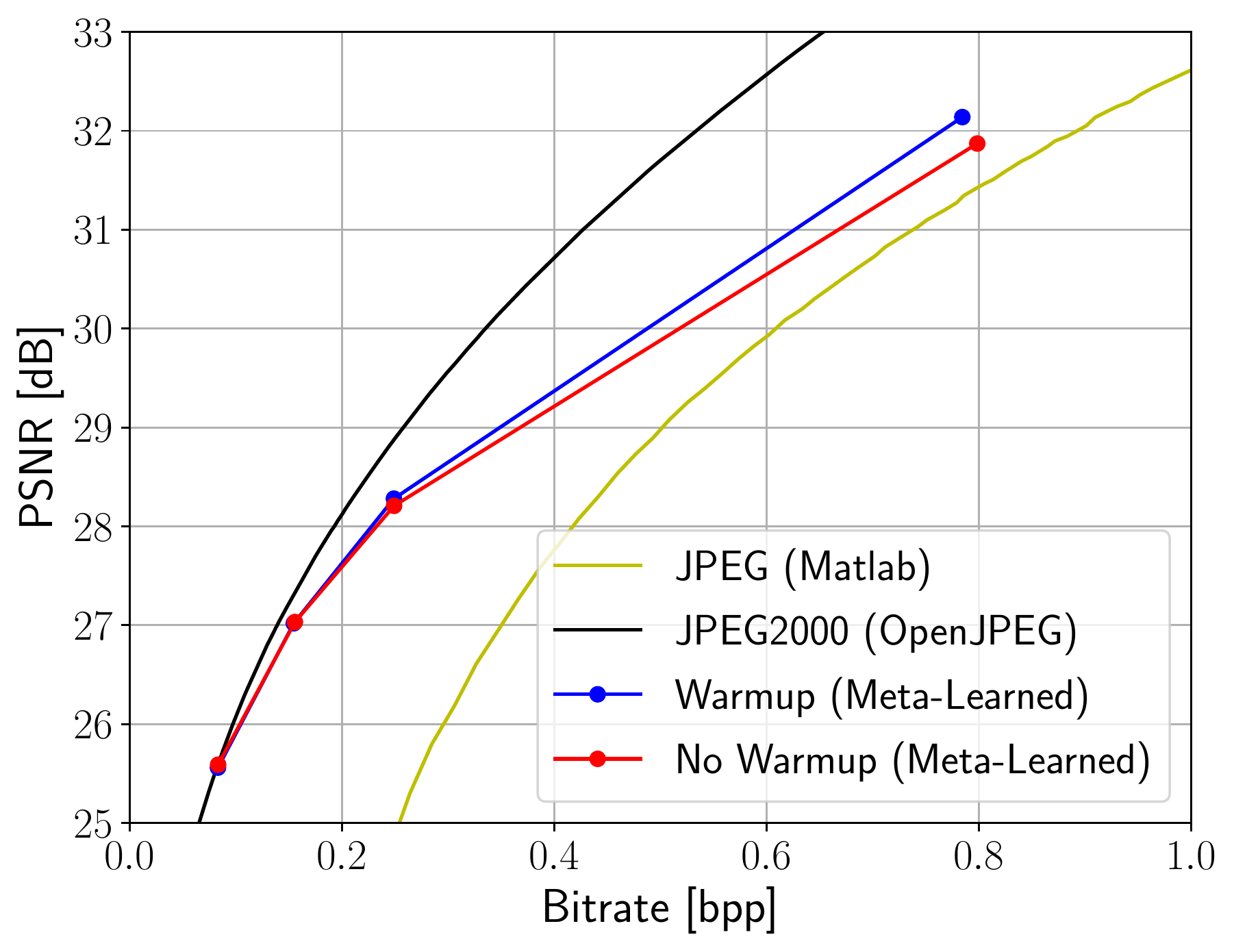}
\caption{Comparing the meta-learned approach evaluated on Kodak with and without a warmup phase in the beginning of the training. We achieve better performance at higher bitrates when using a warmup phase.}
\label{fig:warmup}
\end{figure}

At the beginning of the overfitting phase, we make use of parameter-wise learning rates obtained from meta-learning. Basically, we just run the inner loop once which gets us already close to the final image in just $k=3$ steps as shown in \myalgoref{alg:overfitmeta}. We then continue optimizing with Adam. The momentum terms of the Adam optimizer are uninitialized at this point. We have found that linearly increasing the learning rate in a warmup phase of 100 epochs prevents an initial degradation of reconstruction quality and even improves final performance at higher bitrates (see \myfigref{fig:warmup}).

% \begin{algorithm}
% \begin{algorithmic}[1]
% \end{algorithmic}
% \end{algorithm}

\subsection{Number of Layers and Hidden Dimension}\label{sec:numlayers}
\begin{figure}[!t]
\centering
\begin{subfigure}[b]{.49\linewidth}
\includegraphics[width=\linewidth]{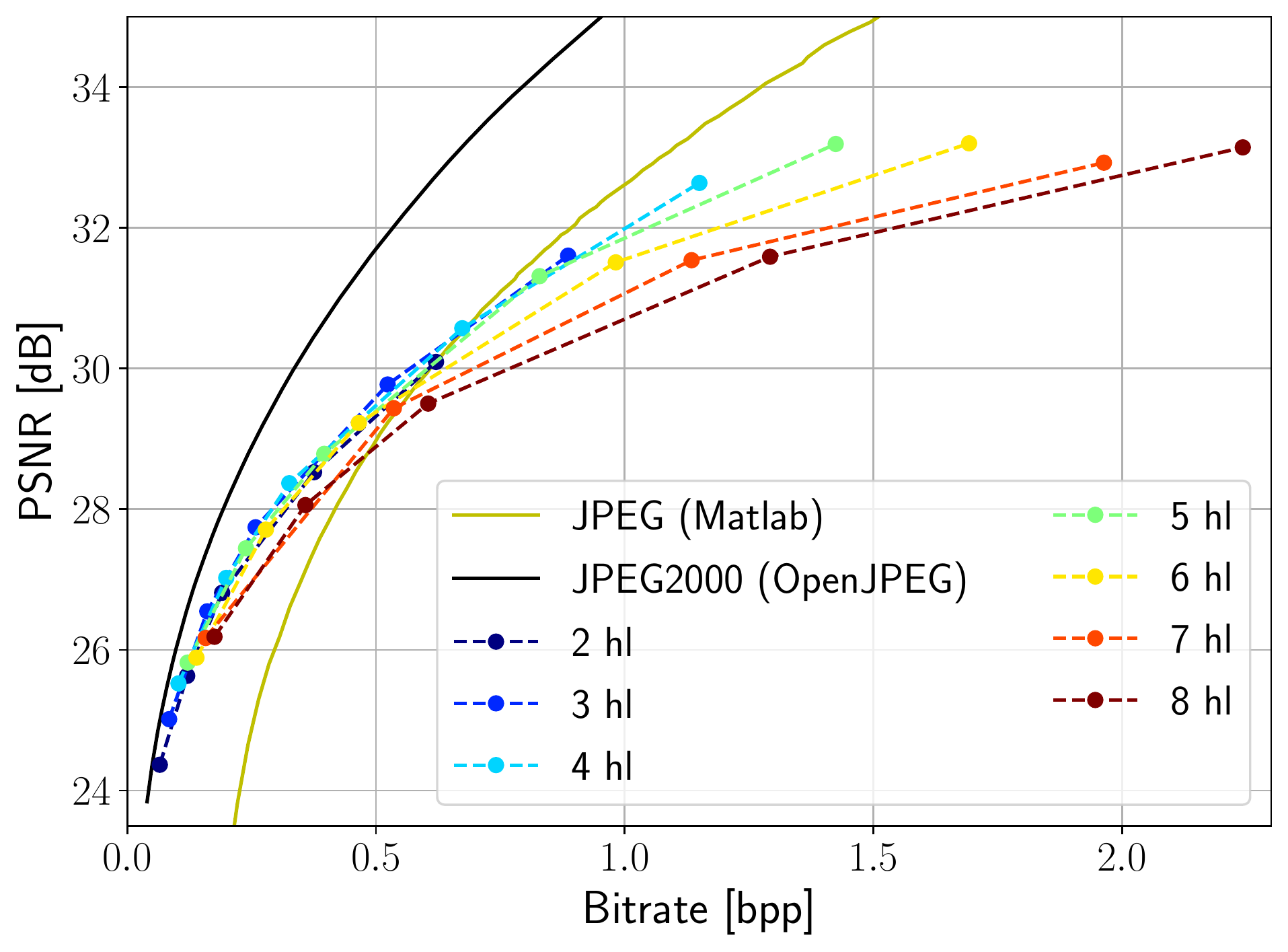}
\caption{Bitwidth $b=7$}
\end{subfigure}
\hfil
\begin{subfigure}[b]{.49\linewidth}
\includegraphics[width=\linewidth]{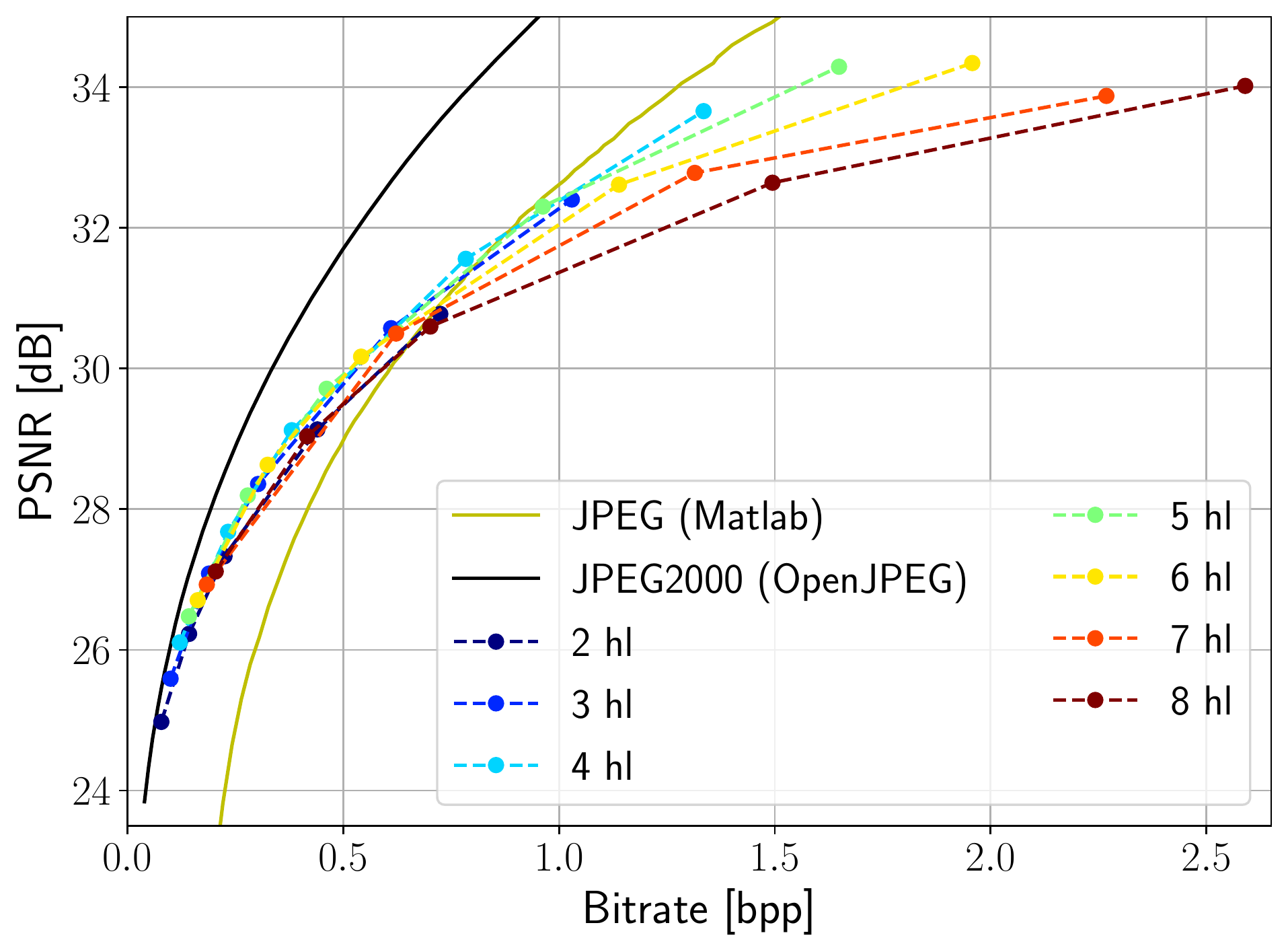}
\caption{Bitwidth $b=8$}
\end{subfigure}
\caption{Comparing compression performance of models with the number of hidden layers (hl) varying between 2-8 and quantization bitwidths of $b=7$ or $b=8$ bits.}
\label{fig:layers}
\end{figure}

Important architecture choices when using \ac{mlp} based networks, are the number of hidden layers and the number of hidden units. Given an \ac{mlp}, \emph{depth} or \emph{width} both directly influence the number of parameters and indirectly impact bitrate. In other words, there are two ways of scaling up the network. We examine the rate-distortion performance for various combinations of hidden units ($M \in \{32, 48, 64, 96, 128\}$) and hidden layers ($\{2, \dots,8\}$) in \myfigref{fig:layers} using our basic approach and $\lambda = 10^{-6}$. 
%The dots connected by a dashed line always show a sweep of hidden layers from 2 to 8, keeping the same hidden dimension.
We can see from both plots that increasing the number of layers eventually leads to diminishing returns: The bitrate keeps increasing while the gain in PSNR is small. The flattening for higher numbers of hidden layers is even more pronounced at the lower bitwidth $b=7$. The quantization noise is stronger here and with increasing depth the noise might get amplified and limit the performance. We conclude that rate-distortion performance scales more gracefully with respect to the width of the model.
We do however notice as well that the lowest setting of 2 hidden layers is typically outperformed by a network with fewer hidden units and more layers.

\subsection{Impact of $L_1$ Regularization.}\label{sec:l1reg}
\begin{figure}
\centering

% \begin{subfigure}[b]{.49\linewidth}
% \includegraphics[width=\linewidth]{img/celeb_l1.pdf}
% \caption{CelebA}
% \end{subfigure}
% \hfil
% \begin{subfigure}[b]{.49\linewidth}
\includegraphics[width=0.7\linewidth]{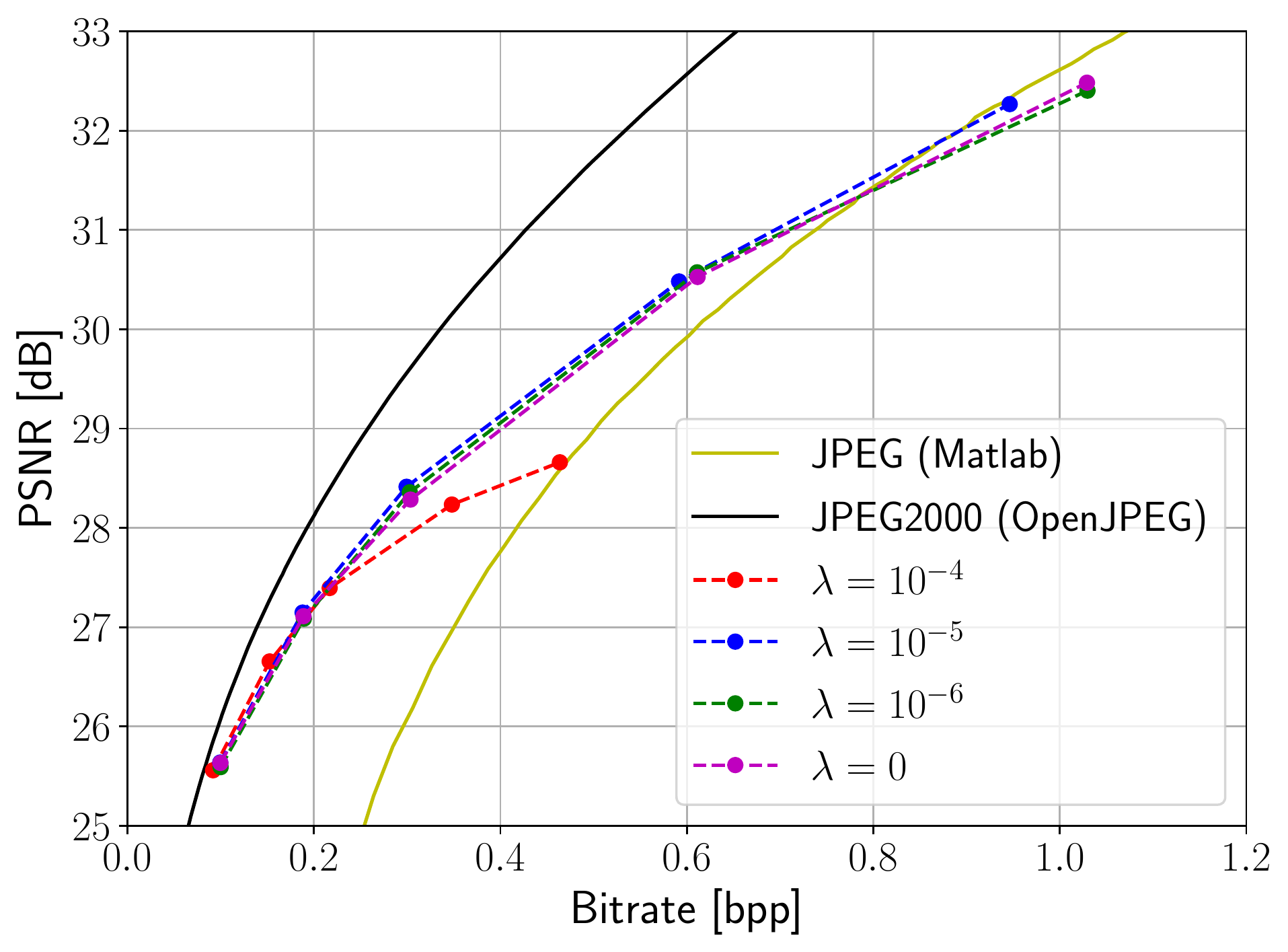}

%\end{subfigure}
\caption{Rate-distortion performance for different $L1$ regularization parameters $\lambda$ evaluated on the Kodak dataset.}
\label{fig:l1reg}
\end{figure}

In this experiment we try to verify whether $L_1$ regularization has a beneficial effect on performance. We train with the default parameters starting from random initializations and vary $\lambda$ within $[0, 10^{-4}]$.\\
In \myfigref{fig:l1reg} we observe that a value of $\lambda = 10
^{-5}$ has better performance at higher bitrates than lower choices for $\lambda$. The performance improvement shows as a reduction in bitrate  which supports the claim that the $L_1$ regularization can lead to a reduction in entropy. Increasing, the regularization strength to $\lambda = 10^{-4}$ restricts the weights too much, resulting in worse performance than $\lambda = 10^{-5}$. Thus, $L_1$ regularization can help to reduce entropy, but needs to be combined with a modification in architecture size to achieve a good rate-distortion tradeoff.

\subsection{Post-Quantization Optimization.}

\begin{figure}[!t]
\centering
% \begin{subfigure}[b]{.49\linewidth}
% \includegraphics[width=\linewidth]{img/celeb_ada.pdf}
% \caption{CelebA}
% \end{subfigure}
% \hfil
% \begin{subfigure}[b]{.49\linewidth}
\includegraphics[width=0.7\linewidth]{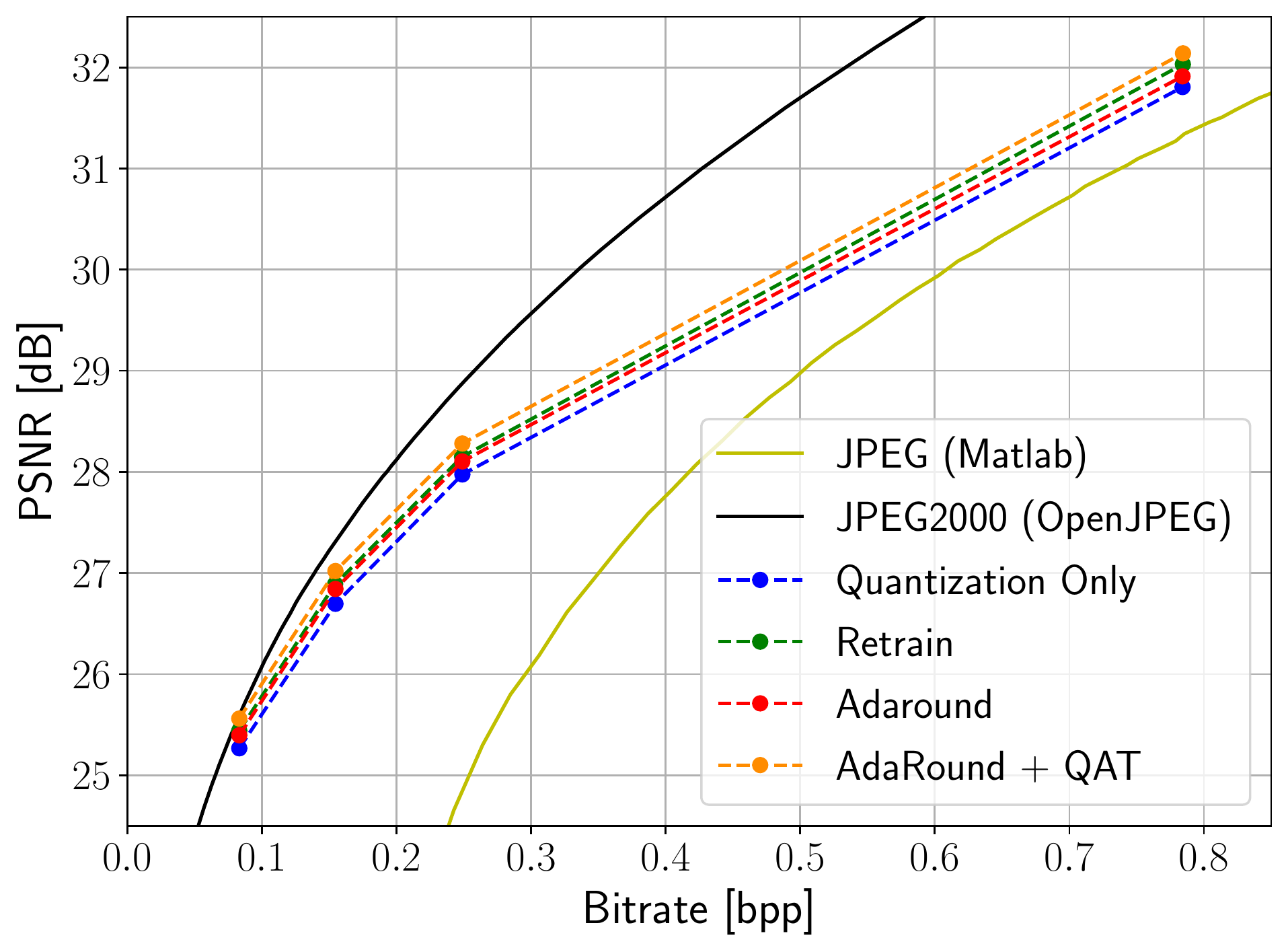}
% \caption{Kodak}
% \end{subfigure}
\caption{Comparison of \ac{qat}, AdaRound and the combination of both to basic quantization on the Kodak dataset.}
\label{fig:ada}
\end{figure}

We compare our meta-learned approach for different post-quantization optimization settings. \myfigref{fig:ada} shows the performance difference evaluated on Kodak. We see that AdaRound and retraining applied on their own lead to a consistent improvement. The best choice throughout the bitrate range is however to apply the methods in conjunction.

% This allows us to reach similar rate-distortion performance as JPEG2000 at the lowest bitrate examined and increases the advantage over JPEG throughout.

\subsection{Influence of Quantization Bitwidth}\label{sec:bitwidth}
\begin{figure}[!t]
\centering
\begin{subfigure}[b]{.49\linewidth}
\includegraphics[width=\linewidth]{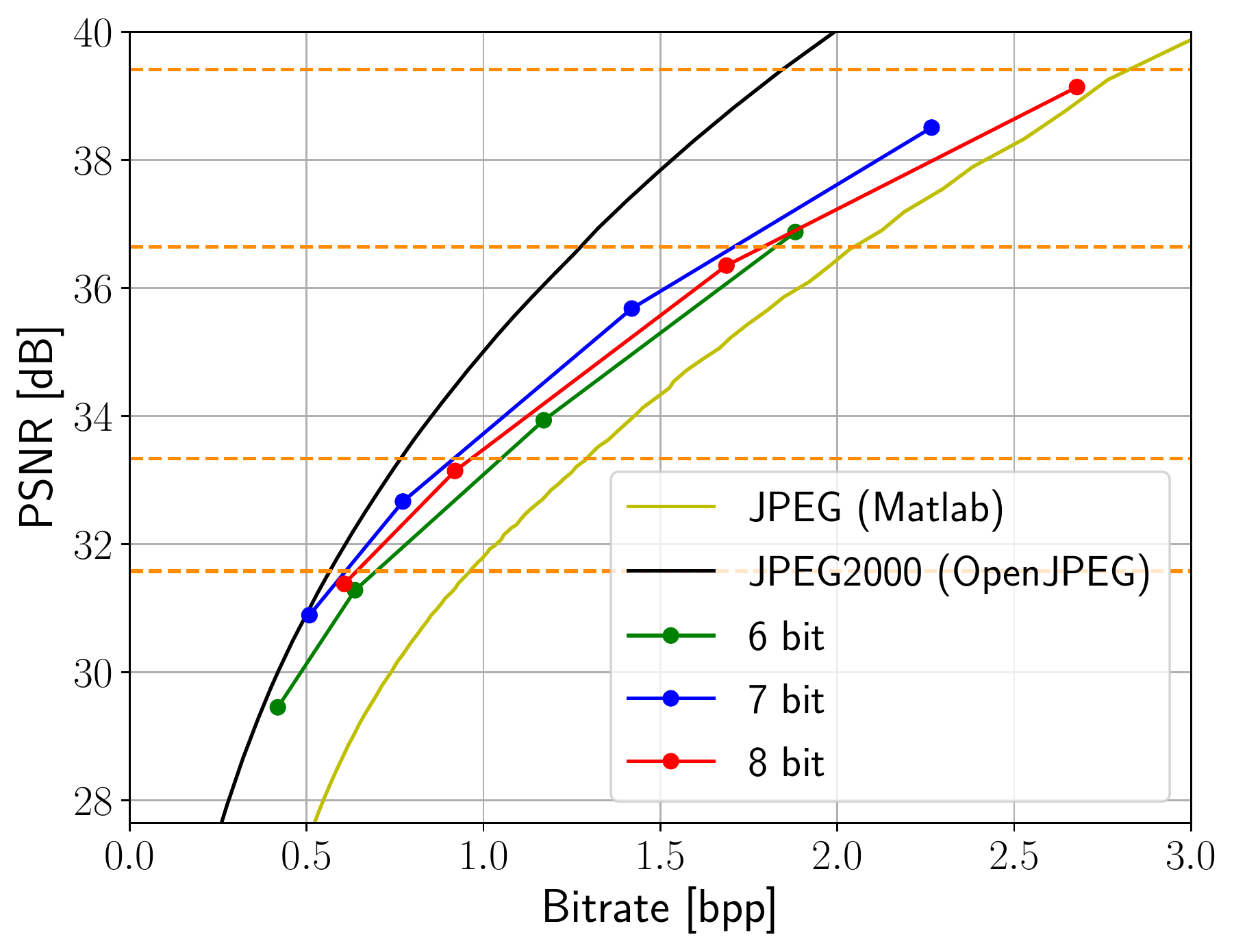}
\caption{CelebA}
\end{subfigure}
\hfil
\begin{subfigure}[b]{.49\linewidth}
\includegraphics[width=\linewidth]{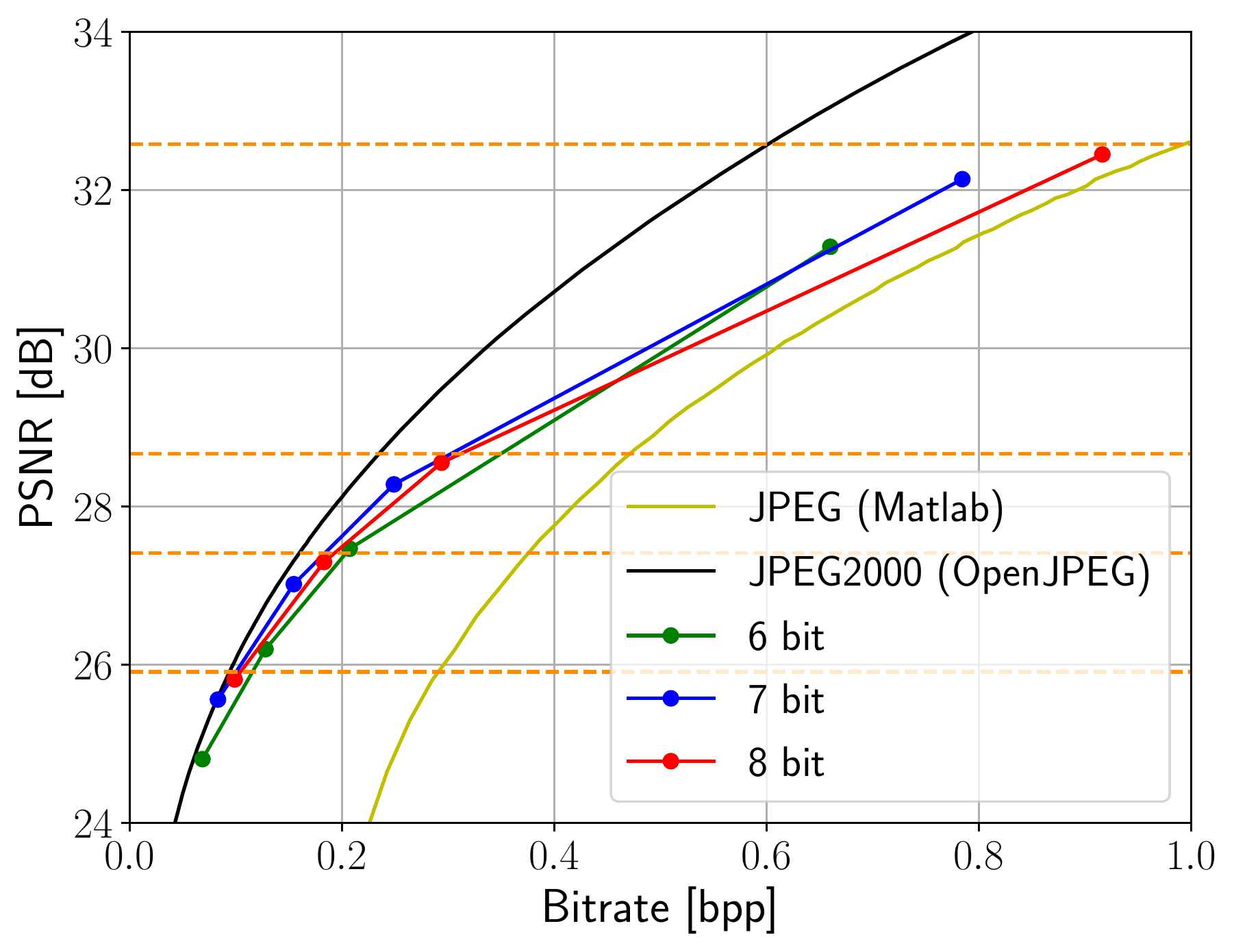}
\caption{Kodak}
\end{subfigure}
\caption{Comparision of different quantization bitwidths for the \emph{meta-learned} approach. The \ac{psnr} achieved by the unquantized models is shown by the dashed horizontal lines. Note that these are not rate-distortion curves and are only supposed to show the distortion introduced by quantization.}
\label{fig:bw}
\end{figure}

\begin{figure}[!t]
\centering
\begin{subfigure}[b]{.49\linewidth}
\includegraphics[width=\linewidth]{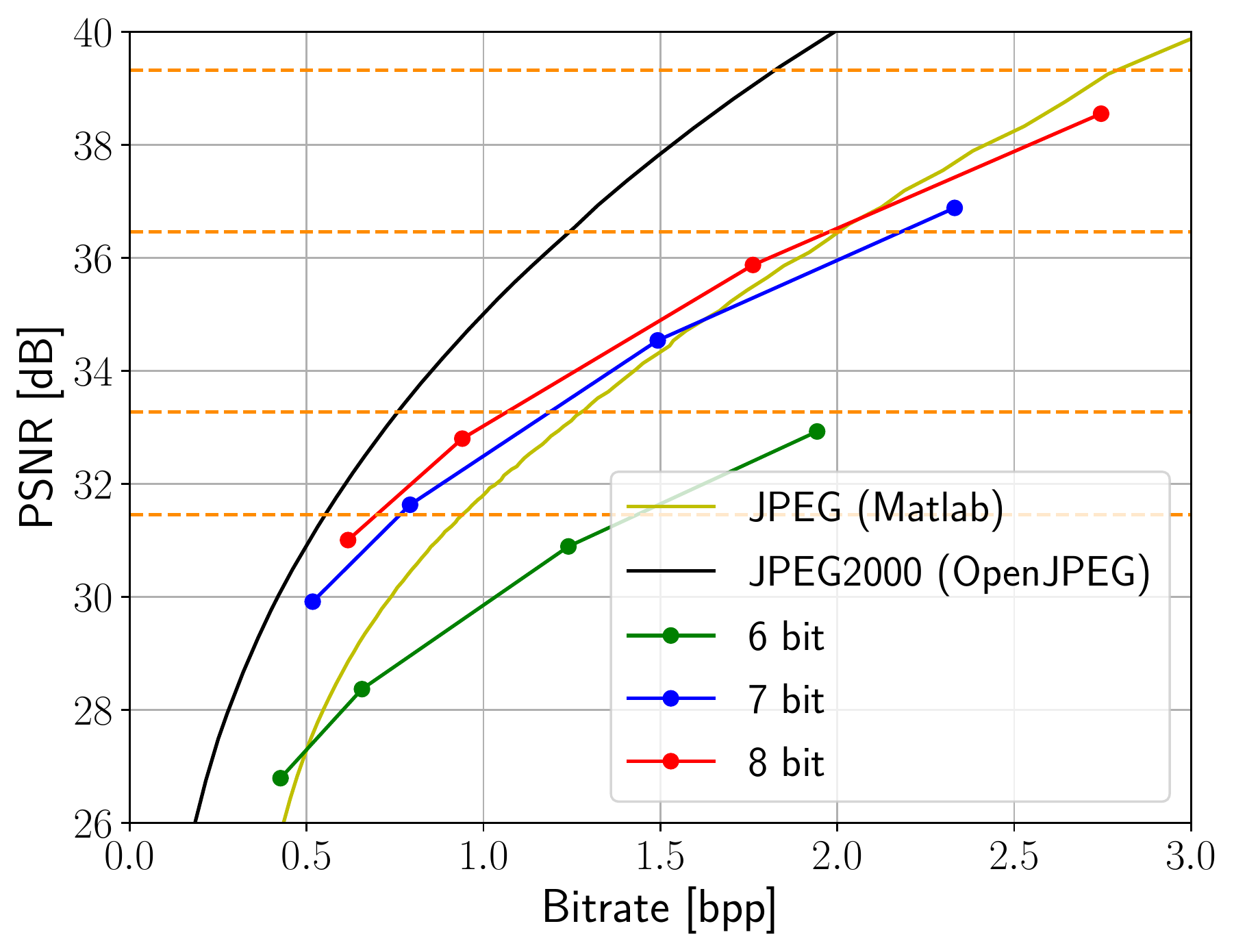}
\caption{CelebA}
\end{subfigure}
\hfil
\begin{subfigure}[b]{.49\linewidth}
\includegraphics[width=\linewidth]{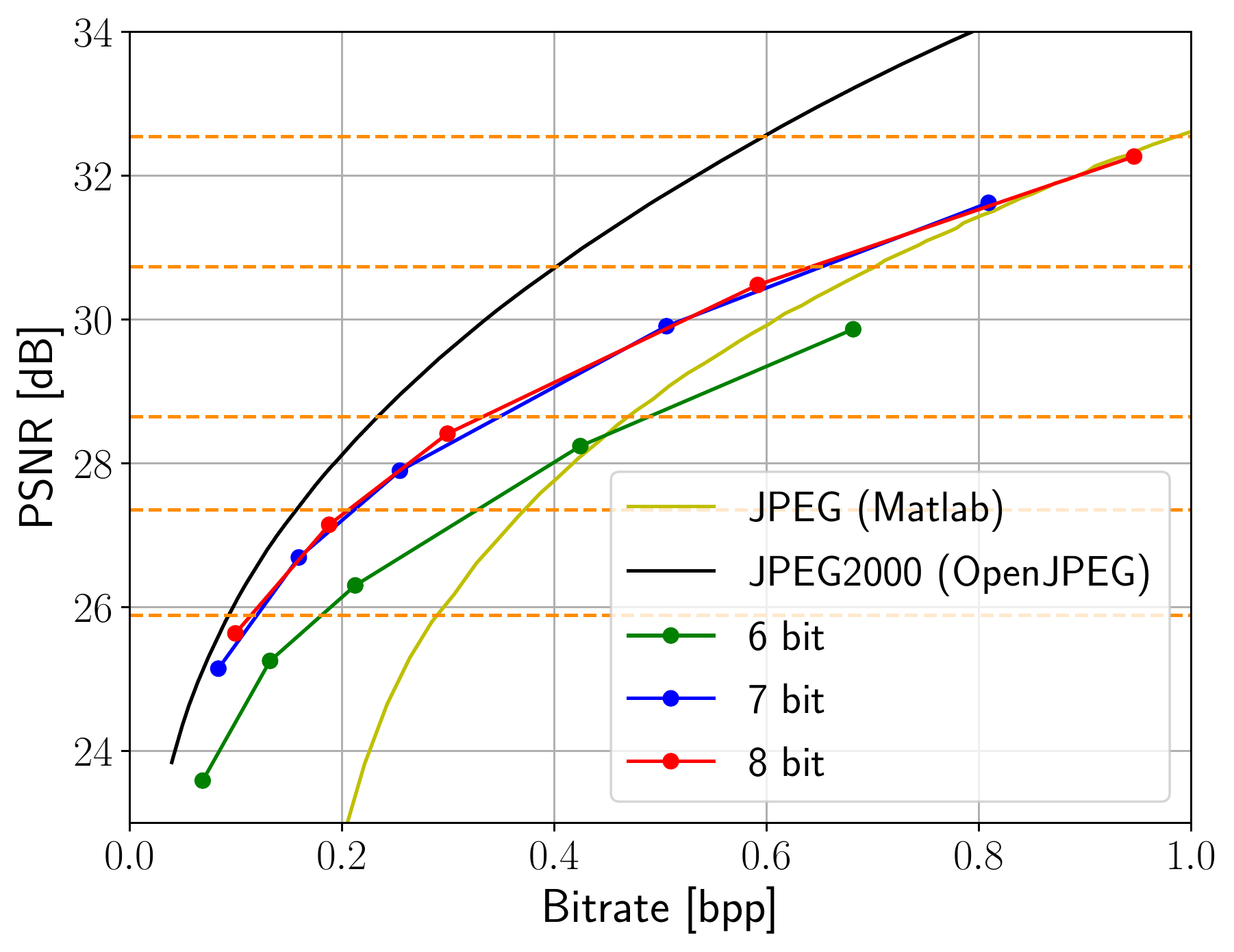}
\caption{Kodak}
\end{subfigure}
\caption{Comparision of different quantization bitwidths for the \emph{basic} approach. The \ac{psnr} achieved by the unquantized models is shown by the dashed horizontal lines. Note that these are not rate-distortion curves and are only supposed to show the distortion introduced by quantization.}
\label{fig:bw_base}
\end{figure}

We show the influence of bitwidth on the rate-distortion performance in \myfigref{fig:bw} for the meta-learned approach and in \myfigref{fig:bw_base} for the basic approach. For the meta-learned approach 7-bit is the best choice for both datasets. For the basic approach however, 8-bit quantization outperforms lower bitwidths. On the Kodak dataset the difference between 7- and 8-bit quantization is quite small nevertheless. We also show the unquantized performance of the 4 \acp{mlp} with variying number of hidden units as dashed horizontal lines. We see that for a bitwidth of 8 we can almost reach unquantized performance for the majority of configurations.

\subsection{Generalization of Meta-Learned Initializations}\label{sec:generalization}
\begin{figure}[!t]
\centering

\includegraphics[width=0.7\linewidth]{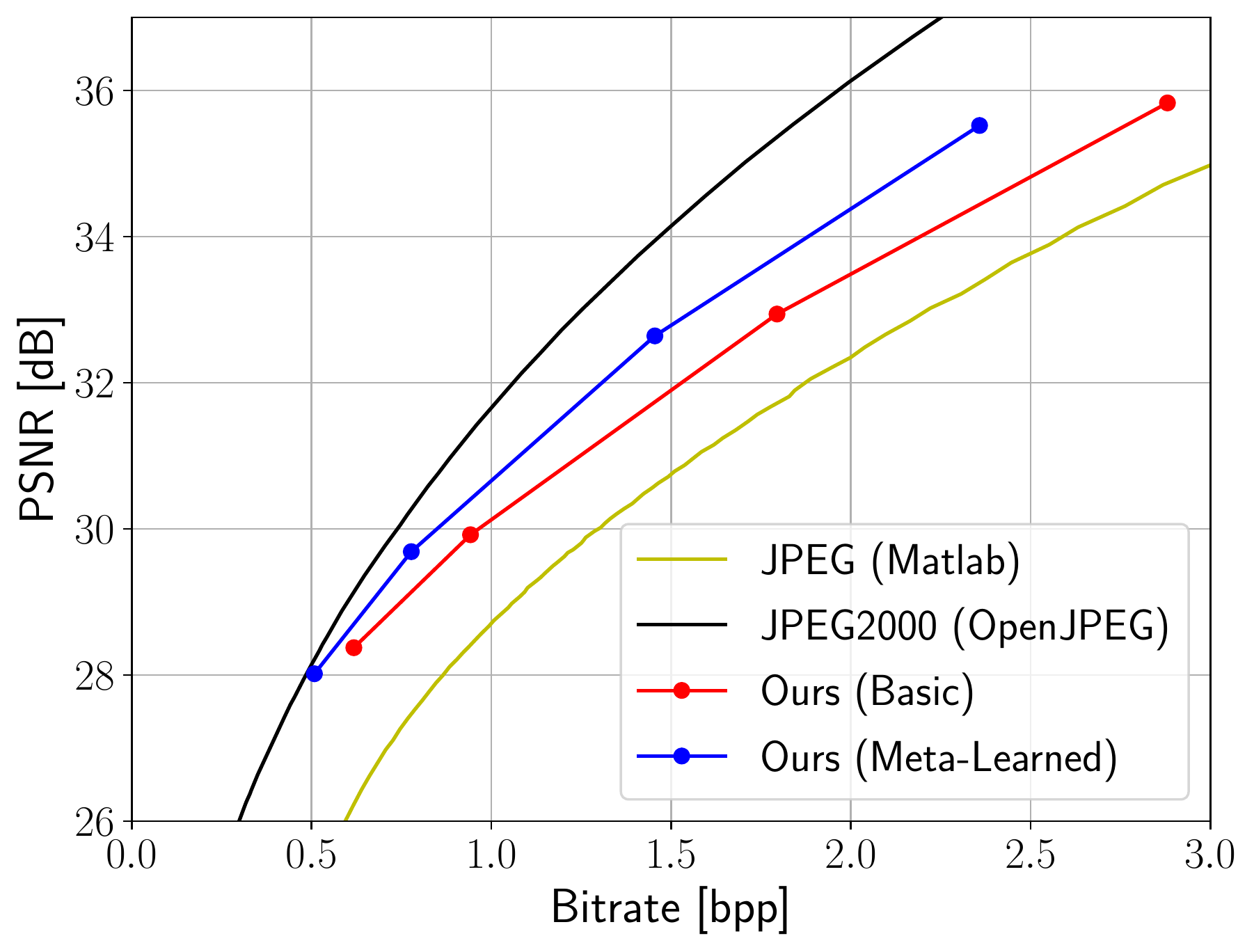}
\caption{Generalization experiment: Using meta-learned initializations trained on CelebA evaluated on (cropped  and resized) Kodak images. The meta-learned initializations provide enough generalization capability to improve compression performance also on out-of-distribution images.}
\label{fig:generalize}
\end{figure}

We want to show that the meta-learned initializations are able to generalize to out-of-distribution images, even if the meta-learning dataset contains only similar images.
To this end, we minimally crop and resize Kodak images to the same resolution and aspect ratio as CelebA ($178 \times 218$) and then compress them using meta-learned initializations obtained from CelebA. In \myfigref{fig:generalize} we show that the meta-learned approach still outperforms the basic approach.

\subsection{Further Runtime Comparison}\label{sec:runtime_jpeg_perf}
In this paper we typically train the models until full convergence, hence we optimize for the best rate-distortion performance. To show that our method can be tuned for fast runtime, we compare the encoding/decoding runtime (in [s]) with JPEG and Xie et al~\cite{xie2021enhanced} in Tab. \ref{rt_comparison}, under the constraint that the rate-distortion performance at least matches JPEG. Thus, we trade-off performance with runtime.\\
%We show that we can outperform JPEG in RD performance within relatively short encoding times.
\begin{table}
\centering
\begin{tabular}{ |p{1.5cm}||p{1.1cm}|p{1.1cm}|p{1.1cm}| p{1.1cm}| }
 \hline
 & JPEG & Xie~\cite{xie2021enhanced}& Ours & Ours*\\
 \hline
Encoding  & 0.0061 &  2.22  & 4.49 & 0.542 \\
Decoding  & 0.0023 &  5.69  & 0.664   & 0.722 \\
 \hline
\end{tabular}
\caption{Comparison of average encoding and decoding speeds in [s] on Kodak. For \acp{inr} we show the time at 0.175 bpp where JPEG is outperformed in terms of PSNR in all cases. 
%The tests were performed on an AMD Ryzen 5900X CPU and an Nvidia RTX 3070 GPU. 
All methods use the same hardware. 
%For our method, training and model evaluation is performed on GPU, while entropy coding is performed on CPU which currently limits better decoding speeds: Loading the weights and evaluating the INR requires only 0.09s. 
\ac{inr} encoding runs on GPU. Everything else, including decoding with \acp{inr} runs on CPU.
Xie [58] only supports CPU inference for encoding and decoding according to the authors. Note that half of our decoding time is entropy coding. Ours* (w. meta-learning) reaches the performance of JPEG approximately 10x faster than Ours (no meta-learning). JPEG is clearly the fastest method, but we can outperform Xie~\cite{xie2021enhanced}, given that the encoding device has a GPU. We deliberately show decoding time for CPU because GPUs are less common on the end-user device and it makes the timings more comparable to Xie~\cite{xie2021enhanced} and JPEG. }\label{rt_comparison}
\end{table}

\subsection{Detailed Training Hyperparameter Overview}\label{sec:hyper}
In this section we provide a complete overview of all hyperparameters. We also mention details specific to the training procedure in the respective subsection.
\subsubsection{Image Compression Hyperparameters}
\paragraph{Architecture.}
We summarize the default architecture hyperparameters in \mytableref{tab:archparam}. We typically evaluate our method for several choices of the hidden dimension $M$. For Kodak and CelebA images at full resolution we use $M \in \{32,48,64,128\} $ and  $M \in \{24, 32, 48, 64\} $ respectively. For Kodak images at half resolution (2x scale) we use $M \in \{8, 16, 32, 48\}$ and also reduce the number of input frequencies to $L=12$. For Kodak images at quarter resolution (4x scale) we use $M \in \{4, 8, 16, 32\}$ and further reduce the number of input frequencies to $L=10$. 
\begin{table}[!t]
\centering
\begin{tabular}{ |p{0.45\linewidth}|p{0.45\linewidth}|}
 \hline
 \multicolumn{2}{|c|}{Architecture Hyperparameters} \\
 \hline
 Description & Value\\
 \hline
 Hidden layers $N$ & $3$\\
 Activation function  & $\sin(30 x)$\\
 Input encoding &  Positional with $\sigma = 1.4$\\
 Input frequencies $L$ & 16 (Kodak), 12 (CelebA)\\
  \hline
\end{tabular}
\caption{Default architecture hyperparameters.}
\label{tab:archparam}
\end{table}

\paragraph{Meta-Learning the Initializations.}

In \mytableref{tab:metaparam} we list the default values of the hyperparameters for meta-learning the initializations. We use a learning rate schedule that halves the learning rate when no improvement has been made in the last ($10 = \text{patience} $) validations. For validation we use a subset of $100$ images sampled from the validation set of the respective dataset, in our case CelebA or DIV2K. We compute the validation loss by running the inner loop optimization for each image in the validation subset. The fact that computing the validation loss involves inner loop training is the reason why we limit the validation set size to 100. When we train the initializations on CelebA we train for 1 epoch. When using the DIV2K dataset we train for 30 epochs because it is a significantly smaller dataset. We finally, save the initialization that achieved the lowest validation loss overall.

\begin{table}[t!]
\centering
\begin{tabular}{ |p{0.45\linewidth}|p{0.45\linewidth}| }
 \hline
 \multicolumn{2}{|c|}{Meta-Learning Hyperparameters} \\
 \hline
 Description & Value\\
 \hline
 Outer loop initial learning rate $\beta$ & $5 \cdot 10^{-5}$\\
 Outer loop batch size  & 1\\
 Outer loop optimizer &   Adam \cite{kingma2015adam}\\
 Epochs &  30 (DIV2K), 1 (CelebA)\\
 \hline
  Inner loop initial learning rate $\alpha_{init}$ &  $10^{-5}$\\
Learning rate type &   \emph{per parameter per step}\\
 Inner loop steps $k$  & $3$\\
 \hline
 Steps until validation & $500$\\
 LR schedule patience & $10$\\
 LR schedule factor & $0.5$\\
  \hline
\end{tabular}
\caption{Default hyperparameters for learning the initializations.}
\label{tab:metaparam}
% \vspace{-2mm}
\end{table}

\begin{table}[t!]
\centering
\begin{tabular}{ |p{0.45\linewidth}|p{0.45\linewidth}| }
 \hline
 \multicolumn{2}{|c|}{Overfitting Hyperparameters} \\
 \hline
 Description & Value\\
 \hline

 Initial learning rate & $5 \cdot 10^{-4}$\\
 Optimizer &   Adam \cite{kingma2015adam}\\
 Epochs & 25000\\
 $L_1$ regularization $\lambda$ & $10^{-5}$\\
 \hline
 Steps until validation & $1$\\
 LR schedule patience & $500$\\
 LR schedule factor & $0.5$\\
 Early stopping epochs & $5000$\\
  \hline
\end{tabular}
\caption{Default hyperparameters for overfitting the \ac{inr}.}
\label{tab:fitparam}
\end{table}

\paragraph{Overfitting.}
The default hyperparameters for the overfitting stage are shown in \mytableref{tab:fitparam}. If not mentioned otherwise we use these hyperparameters for all experiments. In particular, we do not use $L_1$ regularization for experiments at reduced resolution.
Since we are overfitting, we validate on the training image itself. We reduce the learning rate by a factor of $0.5$ if the loss has not improved during the last $500$ epochs. Note that 1 epoch equals 1 optimizer step, in other words, one training batch contains all pixels of the image we overfit. We train for $25000$ epochs to make sure every architecture and configuration we test has the chance to reach convergence. As to be expected, smaller models typically reach peak performance faster. We stop training if the loss has not improved in the last $5000$ epochs to prevent unnecessary computation resource use. In the end, we return the parameters of the model that has achieved the lowest loss during overfitting.

\paragraph{Quantization, Post-Quantization Optimization \& Entropy Coding.}
In \mytableref{tab:quantparam} we show the default values for the quantization and bitstream coding related stages. We emphasize that per default we use the combination of AdaRound and \ac{qat}.
\begin{table}[t!]
\centering

\begin{tabular}{ |p{0.4\linewidth}|p{0.5\linewidth}| }
 \hline
 \multicolumn{2}{|c|}{Quantization Hyperparameters} \\
 \hline
 Description & Value\\
 \hline
 Bitwidth & $7$ (Meta-Learned), $8$ (Basic) \\
 \hline
 Retraining epochs  & 300\\
 Optimizer &   Adam \cite{kingma2015adam}\\
 Retraining learning rate & $10^{-6}$  \\
 \hline
 AdaRound iterations & $1000$\\
 AdaRound regularizer & $10^{-4}$\\
  \hline
   Bitstream coding & arithmetic coding\\
  \hline
\end{tabular}
\caption{Default hyperparameters for quantization, post-quantization optimization and bitstream coding.}
\label{tab:quantparam}
\end{table}

\subsubsection{3D Shape Compression Hyperparameters}
We use a very similar training procedure for the 3D shape compression as for image compression. For images we simply use the pixels of one image as the batch used for overfitting. For 3D however, we need to choose a number of 3D point samples, in our case 100000, for which we first compute the ground truth distance to the surface and then use them to fit our model. We use a subset of 10000 points as a batch for training. Overall, we train for 500 epochs with an initial learning rate of $5 \cdot 10^{-5}$ and the same learning rate schedule as for images. For 3D shape compression we do not use $L_1$ regularization. We use the same architectures as for Kodak, namely $M \in \{32,48,64,128\}$ and $L = 16$ input frequencies. The overfitted models are quantized to 8 bit and we then optimize the weights using 2000 iterations of AdaRound and 50 epochs of retraining with a learning rate of $10^{-7}$.\\
For the Draco baseline we call the encoder with a certain bitwith to quantize the mesh, in particular, the "qp" flag. We vary the bithwidth within $[5, \dots, 12]$. We use the highest compression quality setting "cl" of 10. To make sure we only encode a raw mesh, we set the skip flag for TEXTURE, NORMAL and GENERIC.
% \paragraph{Baselines}
% The conventional methods we report are: the Matlab implementation of JPEG with default settings, the OpenJPEG implementation of JPEG2000 with default settings and BPG with chroma subsampling set to "444". For the baselines using the learned autoencoder approach we use the results published by CompressAI ~\cite{begaint2020compressai} for the Kodak dataset  and use their unofficial implementation and pre-trained models to compute the baselines for the CelebA dataset. The state-of-the art baseline \cite{xie2021enhanced} is not available from CompressAI. We thus report the published results for Kodak from the official repository and evaluate their pre-trained models to obtain the performance on CelebA. For COIN~\cite{coin} we report the results on Kodak published in their work.

\subsection{Additional Image Examples}\label{sec:imagex}
We show additional compression examples to compare our method to the codecs JPEG and JPEG2000. We first show more images at the lowest bitrate, \ie using the lowest hidden dimension, where our method is most competitive to JPEG2000: In \myfigref{fig:imagecomparisokodak04} we evaluate on KODAK using a model with $M=32$ and in \myfigref{fig:imagecomparisonceleb199118} we evaluate on CelebA using a model with $M=24$. We visually confirm a clear advantage over JPEG for all images and similar performance as JPEG2000 at this bitrate. Moreover, we show examples at higher bitrates, \ie using larger hidden dimensions, in \myfigref{fig:imagecomparisokodak20} (Kodak) and \myfigref{fig:imagecomparisonceleb185853} (CelebA). While our method maintains an advantage over JPEG, JPEG2000 outperforms our approach with an increasing advantage towards higher bitrates, where the difference is most apparent in the rendering of fine details.

\renewcommand{\footnotesize}{\fontsize{8pt}{11pt}\selectfont}
\begin{figure*}
    \setlength{\linewidth}{\textwidth}
    \setlength{\hsize}{\linewidth}
	\centering
	\renewcommand{\tabcolsep}{0.0pt}
	\renewcommand{\arraystretch}{1}
	\begin{tabular}{cccc}	
	    \footnotesize original & 
		\footnotesize JPEG & 
		\footnotesize JPEG2000 &
		\footnotesize Ours (Meta-learned)  \\	
		\begin{tikzpicture}[spy using outlines={circle,magenta,magnification=2,size=1.6cm, connect spies}]
			\node{\includegraphics[ width=0.22\linewidth]{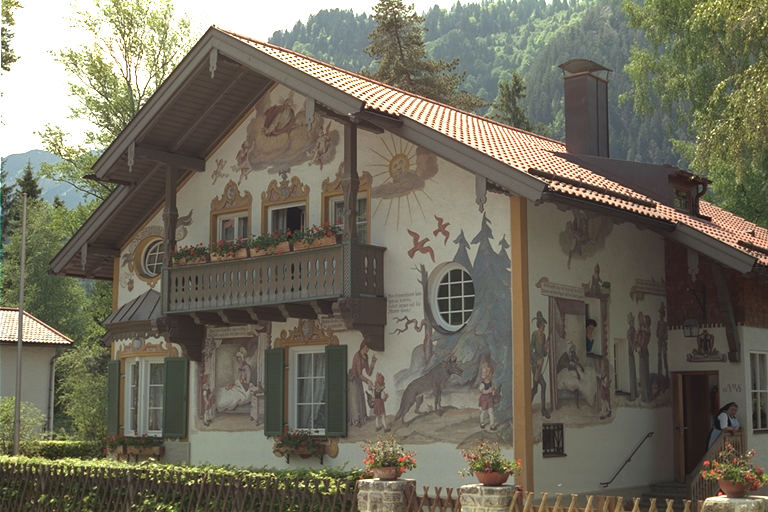}};
			\spy on (-0.5, 0.2) in node [right] at (-1.5, -1.8);%(1.6, 0.7);
		\end{tikzpicture} &
		\begin{tikzpicture}[spy using outlines={circle,magenta,magnification=2,size=1.6cm, connect spies}]
			\node [label={[font=\footnotesize, xshift=0.7cm, yshift=-2.7cm, align=center] $0.182$ bpp\\ $20.78$ dB} ]{\includegraphics[ width=0.22\linewidth]{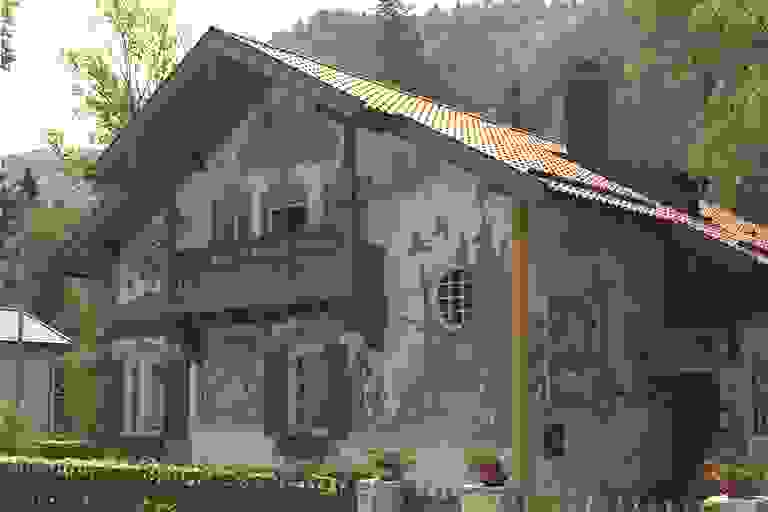}};
			\spy on (-0.5, 0.2) in node [right] at (-1.5, -1.8);%(1.6, 0.7);
		\end{tikzpicture} &			
		\begin{tikzpicture}[spy using outlines={circle,magenta,magnification=2,size=1.6cm, connect spies}]
			\node [label={[font=\footnotesize, xshift=0.7cm, yshift=-2.7cm, align=center] $0.0846$ bpp\\ $22.69$ dB} ]{\includegraphics[width=0.22\linewidth]{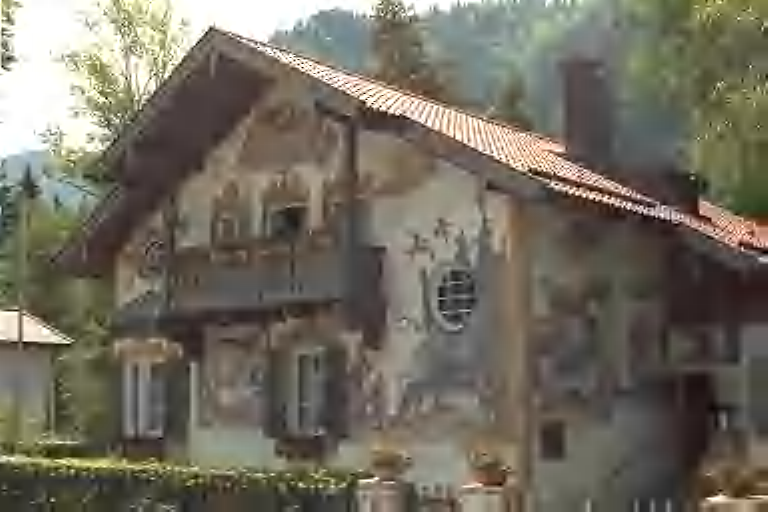}};
			\spy  on (-0.5, 0.2) in node [right] at (-1.5, -1.8);
		\end{tikzpicture} &			
		\begin{tikzpicture}[spy using outlines={circle,magenta,magnification=2,size=1.6cm, connect spies}]
			\node [label={[font=\footnotesize, xshift=0.7cm, yshift=-2.7cm, align=center] $0.0843$ bpp\\ $22.49$ dB} ]{\includegraphics[width=0.22\linewidth]{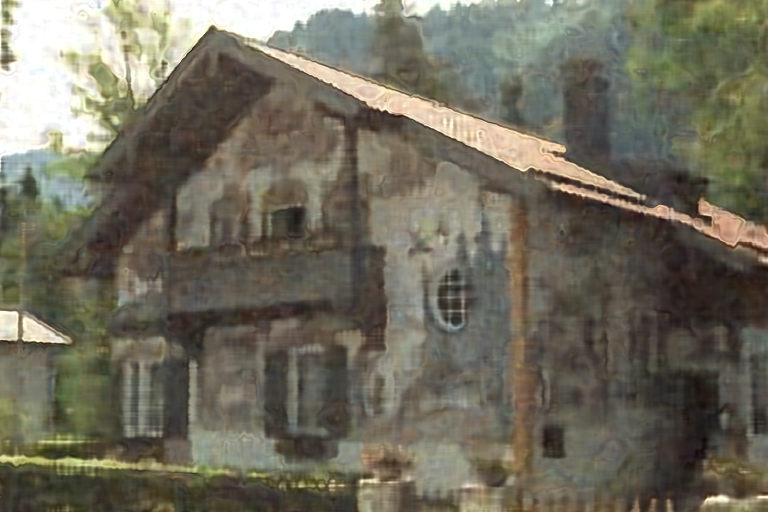}};
			\spy  on (-0.5, 0.2) in node [right] at (-1.5, -1.8);
		\end{tikzpicture} 	\\

		\begin{tikzpicture}[spy using outlines={circle,magenta,magnification=2,size=1.6cm, connect spies}]
			\node{\includegraphics[ width=0.22\linewidth]{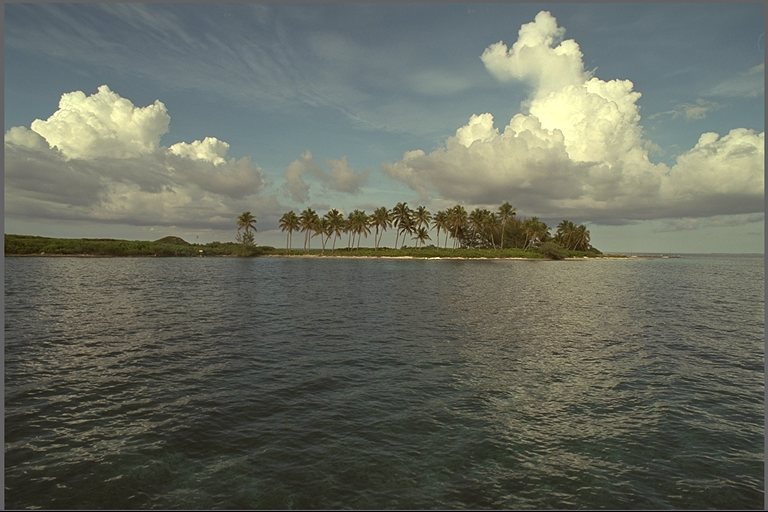}};
			\spy on (-0.5, 0.2) in node [right] at (-1.5, -1.8);%(1.6, 0.7);
		\end{tikzpicture} &
		\begin{tikzpicture}[spy using outlines={circle,magenta,magnification=2,size=1.6cm, connect spies}]
			\node [label={[font=\footnotesize, xshift=0.7cm, yshift=-2.7cm, align=center] $0.150$ bpp\\ $23.13$ dB} ]{\includegraphics[ width=0.22\linewidth]{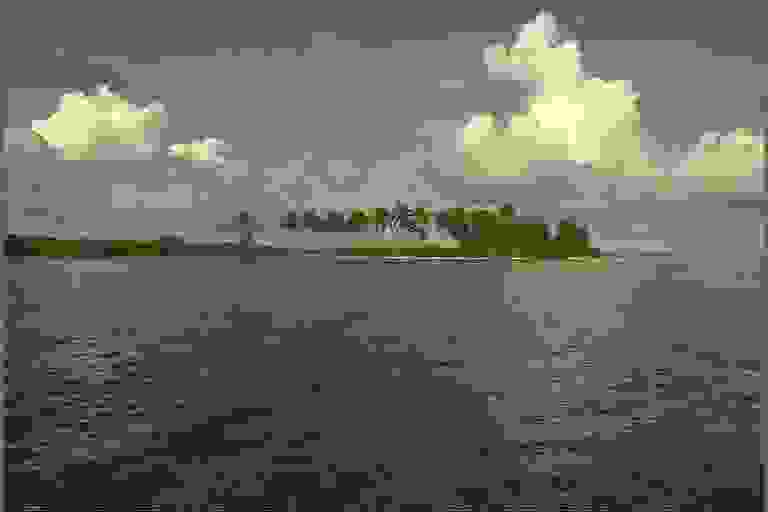}};
			\spy on (-0.5, 0.2) in node [right] at (-1.5, -1.8);%(1.6, 0.7);
		\end{tikzpicture} &			
		\begin{tikzpicture}[spy using outlines={circle,magenta,magnification=2,size=1.6cm, connect spies}]
			\node [label={[font=\footnotesize, xshift=0.7cm, yshift=-2.7cm, align=center] $0.0839$ bpp\\ $27.92$ dB} ]{\includegraphics[width=0.22\linewidth]{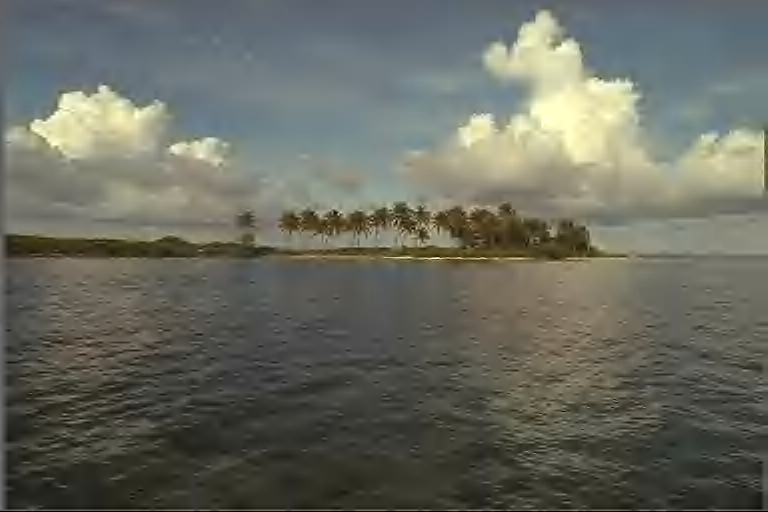}};
			\spy  on (-0.5, 0.2) in node [right] at (-1.5, -1.8);
		\end{tikzpicture} &			
		\begin{tikzpicture}[spy using outlines={circle,magenta,magnification=2,size=1.6cm, connect spies}]
			\node [label={[font=\footnotesize, xshift=0.7cm, yshift=-2.7cm, align=center] $0.0837$ bpp\\ $28.02$ dB} ]{\includegraphics[width=0.22\linewidth]{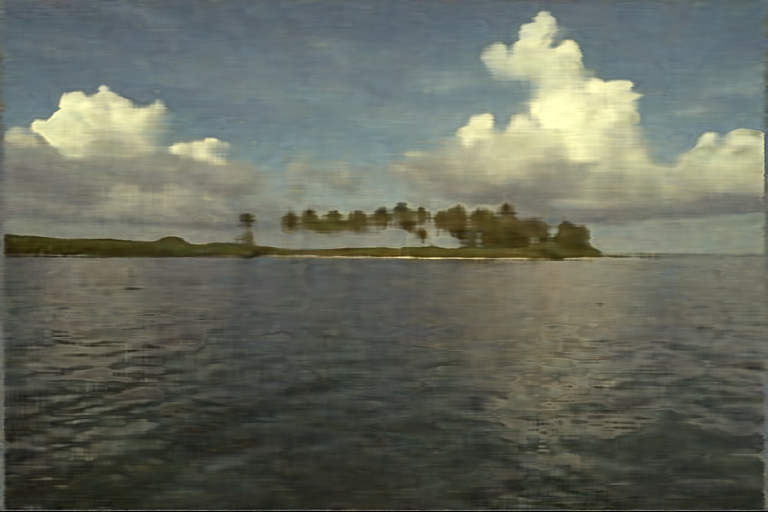}};
			\spy  on (-0.5, 0.2) in node [right] at (-1.5, -1.8);
		\end{tikzpicture} 	\\

		\begin{tikzpicture}[spy using outlines={circle,magenta,magnification=2,size=1.6cm, connect spies}]
			\node{\includegraphics[ width=0.22\linewidth]{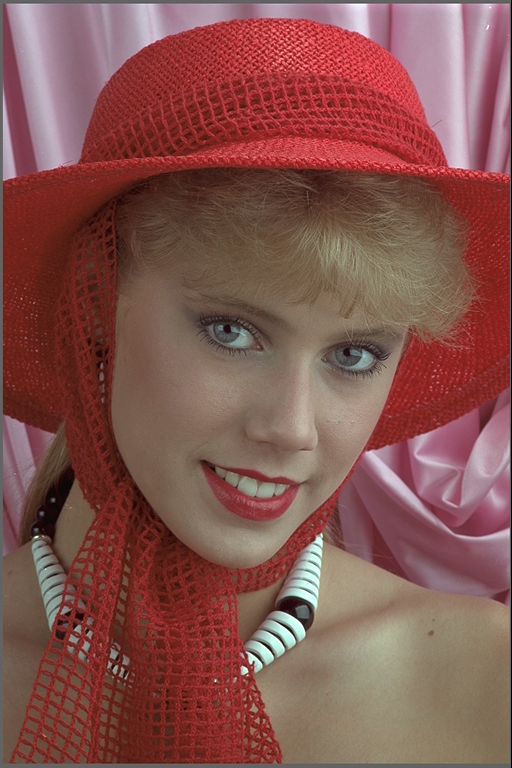}};
			\spy on (-0.3, 0.2) in node [right] at (-1.5, -2.9);%(1.6, 0.7);
		\end{tikzpicture} &
		\begin{tikzpicture}[spy using outlines={circle,magenta,magnification=2,size=1.6cm, connect spies}]
			\node [label={[font=\footnotesize, xshift=0.7cm, yshift=-4.9cm, align=center] $0.158$ bpp\\ $21.24$ dB} ]{\includegraphics[ width=0.22\linewidth]{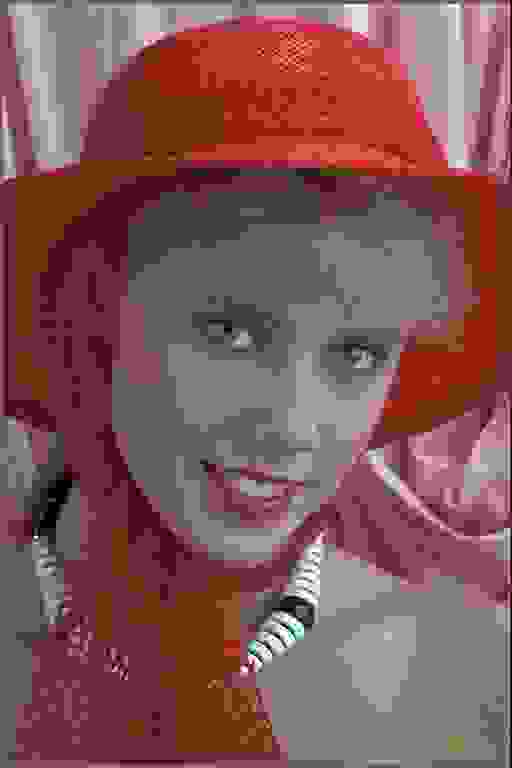}};
			\spy on (-0.3, 0.2) in node [right] at (-1.5, -2.9);%(1.6, 0.7);
		\end{tikzpicture} &			
		\begin{tikzpicture}[spy using outlines={circle,magenta,magnification=2,size=1.6cm, connect spies}]
			\node [label={[font=\footnotesize, xshift=0.7cm, yshift=-4.9cm, align=center] $0.0820$ bpp\\ $27.86$ dB} ]{\includegraphics[width=0.22\linewidth]{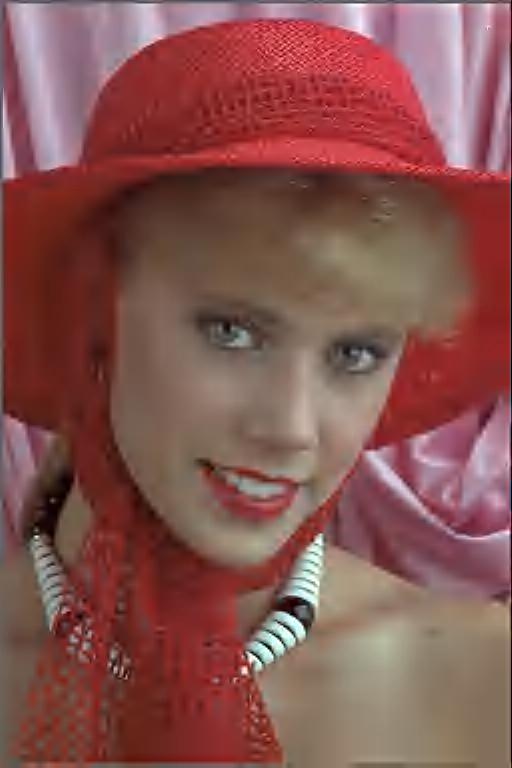}};
			\spy  on (-0.3, 0.2) in node [right] at (-1.5, -2.9);
		\end{tikzpicture} &			
		\begin{tikzpicture}[spy using outlines={circle,magenta,magnification=2,size=1.6cm, connect spies}]
			\node [label={[font=\footnotesize, xshift=0.7cm, yshift=-4.9cm, align=center] $0.0800$ bpp\\ $27.93$ dB} ]{\includegraphics[width=0.22\linewidth]{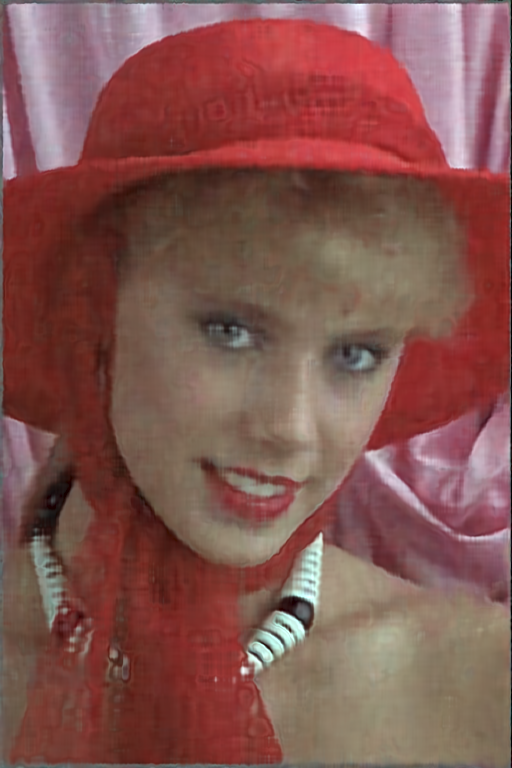}};
			\spy  on (-0.3, 0.2) in node [right] at (-1.5, -2.9);
		\end{tikzpicture} 	\\
	%	\vspace{-5mm}
	\end{tabular}
% 	\vspace{1mm}
\caption{Performance comparison on Kodak using a hidden dimension of $M=32$ for all models.}
\vspace{-5mm}
\label{fig:imagecomparisokodak04}
\end{figure*}

\begin{figure*}
    \setlength{\linewidth}{\textwidth}
    \setlength{\hsize}{\linewidth}
	\centering
	\renewcommand{\tabcolsep}{0.0pt}
	\renewcommand{\arraystretch}{0}
	\begin{tabular}{cccc}	
	    \footnotesize original & 
		\footnotesize JPEG & 
		\footnotesize JPEG2000 &
		\footnotesize Ours (Meta-learned)  \\	
		\begin{tikzpicture}[spy using outlines={circle,magenta,magnification=1.5,size=1.6cm, connect spies}]
			\node{\includegraphics[ width=0.22\linewidth]{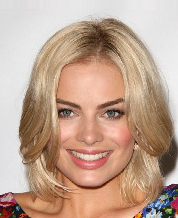}};
			\spy on (-0.3, -0.2) in node [right] at (-1.5, -2.5);%(1.6, 0.7);
		\end{tikzpicture} &
		\begin{tikzpicture}[spy using outlines={circle,magenta,magnification=1.5,size=1.6cm, connect spies}]
			\node [label={[font=\footnotesize, xshift=0.7cm, yshift=-4.2cm, align=center] $0.517$ bpp\\ $25.74$ dB} ]{\includegraphics[ width=0.22\linewidth]{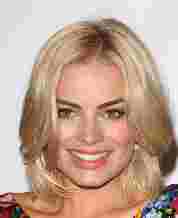}};
			\spy on (-0.3, -0.2) in node [right] at (-1.5, -2.5);%(1.6, 0.7);
		\end{tikzpicture} &			
		\begin{tikzpicture}[spy using outlines={circle,magenta,magnification=1.5,size=1.6cm, connect spies}]
			\node [label={[font=\footnotesize, xshift=0.7cm, yshift=-4.2cm, align=center] $0.522$ bpp\\ $28.31$ dB} ]{\includegraphics[width=0.22\linewidth]{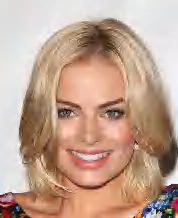}};
			\spy  on (-0.3, -0.2) in node [right] at (-1.5, -2.5);
		\end{tikzpicture} &			
		\begin{tikzpicture}[spy using outlines={circle,magenta,magnification=1.5,size=1.6cm, connect spies}]
			\node [label={[font=\footnotesize, xshift=0.7cm, yshift=-4.2cm, align=center] $0.505$ bpp\\ $27.71$ dB} ]{\includegraphics[width=0.22\linewidth]{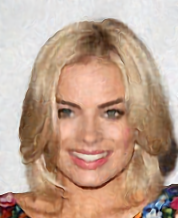}};
			\spy  on (-0.3, -0.2) in node [right] at (-1.5, -2.5);
		\end{tikzpicture} 	\\
	
		\begin{tikzpicture}[spy using outlines={circle,magenta,magnification=1.5,size=1.6cm, connect spies}]
			\node{\includegraphics[ width=0.22\linewidth]{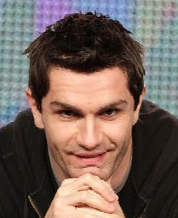}};
			\spy on (-0.3, -0.2) in node [right] at (-1.5, -2.5);
		\end{tikzpicture} &
		\begin{tikzpicture}[spy using outlines={circle,magenta,magnification=1.5,size=1.6cm, connect spies}]
			\node [label={[font=\footnotesize, xshift=0.7cm, yshift=-4.2cm, align=center] $0.529$ bpp\\ $27.02$ dB} ]{\includegraphics[ width=0.22\linewidth]{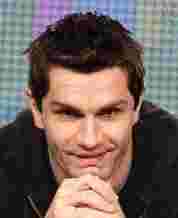}};
			\spy on (-0.3, -0.2) in node [right] at (-1.5, -2.5);%(1.6, 0.7);
		\end{tikzpicture} &			
		\begin{tikzpicture}[spy using outlines={circle,magenta,magnification=1.5,size=1.6cm, connect spies}]
			\node [label={[font=\footnotesize, xshift=0.7cm, yshift=-4.2cm, align=center] $0.520$ bpp\\ $29.90$ dB} ]{\includegraphics[width=0.22\linewidth]{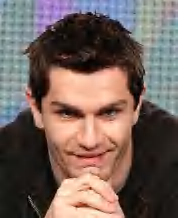}};
			\spy  on (-0.3, -0.2) in node [right] at (-1.5, -2.5);
		\end{tikzpicture} &			
		\begin{tikzpicture}[spy using outlines={circle,magenta,magnification=1.5,size=1.6cm, connect spies}]
			\node [label={[font=\footnotesize, xshift=0.7cm, yshift=-4.2cm, align=center] $0.516$ bpp\\ $30.52$ dB} ]{\includegraphics[width=0.22\linewidth]{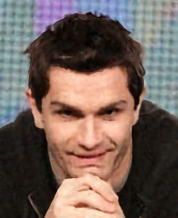}};
			\spy  on (-0.3, -0.2) in node [right] at (-1.5, -2.5);
		\end{tikzpicture} 	\\

		\begin{tikzpicture}[spy using outlines={circle,magenta,magnification=1.5,size=1.6cm, connect spies}]
			\node{\includegraphics[ width=0.22\linewidth]{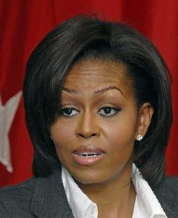}};
			\spy on (-0.3, -0.2) in node [right] at (-1.5, -2.5);%(1.6, 0.7);
		\end{tikzpicture} &
		\begin{tikzpicture}[spy using outlines={circle,magenta,magnification=1.5,size=1.6cm, connect spies}]
			\node [label={[font=\footnotesize, xshift=0.7cm, yshift=-4.2cm, align=center] $0.520$ bpp\\ $28.98$ dB} ]{\includegraphics[ width=0.22\linewidth]{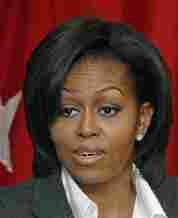}};
			\spy on (-0.3, -0.2) in node [right] at (-1.5, -2.5);%(1.6, 0.7);
		\end{tikzpicture} &			
		\begin{tikzpicture}[spy using outlines={circle,magenta,magnification=1.5,size=1.6cm, connect spies}]
			\node [label={[font=\footnotesize, xshift=0.7cm, yshift=-4.2cm, align=center] $0.533$ bpp\\ $32.82$ dB} ]{\includegraphics[width=0.22\linewidth]{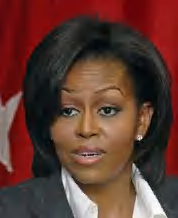}};
			\spy  on (-0.3, -0.2) in node [right] at (-1.5, -2.5);
		\end{tikzpicture} &			
		\begin{tikzpicture}[spy using outlines={circle,magenta,magnification=1.5,size=1.6cm, connect spies}]
			\node [label={[font=\footnotesize, xshift=0.7cm, yshift=-4.2cm, align=center] $0.513$ bpp\\ $33.29$ dB} ]{\includegraphics[width=0.22\linewidth]{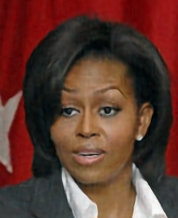}};
			\spy  on (-0.3, -0.2) in node [right] at (-1.5, -2.5);
		\end{tikzpicture} 	\\
	%	\vspace{-5mm}
	\end{tabular}
% 	\vspace{1mm}
\caption{Performance comparison on CelebA using a hidden dimension of $M=24$ for all models.}
\vspace{-5mm}
\label{fig:imagecomparisonceleb199118}
\end{figure*}

%KODAK differen M
\begin{figure*}
    \setlength{\hsize}{\linewidth}
	\centering
	\renewcommand{\tabcolsep}{0.0pt}
	\renewcommand{\arraystretch}{0.2}
	\begin{tabular}{cccc}	
	    \footnotesize original & 
		\footnotesize JPEG & 
		\footnotesize JPEG2000 &
		\footnotesize Ours (Meta-learned)  \\	
		\begin{tikzpicture}[spy using outlines={circle,magenta,magnification=2,size=1.6cm, connect spies}]
			\node{\includegraphics[ width=0.22\linewidth]{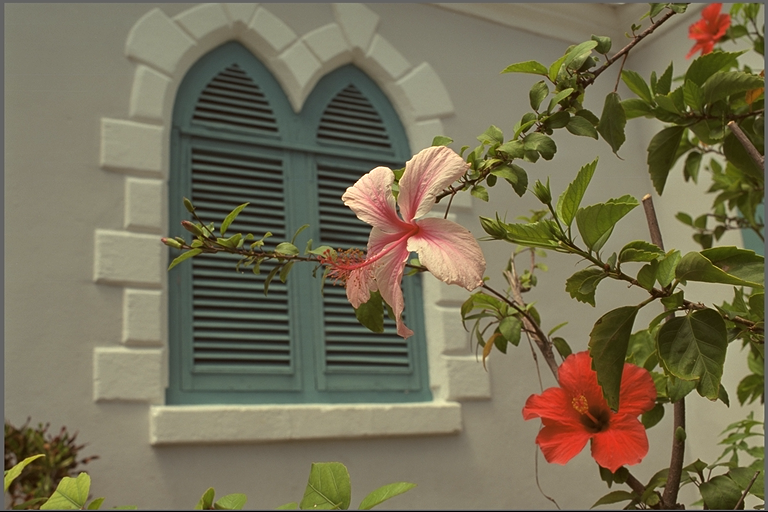}};
			\spy on (-0.5, 0.2) in node [right] at (-1.5, -1.8);%(1.6, 0.7);
		\end{tikzpicture} &
		\begin{tikzpicture}[spy using outlines={circle,magenta,magnification=2,size=1.6cm, connect spies}]
			\node [label={[font=\footnotesize, xshift=0.7cm, yshift=-2.7cm, align=center] $0.171$ bpp\\ $21.82$ dB} ]{\includegraphics[ width=0.22\linewidth]{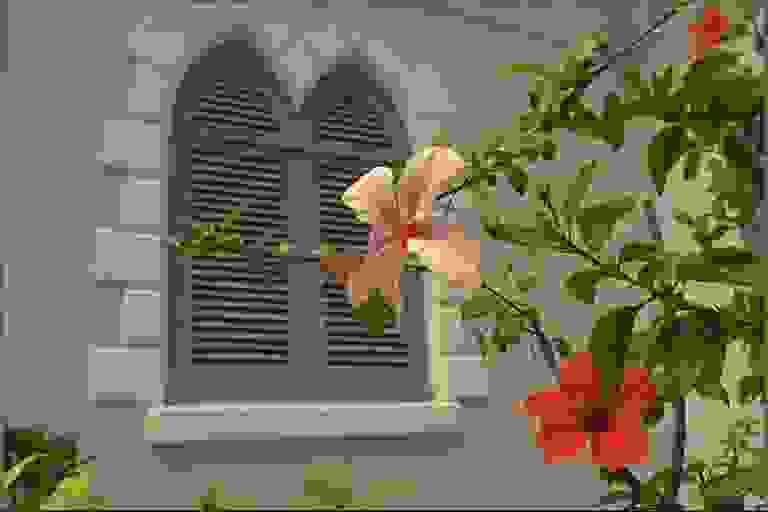}};
			\spy on (-0.5, 0.2) in node [right] at (-1.5, -1.8);%(1.6, 0.7);
		\end{tikzpicture} &			
		\begin{tikzpicture}[spy using outlines={circle,magenta,magnification=2,size=1.6cm, connect spies}]
			\node [label={[font=\footnotesize, xshift=0.7cm, yshift=-2.7cm, align=center] $0.157$ bpp\\ $28.25$ dB} ]{\includegraphics[width=0.22\linewidth]{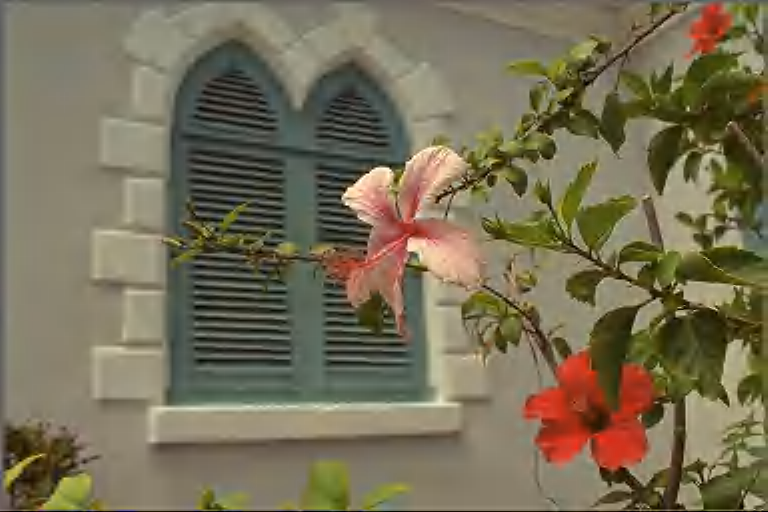}};
			\spy  on (-0.5, 0.2) in node [right] at (-1.5, -1.8);
		\end{tikzpicture} &			
		\begin{tikzpicture}[spy using outlines={circle,magenta,magnification=2,size=1.6cm, connect spies}]
			\node [label={[font=\footnotesize, xshift=0.7cm, yshift=-2.7cm, align=center] $0.156$ bpp\\ $28.05$ dB} ]{\includegraphics[width=0.22\linewidth]{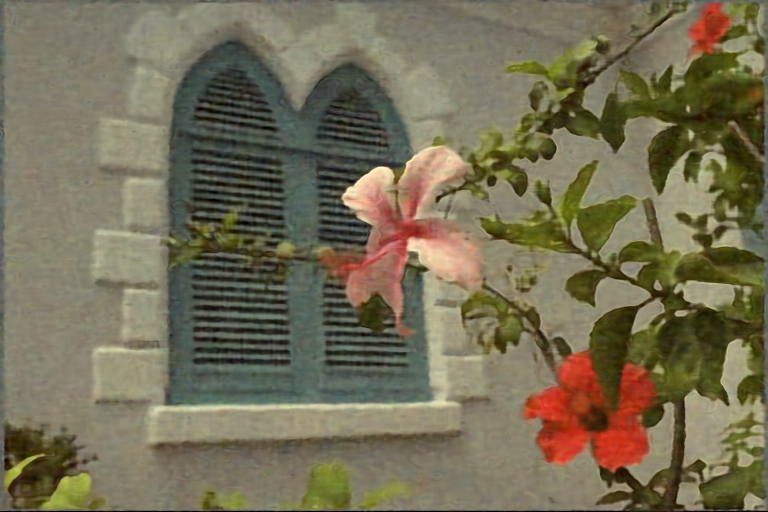}};
			\spy  on (-0.5, 0.2) in node [right] at (-1.5, -1.8);
		\end{tikzpicture} 	\\

		\begin{tikzpicture}[spy using outlines={circle,magenta,magnification=2,size=1.6cm, connect spies}]
			\node{\includegraphics[ width=0.22\linewidth]{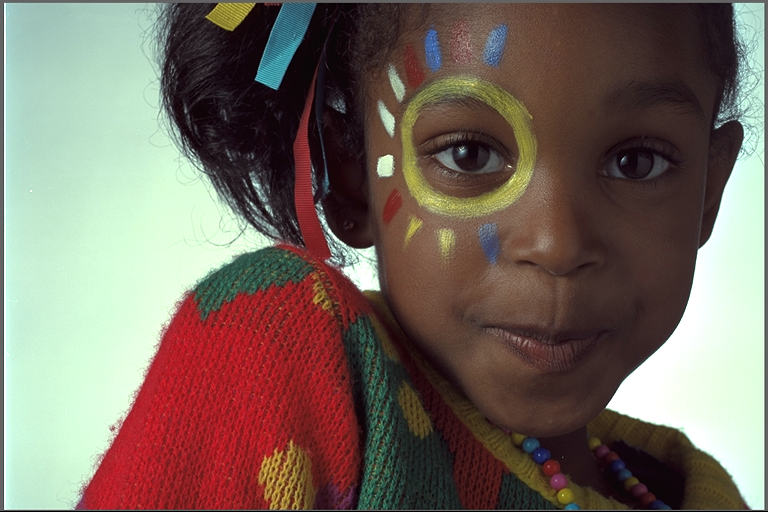}};
			\spy on (-0.5, 0.2) in node [right] at (-1.5, -1.8);%(1.6, 0.7);
		\end{tikzpicture} &
		\begin{tikzpicture}[spy using outlines={circle,magenta,magnification=2,size=1.6cm, connect spies}]
			\node [label={[font=\footnotesize, xshift=0.7cm, yshift=-2.7cm, align=center] $0.258$ bpp\\ $27.72$ dB} ]{\includegraphics[ width=0.22\linewidth]{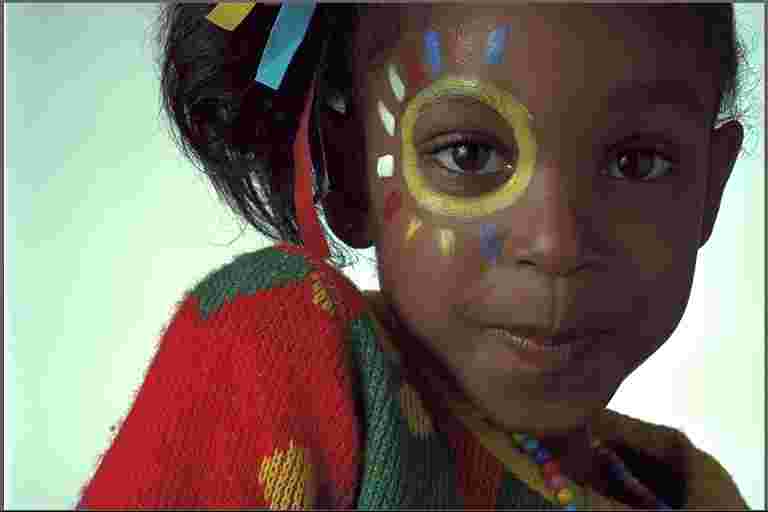}};
			\spy on (-0.5, 0.2) in node [right] at (-1.5, -1.8);%(1.6, 0.7);
		\end{tikzpicture} &			
		\begin{tikzpicture}[spy using outlines={circle,magenta,magnification=2,size=1.6cm, connect spies}]
			\node [label={[font=\footnotesize, xshift=0.7cm, yshift=-2.7cm, align=center] $0.253$ bpp\\ $31.22$ dB} ]{\includegraphics[width=0.22\linewidth]{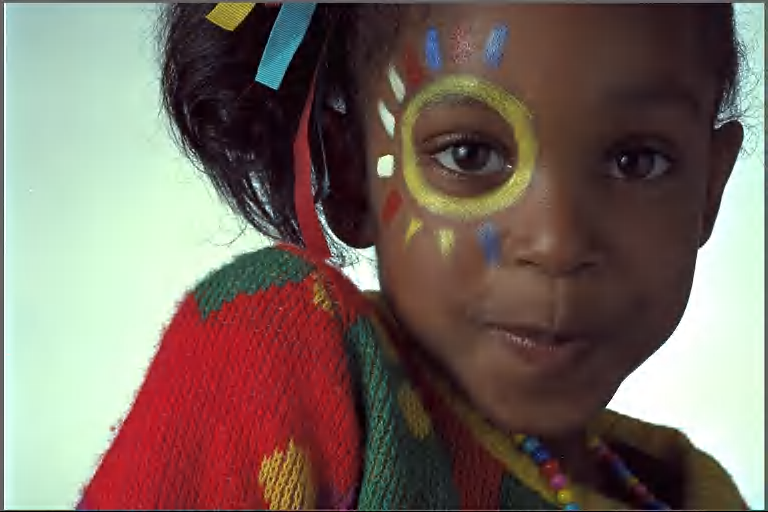}};
			\spy  on (-0.5, 0.2) in node [right] at (-1.5, -1.8);
		\end{tikzpicture} &			
		\begin{tikzpicture}[spy using outlines={circle,magenta,magnification=2,size=1.6cm, connect spies}]
			\node [label={[font=\footnotesize, xshift=0.7cm, yshift=-2.7cm, align=center] $0.250$ bpp\\ $30.74$ dB} ]{\includegraphics[width=0.22\linewidth]{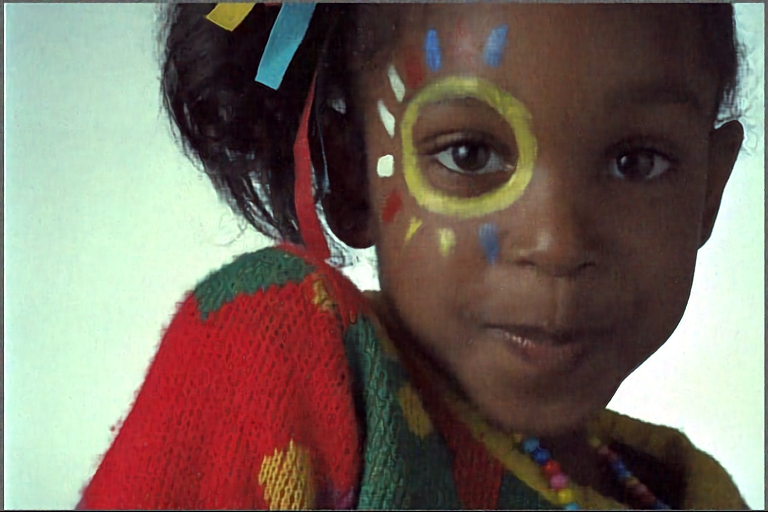}};
			\spy  on (-0.5, 0.2) in node [right] at (-1.5, -1.8);
		\end{tikzpicture} 	\\

		\begin{tikzpicture}[spy using outlines={circle,magenta,magnification=2,size=1.6cm, connect spies}]
			\node{\includegraphics[ width=0.22\linewidth]{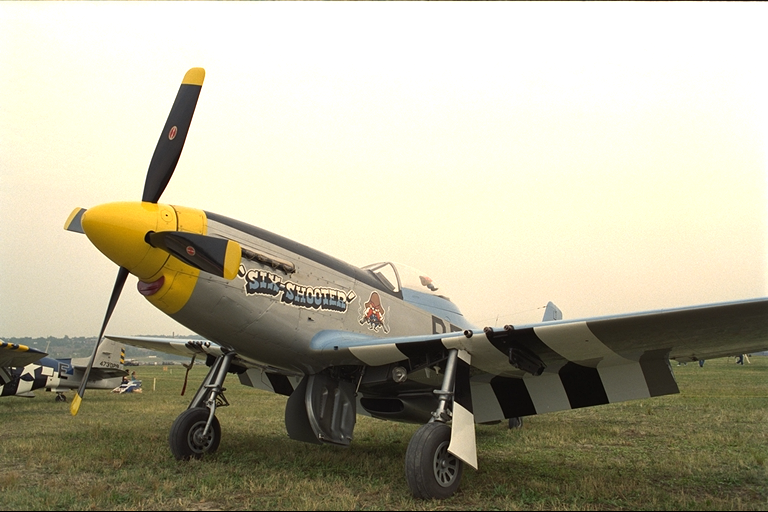}};
			\spy on (-0.7, 0.0) in node [right] at (-1.5, -1.8);%(1.6, 0.7);
		\end{tikzpicture} &
		\begin{tikzpicture}[spy using outlines={circle,magenta,magnification=2,size=1.6cm, connect spies}]
			\node [label={[font=\footnotesize, xshift=0.7cm, yshift=-2.7cm, align=center] $0.740$ bpp\\ $34.42$ dB} ]{\includegraphics[ width=0.22\linewidth]{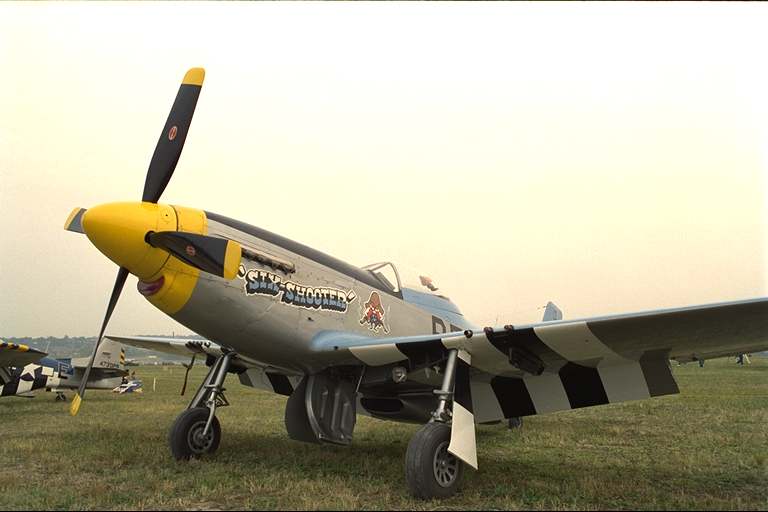}};
			\spy on (-0.7, 0.0) in node [right] at (-1.5, -1.8);%(1.6, 0.7);
		\end{tikzpicture} &			
		\begin{tikzpicture}[spy using outlines={circle,magenta,magnification=2,size=1.6cm, connect spies}]
			\node [label={[font=\footnotesize, xshift=0.7cm, yshift=-2.7cm, align=center] $0.750$ bpp\\ $37.21$ dB} ]{\includegraphics[width=0.22\linewidth]{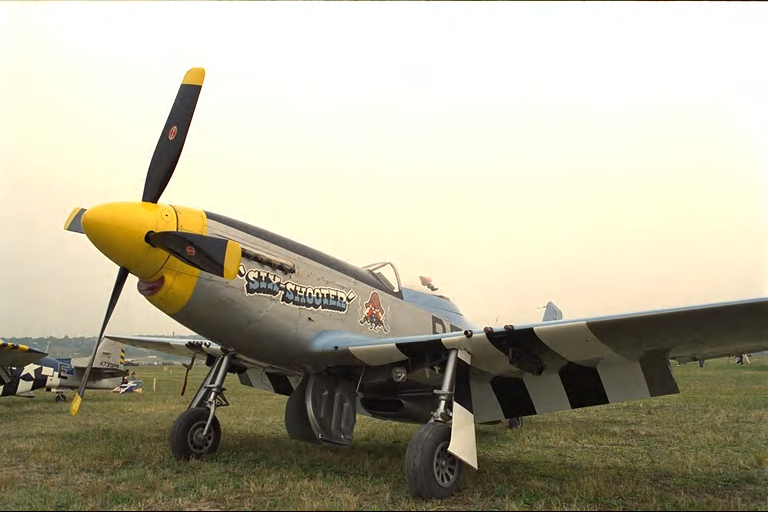}};
			\spy  on (-0.7, 0.0) in node [right] at (-1.5, -1.8);
		\end{tikzpicture} &			
		\begin{tikzpicture}[spy using outlines={circle,magenta,magnification=2,size=1.6cm, connect spies}]
			\node [label={[font=\footnotesize, xshift=0.7cm, yshift=-2.7cm, align=center] $0.729$ bpp\\ $35.03$ dB} ]{\includegraphics[width=0.22\linewidth]{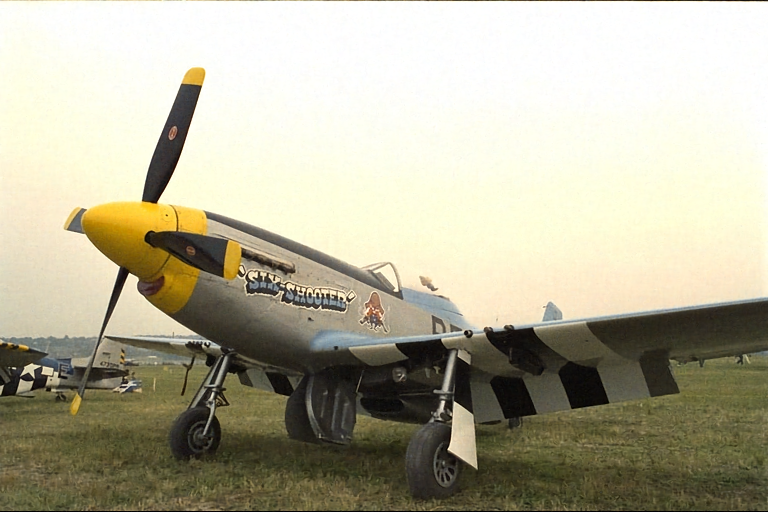}};
			\spy  on (-0.7, 0.0) in node [right] at (-1.5, -1.8);
		\end{tikzpicture} 	\\
	%	\vspace{-5mm}
	\end{tabular}
% 	\vspace{1mm}
\caption{Performance comparison on Kodak using a hidden dimensions of $M=48$ (top),  $M=64$ (middle),  $M=128$ (bottom).}
\vspace{-5mm}
\label{fig:imagecomparisokodak20}
\end{figure*}

%Celeb different M
\begin{figure*}
    \setlength{\linewidth}{\textwidth}
    \setlength{\hsize}{\linewidth}
	\centering
	\renewcommand{\tabcolsep}{0.0pt}
	\renewcommand{\arraystretch}{0}
	\begin{tabular}{cccc}	
	    \footnotesize original & 
		\footnotesize JPEG & 
		\footnotesize JPEG2000 &
		\footnotesize Ours (Meta-learned)  \\	
		\begin{tikzpicture}[spy using outlines={circle,magenta,magnification=1.5,size=1.6cm, connect spies}]
			\node{\includegraphics[ width=0.22\linewidth]{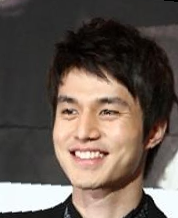}};
			\spy on (-0.3, -0.2) in node [right] at (-1.5, -2.5);%(1.6, 0.7);
		\end{tikzpicture} &
		\begin{tikzpicture}[spy using outlines={circle,magenta,magnification=1.5,size=1.6cm, connect spies}]
			\node [label={[font=\footnotesize, xshift=0.7cm, yshift=-4.2cm, align=center] $0.767$ bpp\\ $34.27$ dB} ]{\includegraphics[ width=0.22\linewidth]{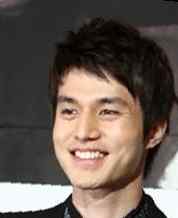}};
			\spy on (-0.3, -0.2) in node [right] at (-1.5, -2.5);%(1.6, 0.7);
		\end{tikzpicture} &			
		\begin{tikzpicture}[spy using outlines={circle,magenta,magnification=1.5,size=1.6cm, connect spies}]
			\node [label={[font=\footnotesize, xshift=0.7cm, yshift=-4.2cm, align=center] $0.775$ bpp\\ $37.50$ dB} ]{\includegraphics[width=0.22\linewidth]{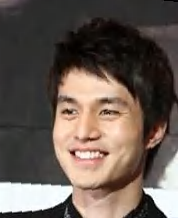}};
			\spy  on (-0.3, -0.2) in node [right] at (-1.5, -2.5);
		\end{tikzpicture} &			
		\begin{tikzpicture}[spy using outlines={circle,magenta,magnification=1.5,size=1.6cm, connect spies}]
			\node [label={[font=\footnotesize, xshift=0.7cm, yshift=-4.2cm, align=center] $0.753$ bpp\\ $36.83$ dB} ]{\includegraphics[width=0.22\linewidth]{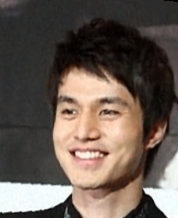}};
			\spy  on (-0.3, -0.2) in node [right] at (-1.5, -2.5);
		\end{tikzpicture} 	\\

		\begin{tikzpicture}[spy using outlines={circle,magenta,magnification=1.5,size=1.6cm, connect spies}]
			\node{\includegraphics[ width=0.22\linewidth]{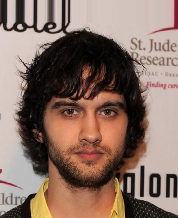}};
			\spy on (-0.3, -0.2) in node [right] at (-1.5, -2.5);%(1.6, 0.7);
		\end{tikzpicture} &
		\begin{tikzpicture}[spy using outlines={circle,magenta,magnification=1.5,size=1.6cm, connect spies}]
			\node [label={[font=\footnotesize, xshift=0.7cm, yshift=-4.2cm, align=center] $1.49$ bpp\\ $31.04$ dB} ]{\includegraphics[ width=0.22\linewidth]{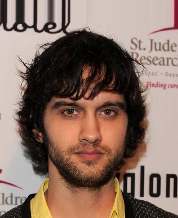}};
			\spy on (-0.3, -0.2) in node [right] at (-1.5, -2.5);%(1.6, 0.7);
		\end{tikzpicture} &			
		\begin{tikzpicture}[spy using outlines={circle,magenta,magnification=1.5,size=1.6cm, connect spies}]
			\node [label={[font=\footnotesize, xshift=0.7cm, yshift=-4.2cm, align=center] $1.50$ bpp\\ $34.87$ dB} ]{\includegraphics[width=0.22\linewidth]{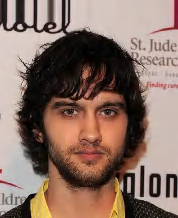}};
			\spy  on (-0.3, -0.2) in node [right] at (-1.5, -2.5);
		\end{tikzpicture} &			
		\begin{tikzpicture}[spy using outlines={circle,magenta,magnification=1.5,size=1.6cm, connect spies}]
			\node [label={[font=\footnotesize, xshift=0.7cm, yshift=-4.2cm, align=center] $1.47$ bpp\\ $33.39$ dB} ]{\includegraphics[width=0.22\linewidth]{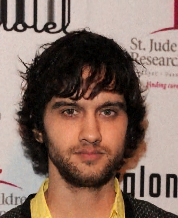}};
			\spy  on (-0.3, -0.2) in node [right] at (-1.5, -2.5);
		\end{tikzpicture} 	\\

		\begin{tikzpicture}[spy using outlines={circle,magenta,magnification=1.5,size=1.6cm, connect spies}]
			\node{\includegraphics[ width=0.22\linewidth]{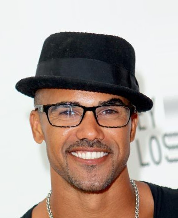}};
			\spy on (-0.3, -0.2) in node [right] at (-1.5, -2.5);%(1.6, 0.7);
		\end{tikzpicture} &
		\begin{tikzpicture}[spy using outlines={circle,magenta,magnification=1.5,size=1.6cm, connect spies}]
			\node [label={[font=\footnotesize, xshift=0.7cm, yshift=-4.2cm, align=center] $2.31$ bpp\\ $37.38$ dB} ]{\includegraphics[ width=0.22\linewidth]{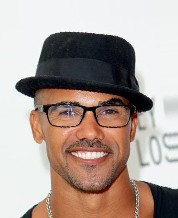}};
			\spy on (-0.3, -0.2) in node [right] at (-1.5, -2.5);%(1.6, 0.7);
		\end{tikzpicture} &			
		\begin{tikzpicture}[spy using outlines={circle,magenta,magnification=1.5,size=1.6cm, connect spies}]
			\node [label={[font=\footnotesize, xshift=0.7cm, yshift=-4.2cm, align=center] $2.40$ bpp\\ $41.91$ dB} ]{\includegraphics[width=0.22\linewidth]{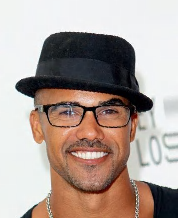}};
			\spy  on (-0.3, -0.2) in node [right] at (-1.5, -2.5);
		\end{tikzpicture} &			
		\begin{tikzpicture}[spy using outlines={circle,magenta,magnification=1.5,size=1.6cm, connect spies}]
			\node [label={[font=\footnotesize, xshift=0.7cm, yshift=-4.2cm, align=center] $2.23$ bpp\\ $39.34$ dB} ]{\includegraphics[width=0.22\linewidth]{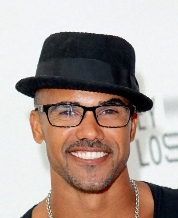}};
			\spy  on (-0.3, -0.2) in node [right] at (-1.5, -2.5);
		\end{tikzpicture} 	\\
	%	\vspace{-5mm}
	\end{tabular}
% 	\vspace{1mm}
\caption{Performance comparison on CelebA using a hidden dimensions of $M=32$ (top),  $M=48$ (middle),  $M=64$ (bottom).}
\vspace{-5mm}
\label{fig:imagecomparisonceleb185853}
\end{figure*}

\subsection{Visualization of compressed 3D shapes}\label{sec:3dshapevisual}
\begin{figure*}
    \setlength{\hsize}{\linewidth}
	\centering
	\renewcommand{\tabcolsep}{0.0pt}
	\renewcommand{\arraystretch}{0.1}
	\begin{tabular}{ccccc}	
	    \footnotesize original & 
		\footnotesize  Draco (7 bit) & 
		\footnotesize  Draco (6 bit) & 
		\footnotesize Ours ($M = 128$) & 
		\footnotesize Ours ($M=64$) \\	
%statuette		
		\begin{tikzpicture}[spy using outlines={circle,magenta,magnification=1.5,size=1.2cm, connect spies}]
			\node [label={[font=\footnotesize, xshift=0cm, yshift=-5cm, align=center] 407.5 MB} ]{\includegraphics[ width=0.19\linewidth, trim={10cm 0 10cm 0},clip]{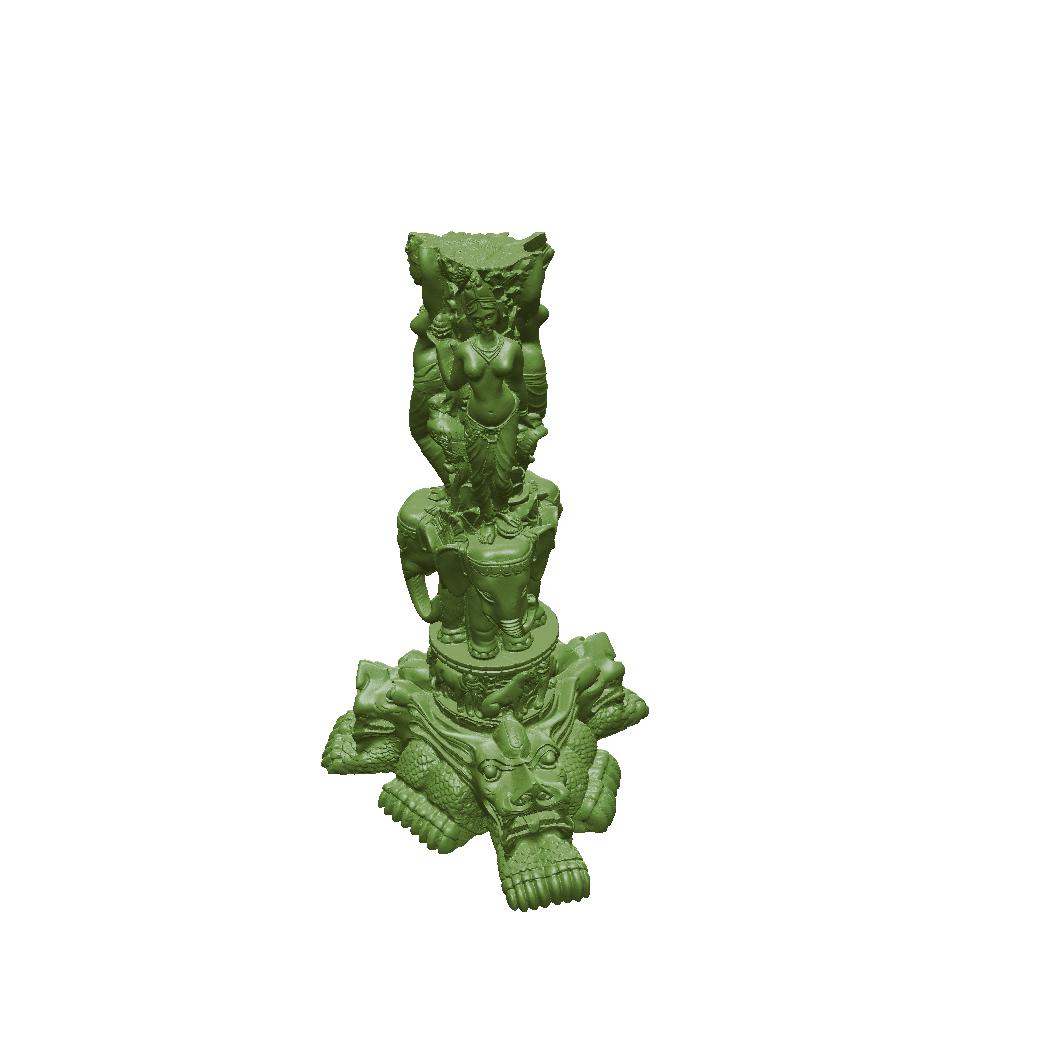}};
% 			\spy on (-0.5, 0.2) in node [right] at (-1.3, -1.8);%(1.6, 0.7);
		\end{tikzpicture} &
		\begin{tikzpicture}[spy using outlines={circle,magenta,magnification=2,size=1.6cm, connect spies}]
			\node [label={[font=\footnotesize, xshift=0cm, yshift=-5cm, align=center] 2.2 MB} ]{\includegraphics[ width=0.19\linewidth, trim={10cm 0 10cm 0},clip]{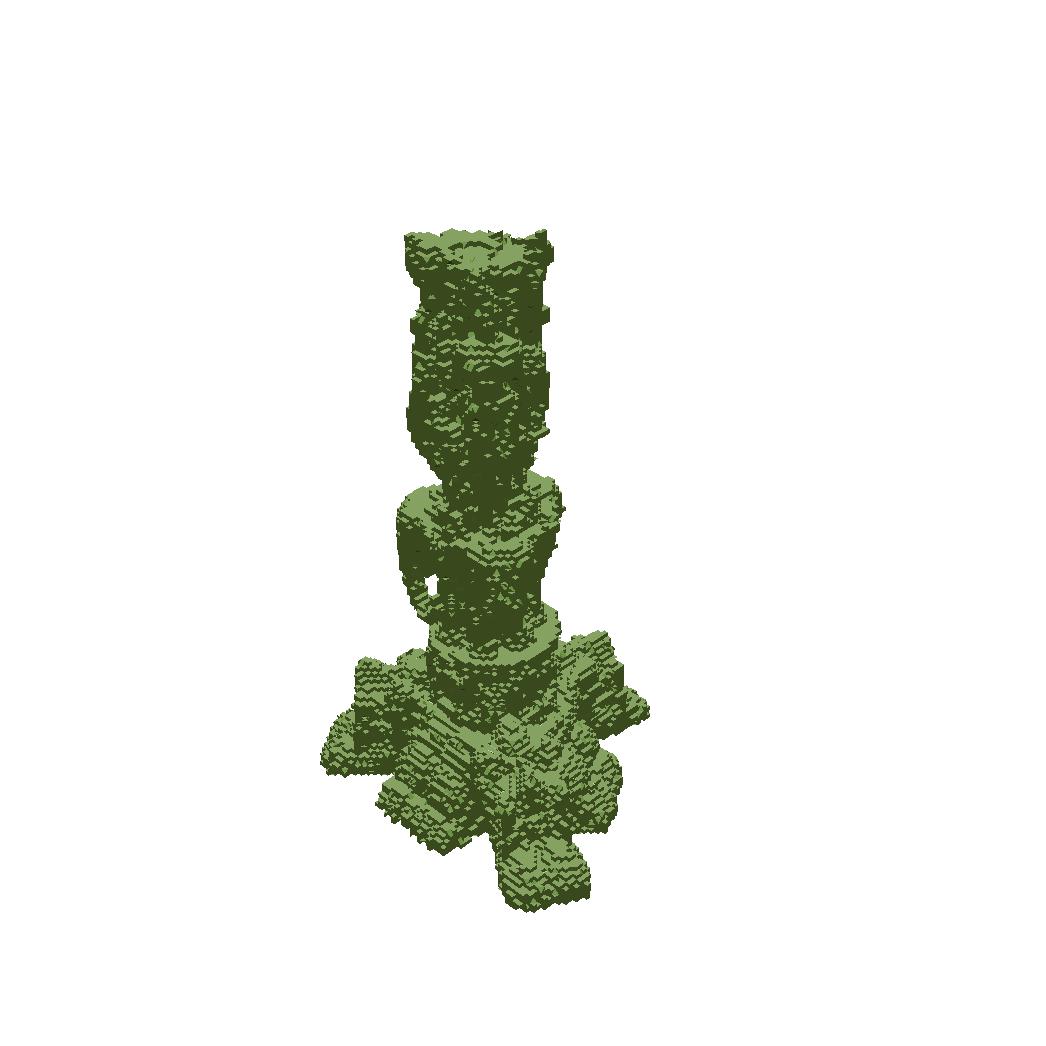}};
% 			\spy on (-0.5, 0.2) in node [right] at (-1.3, -1.8);%(1.6, 0.7);
		\end{tikzpicture} &	
		\begin{tikzpicture}[spy using outlines={circle,magenta,magnification=2,size=1.6cm, connect spies}]
			\node [label={[font=\footnotesize, xshift=0cm, yshift=-5cm, align=center] 1.9 MB} ]{\includegraphics[ width=0.19\linewidth, trim={10cm 0 10cm 0},clip]{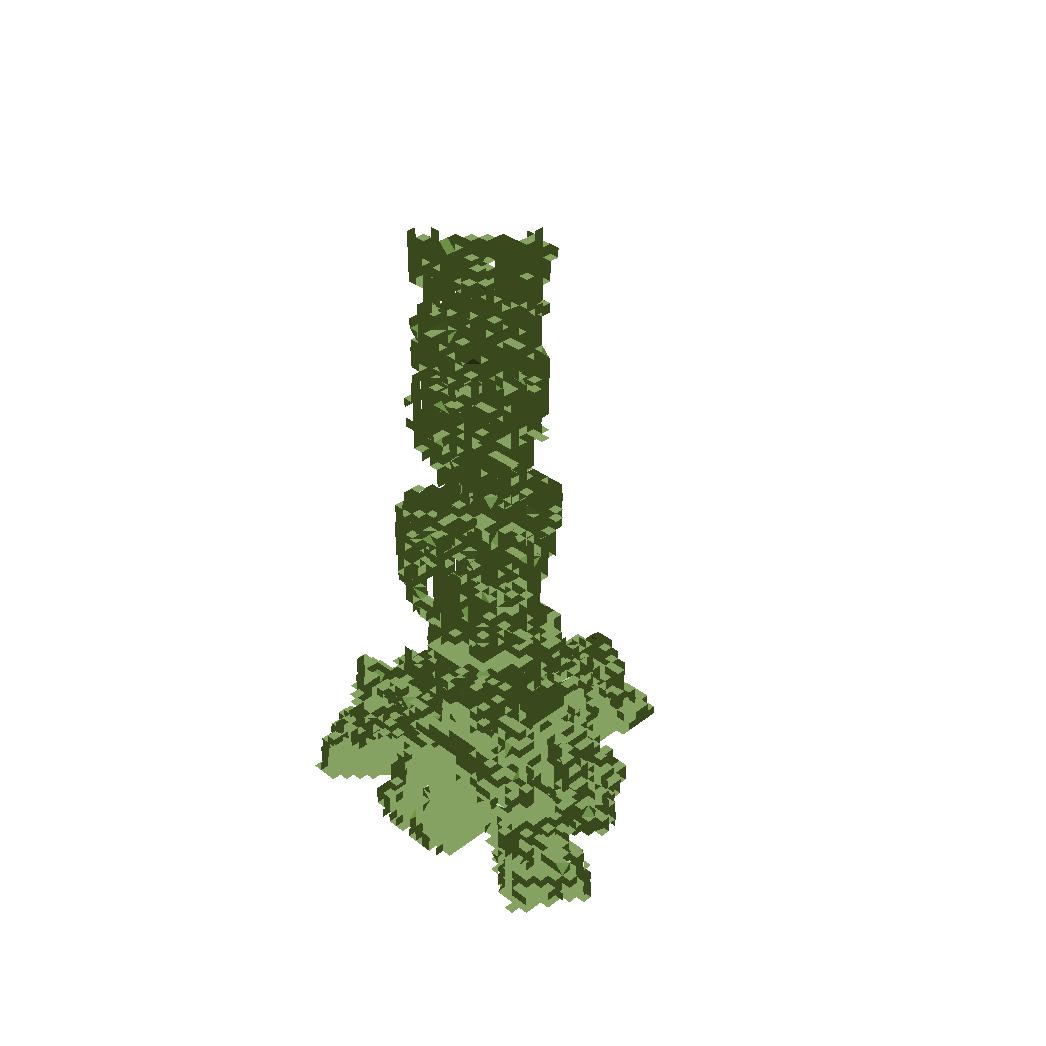}};
% 			\spy on (-0.5, 0.2) in node [right] at (-1.3, -1.8);%(1.6, 0.7);
		\end{tikzpicture} &	
		\begin{tikzpicture}[spy using outlines={circle,magenta,magnification=2,size=1.6cm, connect spies}]
			\node [label={[font=\footnotesize, xshift=0cm, yshift=-5cm, align=center] 57.0 KB} ]{\includegraphics[ width=0.19\linewidth, trim={4.8cm 0 5cm 0},clip]{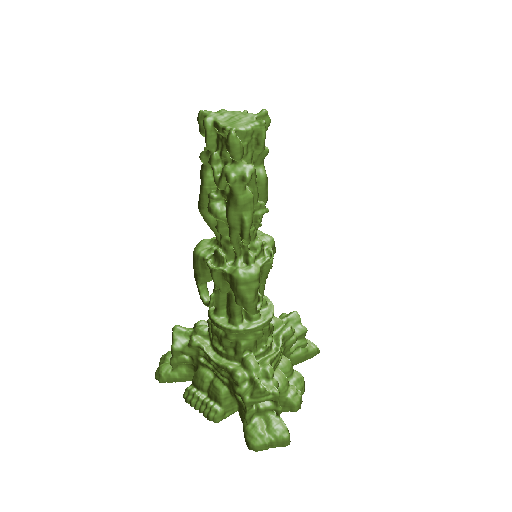}};
% 			\spy  on (-0.5, 0.2) in node [right] at (-1.3, -1.8);
		\end{tikzpicture} &	
		\begin{tikzpicture}[spy using outlines={circle,magenta,magnification=2,size=1.6cm, connect spies}]
			\node [label={[font=\footnotesize, xshift=0cm, yshift=-5cm, align=center] 16.9 KB} ]{\includegraphics[ width=0.19\linewidth, trim={4.8cm 0 5cm 0},clip]{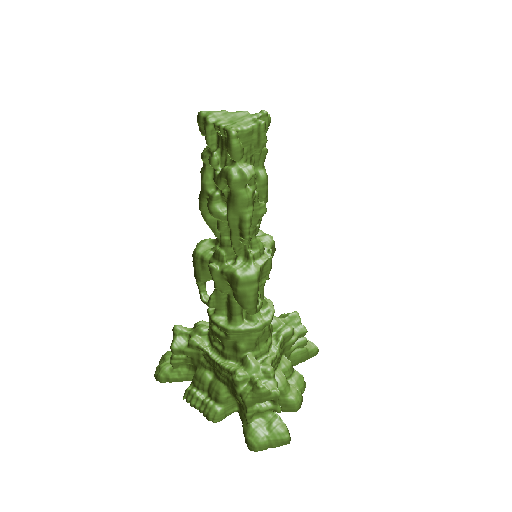}};
% 			\spy  on (-0.5, 0.2) in node [right] at (-1.3, -1.8);
		\end{tikzpicture}	\\
%armadillo		
		\begin{tikzpicture}[spy using outlines={circle,magenta,magnification=1.5,size=1.2cm, connect spies}]
			\node [label={[font=\footnotesize, xshift=0cm, yshift=-4.2cm, align=center]  3.4 MB} ]{\includegraphics[ width=0.19\linewidth, trim={7cm 0 10cm 0},clip]{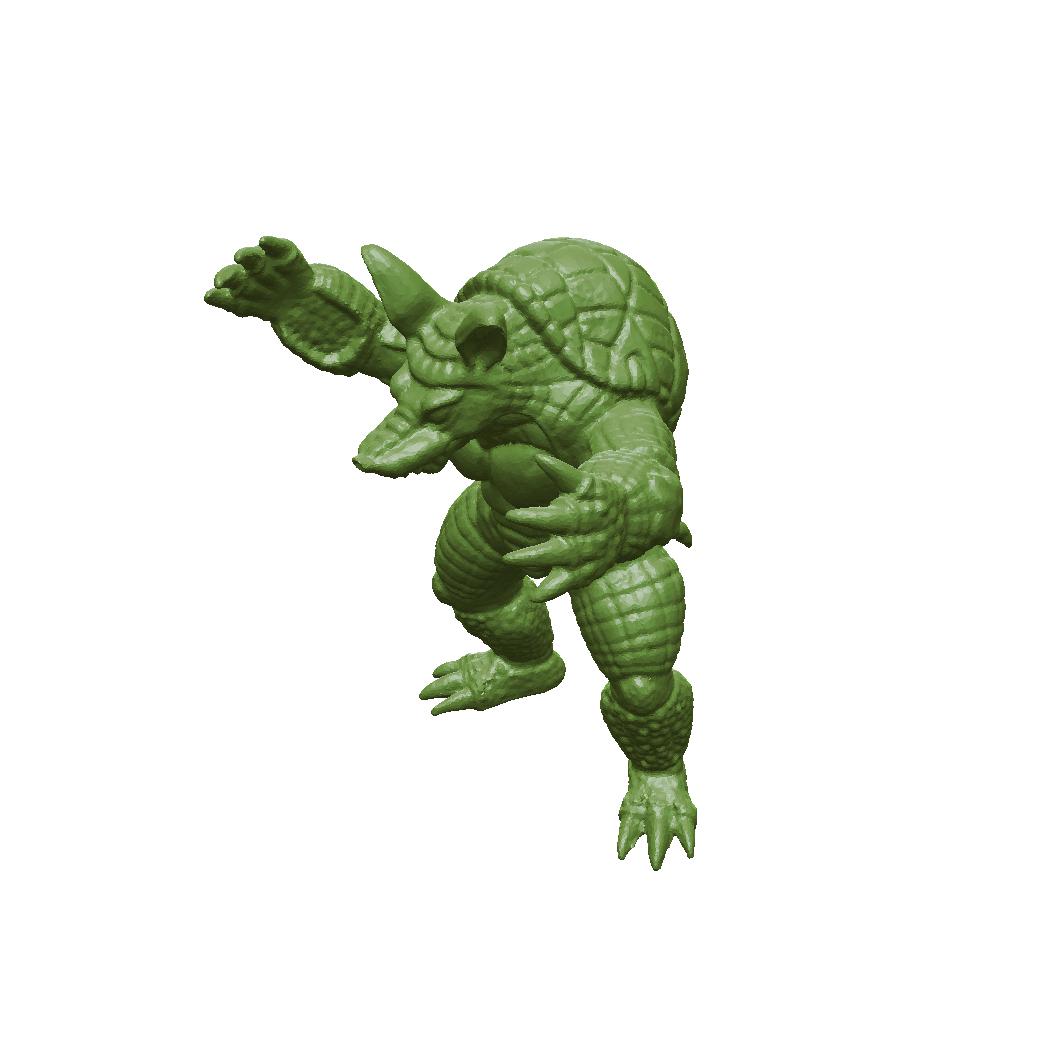}};
% 			\spy on (-0.5, 0.2) in node [right] at (-1.3, -1.8);%(1.6, 0.7);
		\end{tikzpicture} &
		\begin{tikzpicture}[spy using outlines={circle,magenta,magnification=2,size=1.6cm, connect spies}]
			\node [label={[font=\footnotesize, xshift=0cm, yshift=-4.2cm, align=center] 54.8 KB} ]{\includegraphics[ width=0.19\linewidth, trim={7cm 0 10cm 0},clip]{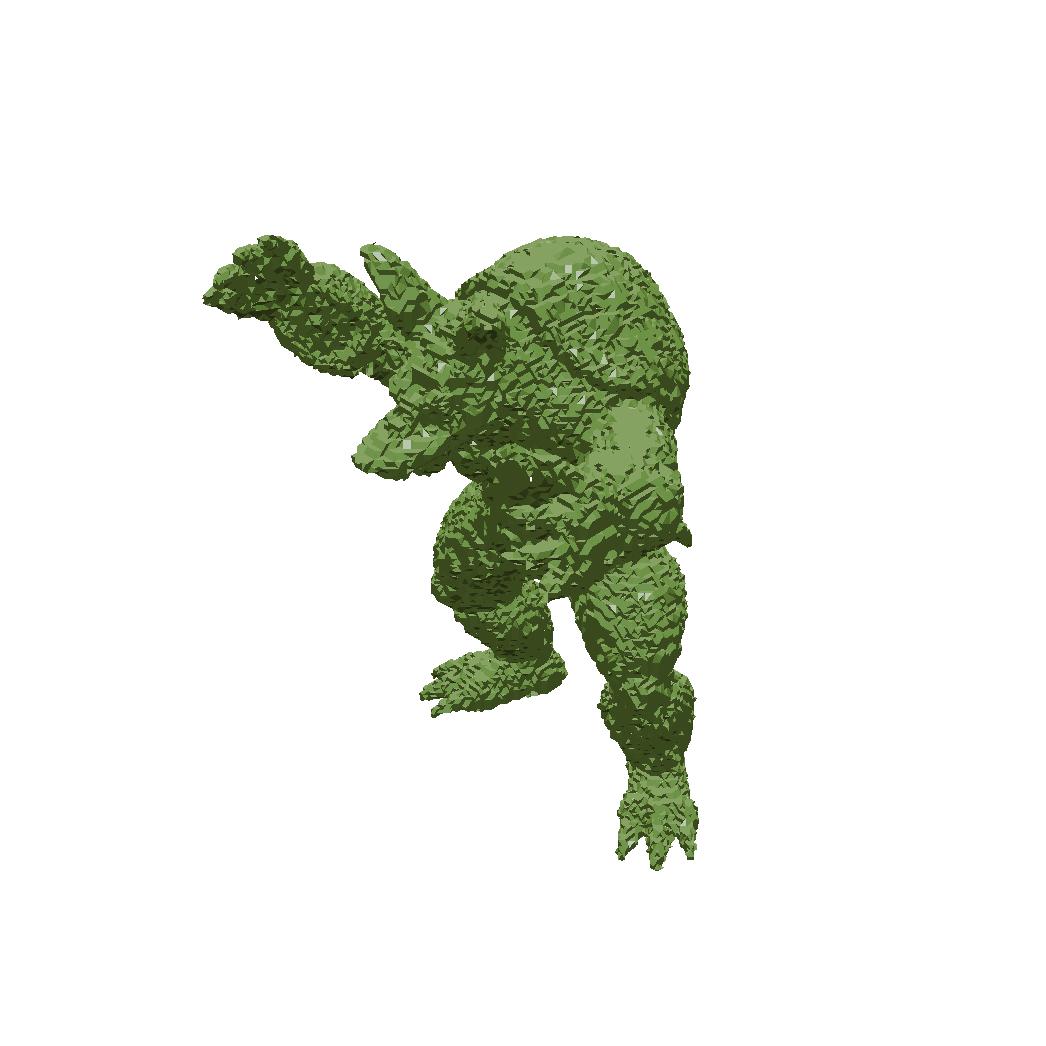}};
% 			\spy on (-0.5, 0.2) in node [right] at (-1.3, -1.8);%(1.6, 0.7);
		\end{tikzpicture} &	
		\begin{tikzpicture}[spy using outlines={circle,magenta,magnification=2,size=1.6cm, connect spies}]
			\node [label={[font=\footnotesize, xshift=0cm, yshift=-4.2cm, align=center] 45.9 KB} ]{\includegraphics[ width=0.19\linewidth, trim={7cm 0 10cm 0},clip]{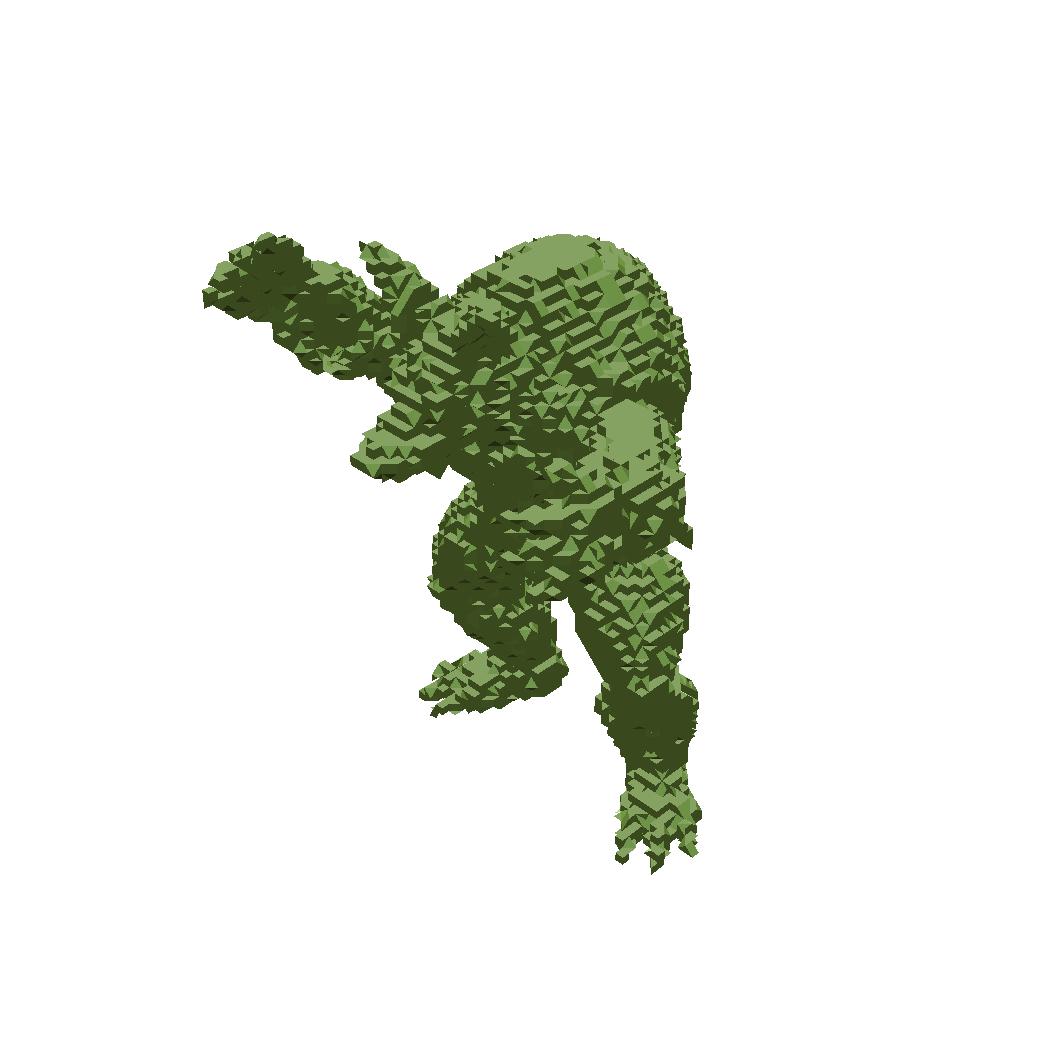}};
% 			\spy on (-0.5, 0.2) in node [right] at (-1.3, -1.8);%(1.6, 0.7);
		\end{tikzpicture} &
		\begin{tikzpicture}[spy using outlines={circle,magenta,magnification=2,size=1.6cm, connect spies}]
			\node [label={[font=\footnotesize, xshift=0cm, yshift=-4.2cm, align=center] 57.4 KB} ]{\includegraphics[ width=0.19\linewidth, trim={3.5cm 0 5cm 0},clip]{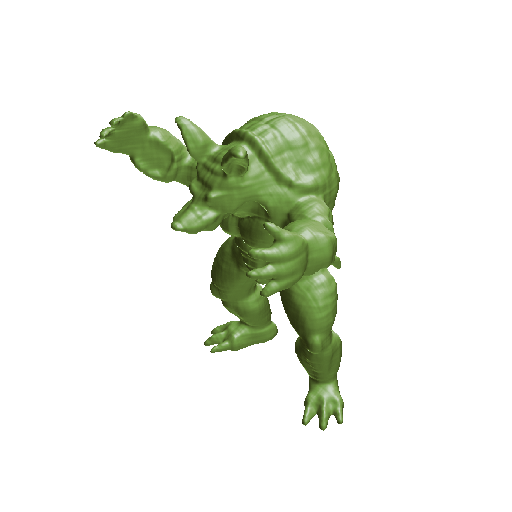}};
% 			\spy  on (-0.5, 0.2) in node [right] at (-1.3, -1.8);
		\end{tikzpicture} &
		\begin{tikzpicture}[spy using outlines={circle,magenta,magnification=2,size=1.6cm, connect spies}]
			\node [label={[font=\footnotesize, xshift=0cm, yshift=-4.2cm, align=center] 17.2 KB} ]{\includegraphics[ width=0.19\linewidth, trim={3.5cm 0 5cm 0},clip]{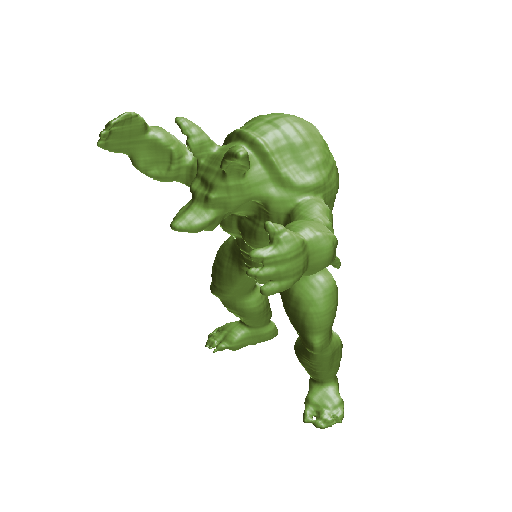}};
% 			\spy  on (-0.5, 0.2) in node [right] at (-1.3, -1.8);
		\end{tikzpicture}	\\
%dragon
		\begin{tikzpicture}[spy using outlines={circle,magenta,magnification=1.5,size=1.2cm, connect spies}]
			\node [label={[font=\footnotesize, xshift=0cm, yshift=-3.5cm, align=center] 32.8 MB} ]{\includegraphics[ width=0.19\linewidth, trim={6.5cm 0 7cm 0},clip]{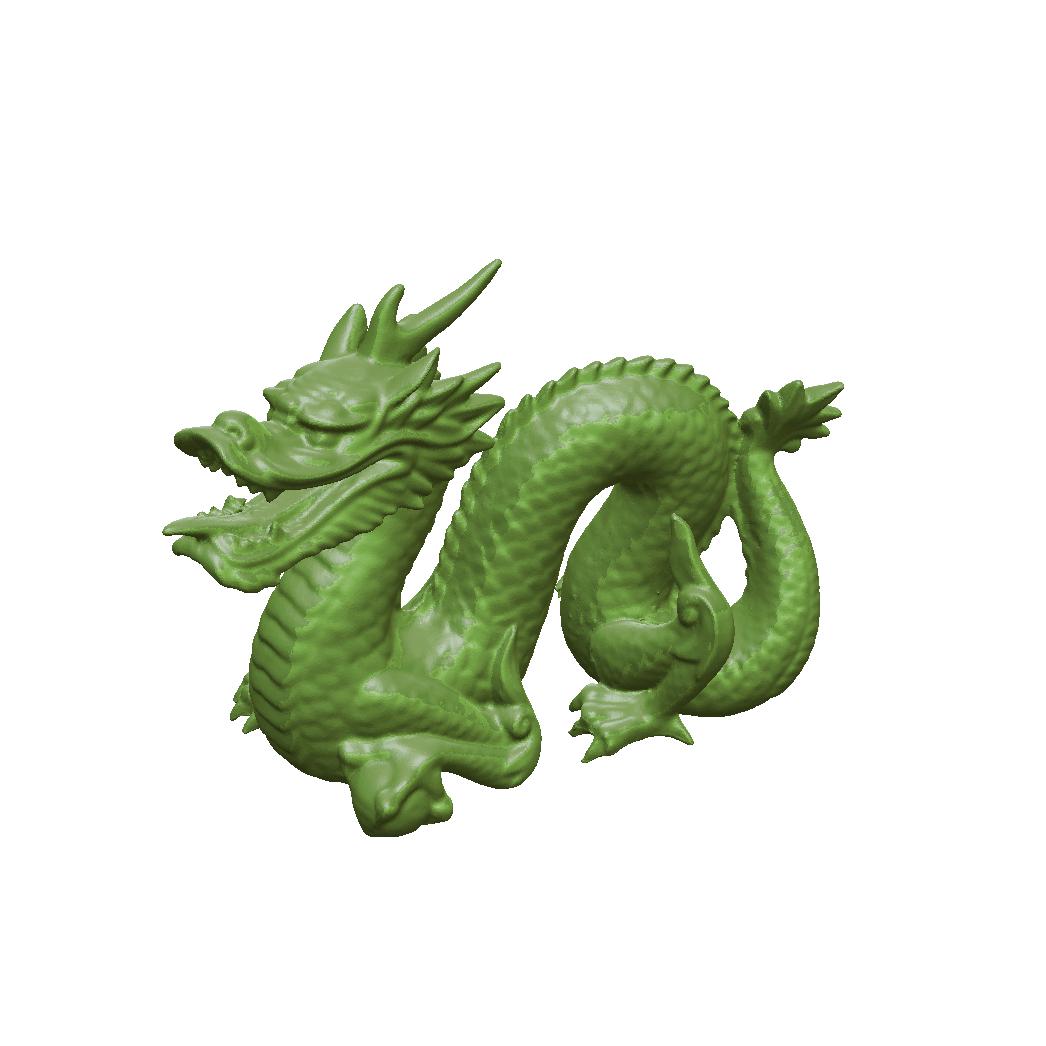}};
% 			\spy on (-0.5, 0.2) in node [right] at (-1.3, -1.8);%(1.6, 0.7);
		\end{tikzpicture} &
		\begin{tikzpicture}[spy using outlines={circle,magenta,magnification=2,size=1.6cm, connect spies}]
			\node [label={[font=\footnotesize, xshift=0cm, yshift=-3.5cm, align=center] 283.0 KB} ]{\includegraphics[ width=0.19\linewidth, trim={6.5cm 0 7cm 0},clip]{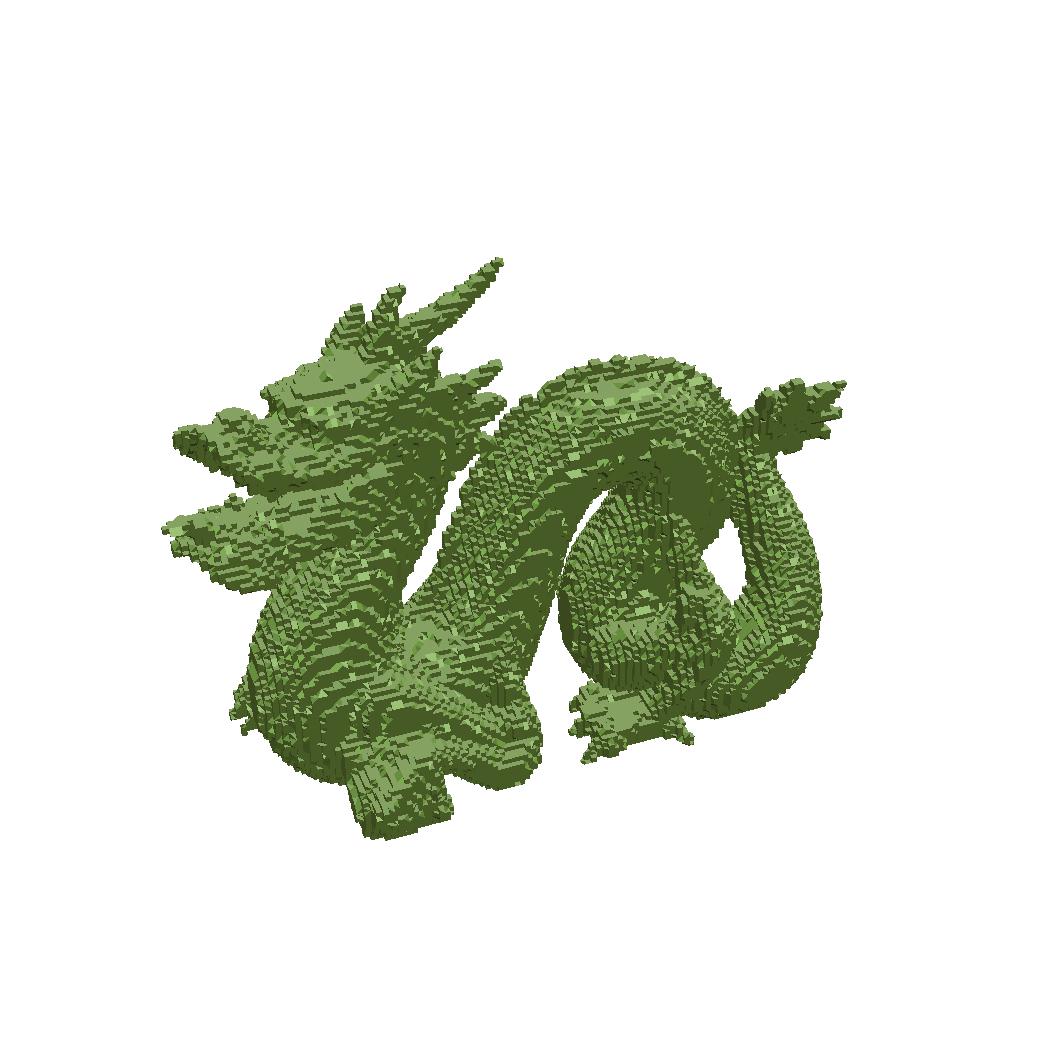}};
% 			\spy on (-0.5, 0.2) in node [right] at (-1.3, -1.8);%(1.6, 0.7);
		\end{tikzpicture} &	
		\begin{tikzpicture}[spy using outlines={circle,magenta,magnification=2,size=1.6cm, connect spies}]
			\node [label={[font=\footnotesize, xshift=0cm, yshift=-3.5cm, align=center] 231.7 KB} ]{\includegraphics[ width=0.19\linewidth, trim={6.5cm 0 7cm 0},clip]{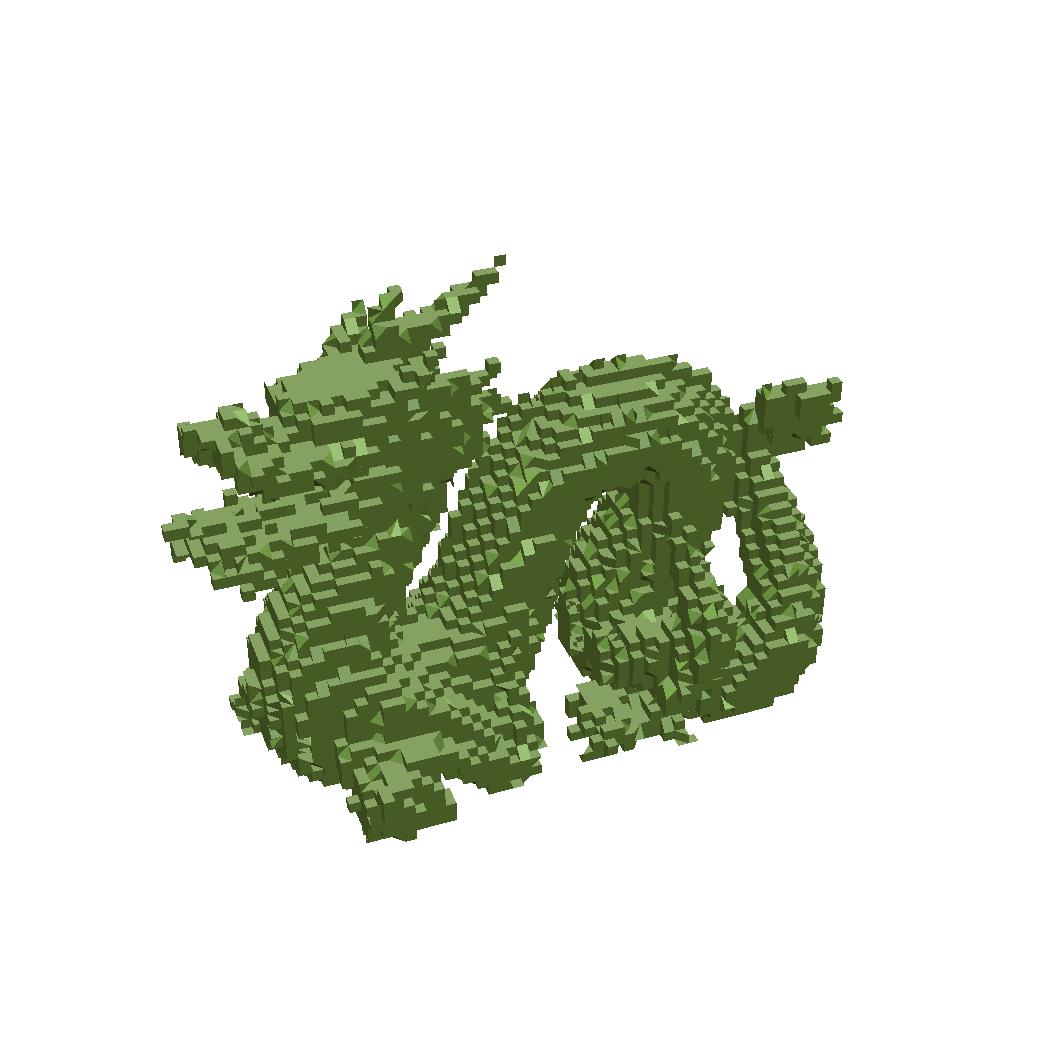}};
% 			\spy on (-0.5, 0.2) in node [right] at (-1.3, -1.8);%(1.6, 0.7);
		\end{tikzpicture} &
		\begin{tikzpicture}[spy using outlines={circle,magenta,magnification=2,size=1.6cm, connect spies}]
			\node [label={[font=\footnotesize, xshift=0cm, yshift=-3.5cm, align=center] 54.8 KB} ]{\includegraphics[ width=0.19\linewidth, trim={3.25cm 0 3.5cm 0},clip]{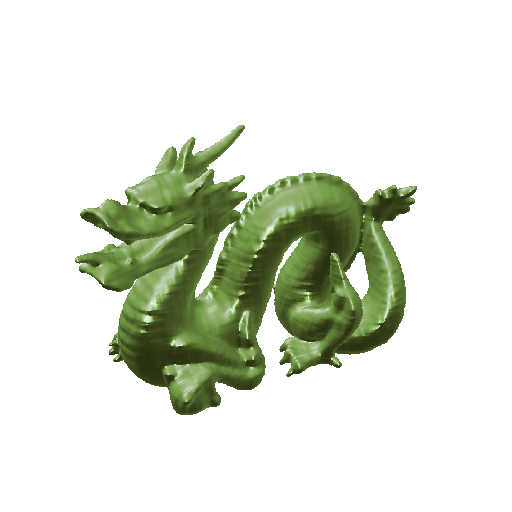}};
% 			\spy  on (-0.5, 0.2) in node [right] at (-1.3, -1.8);
		\end{tikzpicture} &
		\begin{tikzpicture}[spy using outlines={circle,magenta,magnification=2,size=1.6cm, connect spies}]
			\node [label={[font=\footnotesize, xshift=0cm, yshift=-3.5cm, align=center] 16.6 KB} ]{\includegraphics[ width=0.19\linewidth, trim={3.25cm 0 3.5cm 0},clip]{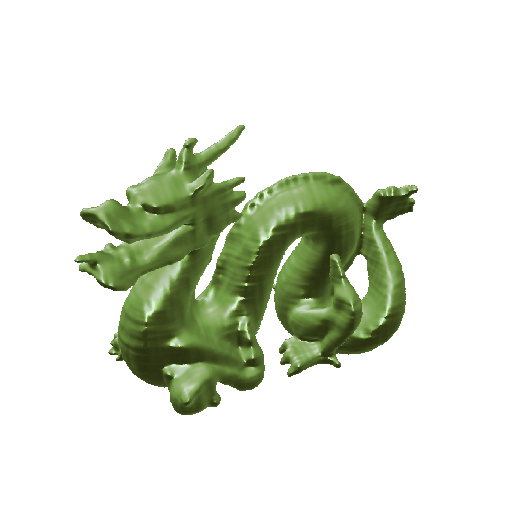}};
% 			\spy  on (-0.5, 0.2) in node [right] at (-1.3, -1.8);
		\end{tikzpicture}	\\
	%	\vspace{-5mm}
	\end{tabular}
% 	\vspace{1mm}
\caption{Visual comparsion of the mesh compression algorithm Draco compared to our method applied to 3D shape compression. We compare 2 quantization setting for Draco, namely 6 and 7 bit, and two hidden dimensions $M=64, 128$ using our method. Our method shows a significantly smoother surface reconstruction and better detail at similar or lower storage.}
\vspace{-5mm}
\label{fig:3dcomparison}
\end{figure*}

We demonstrate the effectiveness of \acp{inr} for 3D shape compression by visual comparison in \myfigref{fig:3dcomparison}. The SDFs learned by the \acp{inr} in general render a much smoother surface than the mesh compression algorithm Draco, while being very storage efficient. Draco introduces significant surface noise making the compressed shape very rough. The reduction of information for \acp{inr} is more faithful in that the encoded shape is a simplified version of the original: The smaller model with $M=64$, that requires only roughly a quarter of the storage of the model with $M=128$, still looks very much like the original with certain details smoothed out. This makes it visually much more pleasing than the rough looking Draco compression, where the details are lost through surface noise.
%3D shape

\end{document}